\documentclass[11pt,a4paper]{article}
\usepackage{jheppub}
\usepackage{amsmath}
\allowdisplaybreaks[4]
\usepackage{amssymb}
\usepackage{graphicx}
\usepackage{booktabs}
\usepackage{bm}
\usepackage{multirow}
\usepackage{psfrag}
\usepackage[normalem]{ulem}
\usepackage{xcolor}
\usepackage{overpic}
\usepackage[utf8x]{inputenc}
\makeatletter
\def\@fpheader{~}
\makeatother

\newcommand{\spac}{{\hspace{0.3mm}}}

\title{Factorization of Non-Global LHC Observables and Resummation of Super-Leading Logarithms}

\author[a]{Thomas Becher,}
\author[b,c]{Matthias Neubert,}
\author[d]{Ding Yu Shao,}
\author[b]{and Michel Stillger}
\affiliation[a]{Albert Einstein Center for Fundamental Physics, Institut f\"ur Theoretische Physik,\\ 
Universit\"at Bern, Sidlerstrasse 5, CH-3012 Bern, Switzerland}
\affiliation[b]{PRISMA Cluster of Excellence {\em \&} Mainz Institute for Theoretical Physics,\\ 
Johannes Gutenberg University, 55099 Mainz, Germany}
\affiliation[c]{Department of Physics, LEPP, Cornell University, Ithaca, NY 14853, U.S.A.}
\affiliation[d]{Department of Physics, Center for Field Theory and Particle Physics {\em \&} Key Laboratory of Nuclear 
Physics and Ion-beam Application (MOE), Fudan University, Shanghai, 200433, China}

\emailAdd{becher@itp.unibe.ch}
\emailAdd{matthias.neubert@uni-mainz.de}
\emailAdd{dyshao@fudan.edu.cn}
\emailAdd{m.stillger@uni-mainz.de}

\date{July 12, 2023}

\preprint{\begin{flushright}
MITP-23-032\\ 
July 12, 2023
\end{flushright}}

\abstract{We present a systematic formalism based on a factorization theorem in soft-collinear effective theory to describe non-global observables at hadron colliders, such as gap-between-jets cross sections. The cross sections are factorized into convolutions of hard functions, capturing the dependence on the partonic center-of-mass energy $\sqrt{\hat s}$, and low-energy matrix elements, which are sensitive to the low scale $Q_0\ll\sqrt{\hat s}$ characteristic of the veto imposed on energetic emissions into the gap between the jets. The scale evolution of both objects is governed by a renormalization-group equation, which we derive at one-loop order. By solving the evolution equation for the hard functions for arbitrary $2\to M$ jet processes in the leading logarithmic approximation, we accomplish for the first time the all-order resummation of the so-called ``super-leading logarithms'' discovered in 2006, thereby solving an old problem of quantum field theory. We study the numerical size of the corresponding effects for different partonic scattering processes and explain why they are sizable for $pp\to 2\,\text{jets}$ processes, but suppressed in $H/Z$ and $H/Z$\,+\,jet production. The super-leading logarithms are given by an alternating series, whose individual terms can be much larger than the resummed result, even in very high orders of the loop expansion. Resummation is therefore essential to control these effects. We find that the asymptotic fall-off of the resummed series is much weaker than for standard Sudakov form factors.}

\begin{document}

\maketitle

\section{Introduction}

Jet observables are a crucial tool to extract information about the underlying hard interactions in high-energy processes and are used in a wide range of physics analyses. The definition of an $M$-jet cross section involves clustering energetic radiation into jets and a veto criterion ensuring that the remaining radiation outside the jets is soft. This veto leads to an intricate pattern of logarithmically enhanced corrections in perturbation theory. 

The simplest observable to discuss the effects of the veto are gap-between-jets cross sections, where one vetoes radiation above a low scale $Q_0$ in a region outside the jets, which themselves carry large energy $Q\gg Q_0$, with $Q$ of order the center-of-mass energy. For $e^+ e^-$ collisions with large jet radius $R\sim 1$ the leading logarithmic (LL) effects in such observables are of the form $\alpha_s^n L^n$, where $L\sim\ln(Q/Q_0)$. Leading logarithms arise when soft gluons are emitted from the primary hard partons produced in the collision, but Dasgupta and Salam observed that also soft gluons emitted off secondary emissions inside jets produce leading-logarithmic contributions~\cite{Dasgupta:2001sh}. They called the large logarithms arising from the secondary emissions non-global logarithms (NGLs), since they arise whenever an observable vetoes soft radiation only in a restricted phase-space region rather than globally. Even at leading-logarithmic  order, the non-global contributions have a quite complicated structure. In the large-$N_c$ limit, they can be resummed using a dedicated parton shower or by solving the BMS equation, a non-linear integral equation derived by Banfi, Marchesini and Smye~\cite{Banfi:2002hw}. A generalization of this equation to finite $N_c$ was obtained in~\cite{Weigert:2003mm} based on a mapping between the JIMWLK~\cite{Jalilian-Marian:1997qno,Weigert:2000gi,Ferreiro:2001qy} and BK~\cite{Balitsky:1995ub,Kovchegov:1999ua} evolution equations for small-$x$ dynamics. Using this approach, numerical results for NGLs at $N_c=3$ were obtained in~\cite{Hatta:2013iba,Hagiwara:2015bia, Hatta:2020wre}. Over the past few years, also the parton-shower method for performing the resummation has been formulated at the amplitude level and extended to finite $N_c$~\cite{Platzer:2013fha,AngelesMartinez:2018cfz,Forshaw:2019ver}, so that by now numerical results for NGLs at $N_c=3$ are also available in this framework~\cite{AngelesMartinez:2018cfz,Forshaw:2019ver,DeAngelis:2020rvq}, following earlier approximate treatments of effects of subleading order in $1/N_c$~\cite{Nagy:2015hwa,Nagy:2019bsj,Nagy:2019rwb}. Recently, also first resummations of subleading NGLs in the large-$N_c$ limit were achieved~\cite{Banfi:2021xzn,Banfi:2021owj,Becher:2021urs,Becher:2023vrh}.

At hadron colliders, there is a second mechanism that produces an interesting set of logarithmically enhanced contributions, related to the presence of complex phase factors with non-trivial color structure, which can prevent the cancellation of soft$\,+\,$collinear contributions among real and virtual corrections. This non-cancellation leads to double-logarithmic effects starting at four-loop order, i.e., one finds that the leading-logarithmic terms in the perturbative series  are of the form $\alpha_s^{3+n} L^{3+2n}$. The existence of these double-logarithmic corrections appears surprising at first sight, since they are not tied to a small angular scale but arise in an observable that superficially only involves wide-angle soft dynamics. Forshaw, Kyrieleis and Seymour~\cite{Forshaw:2006fk}, who discovered this effect, therefore called these terms ``super-leading" logarithms (SLLs). The complex phase factors responsible for the SLLs also induce collinear factorization violation in processes with both incoming and outgoing partons~\cite{Catani:2011st,Forshaw:2012bi,Schwartz:2017nmr}. These effects vanish in the large-$N_c$ limit and are not captured by traditional probabilistic parton showers. Gap-between-jets cross sections at hadron colliders are therefore examples of observables where traditional showers do not even capture the leading double-logarithmic terms. Due to the complicated color structures involved, the early papers on SLLs have restricted themselves to the computation of the four-loop $\alpha_s^4 L^5$ term for the $q q'\to q q'$ partonic contribution to dijet production~\cite{Forshaw:2006fk,Forshaw:2008cq}. In later work, additional partonic channels were analyzed, and also the five-loop $\alpha_s^5 L^7$ terms were extracted~\cite{Keates:2009dn}. A numerical analysis for the $t$-channel diagrams including a partial resummation of some higher-order contributions found that the effect of SLLs on the gaps-between-dijets cross sections could amount to as much as 15\%~\cite{Forshaw:2009fz}. However, a complete phenomenological analysis of SLLs including interference effects of different Feynman diagrams in the Born-level amplitude has never been performed. More importantly, the all-order structure of the SLLs remained completely unknown. In parameter regions where the SLLs give rise to significant corrections to physical cross sections, the large double logarithms are numerically important, and hence an all-order resummation of their contributions becomes mandatory.

In~\cite{Becher:2015hka,Becher:2016mmh} we have developed an effective field-theory framework for non-global observables at $e^+e^-$ colliders and used it to derive factorization theorems for a variety of relevant jet cross sections. The framework is based on soft-collinear effective theory (SCET)~\cite{Bauer:2001yt,Bauer:2002nz,Beneke:2002ph} and factorizes the cross sections into hard and soft functions. The hard functions correspond to the squared $e^+ e^-$ scattering amplitudes into $m$ hard partons at fixed directions inside the final-state jets. The soft functions are given by matrix elements of Wilson lines along these directions. In the effective theory, the NGLs can be resummed by solving renormalization-group (RG) equations. This approach has provided a new way to think about NGLs by showing that they -- like any other large logarithms -- result from RG evolution between two hierarchical energy scales. It has therefore reformulated this complicated problem in a language familiar from other applications of effective field theories. Our RG equation is structurally simpler than the non-linear BMS integral equation. Yet, its solution is still a challenging task, since the anomalous dimension entering the evolution equation is an operator not only in the large color space of the incoming and outgoing particles, but also in the infinite space of particle multiplicities.

A similar factorization formula holds for hadron-collider observables~\cite{Balsiger:2018ezi,Becher:2021zkk}. The main complication is that the hard functions then also involve two incoming hard partons, and the low-energy matrix elements contain collinear fields associated with these initial-state partons. While these modifications seem obvious, they have profound effects on the structure of the anomalous dimension and the associated RG evolution. In the lepton-collider case, the individual hard functions involve collinear singularities, but the singular terms cancel among different hard functions when adding up the contributions from different multiplicities~\cite{Becher:2021urs}. In contrast, in the hadron-collider case the presence of the initial-state partons leads to a non-trivial collinear RG evolution. A second important difference is the appearance of complex phase factors in the soft anomalous dimension, which result from the exchange of Glauber gluons.\footnote{In the literature, the phase terms are sometimes called Coulomb phases.} 
These cancel by color conservation for $e^+e^-$ collisions, but they give rise to non-vanishing effects for processes with two color-charged initial-state partons. The Glauber phases spoil the cancellation of soft$\,+\,$collinear terms in the evolution, which leads to double-logarithmic corrections in higher orders -- the aforementioned SLLs. Also, in the presence of Glauber phases, the collinear anomalous dimension involves real and virtual parts with different color structures, rather than the usual parton distribution function (PDF) evolution. In the first part of this paper, we present the factorization theorem for non-global hadron-collider observables (Section~\ref{sec:factorization}) and then provide a detailed derivation of the one-loop anomalous dimension, with a particular focus on its collinear part (Section~\ref{sec:gamma}).

By iterating the one-loop anomalous-dimension matrix $n$ times, one can calculate the leading-logarithmic terms at the $n$-th order in perturbation theory. Due to some key properties of different parts of the anomalous dimension, only a specific subset of contributions to these products generates the SLLs in the leading-logarithmic approximation, as we show in Section~\ref{sec:lltraces}. In~\cite{Becher:2021zkk} we have analyzed the relevant color structures for arbitrary partonic scattering processes in which the two colliding partons are quarks or anti-quarks and obtained a closed formula for the SLLs at the $n$-th order in perturbation theory. The result was written in terms of four color structures depending on the Born-level hard functions for the process. Furthermore, we found that the $n$ dependence was simple enough that (ignoring the running of the strong coupling) the perturbative series could be resummed into a closed-form expression. The simplest example is the scattering of two quarks with different flavors, $q q'\to q q'$, mediated by a color-singlet exchange in the $t$-channel, with a gap of size $\Delta Y$ at central rapidity in which the radiation is vetoed. For this case, the resummation of the infinite tower of SLL contributions to the partonic cross section leads to the result~\cite{Becher:2021zkk}
\begin{equation}\label{eq:singletres}
   \hat\sigma_{qq'\to qq'}^{\rm SLL} 
   = - \hat{\sigma}_{qq'\to qq'}\,\frac{4C_F}{3} \left( \frac{\alpha_s}{\pi} \right)^3 \pi^2 L^3\spac
    \Delta Y\,{}_2F_2\big(1,1;2,\mbox{$\frac52$};-w\big) \,,
\end{equation}
where $\hat{\sigma}_{qq'\to qq'}$ is the Born cross section, and $w=\frac{N_c\alpha_s}{\pi}\,L^2$ encodes the double-logarithmic dependence. There are several general features of this result worth pointing out:
\begin{enumerate}
\item 
Using ${}_2F_2(1,1;2,\mbox{$\frac52$};0)=1$, one finds that the Glauber exchange mechanism yields a first contribution already at three-loop (not four-loop) order. At this order there are three powers of $L$, so the three-loop term is not ``super-leading'' in the strict sense of the word. Nevertheless, the presence of the squared Glauber phase $|i\pi|^2$ implies a sizable enhancement factor.
\item 
The series expansion of the hypergeometric function in the variable $w$ has alternating signs. As we will show, this is a general feature of the series of SLLs. For realistic values $w\gtrsim\mathcal{O}(1)$, this leads to significant cancellations between different terms in the series. Indeed, in the asymptotic limit $w\gg 1$ one finds 
\begin{equation}
   {}_2F_2\big(1,1;2,\mbox{$\frac52$};-w\big) \sim \frac{\ln w}{w} \,,
\end{equation}
corresponding to a fall-off of the resummed series. Interestingly, this fall-off is much weaker than for the standard Sudakov form factor, which in the variable $w$ takes the form $e^{-c w}$ with some constant $c$.
\item 
The color factor in front of the hypergeometric function is $C_F\sim N_c$, so that the SLLs are suppressed by $1/N_c^2$ in the large-$N_c$ limit. (Recall that $N_c\spac\alpha_s\sim 1$ remains finite in this limit.) This suppression can be understood by noting that the Glauber phases are an interference effect associated with a non-trivial operator in color space, whose contributions are thus absent in the large-$N_c$ limit.
\item 
For reasonable values of parameters one finds that the variables $w=\frac{N_c\alpha_s}{\pi}\,L^2$ and $w_\pi=\frac{N_c\alpha_s}{\pi}\,\pi^2$ are both of $\mathcal{O}(1)$. The result~\eqref{eq:singletres} then behaves like a one-loop correction $\sim\frac{\alpha_s L}{N_c}$ to the partonic cross section. The potentially sizable effect of the complex phase terms in double-logarithmic observables has been pointed out long ago in the context of Drell--Yan and Higgs production~\cite{Magnea:1990zb,Ahrens:2008qu,Ahrens:2009cxz}. In future work, it will be interesting to consider the impact of higher-order terms in the variable $w_\pi$. 
\end{enumerate}

The quark-initiated processes considered in~\cite{Becher:2021zkk} are relatively simple, since arbitrary products of color generators in the fundamental representation can be reduced to structures linear in the generators. In the present paper, we discuss the general case of a hard-scattering process with arbitrary initial-state partons (quarks, anti-quarks, gluons, or even exotic objects transforming in different representations of $SU(N_c)$). In Section~\ref{sec:integrals_resummation}, we carry out the all-order resummation for the general case and show that (at fixed coupling $\alpha_s$) this leads certain Kamp\'e de F\'eriet functions, for which we provide several explicit representations. The relevant color algebra is much more involved in the general case, but we prove in Section~\ref{sec:traces} that, for any hard process, the SLLs can be expressed in terms of seven linear combinations of ten basic color traces. A full phenomenological analysis of SLLs is beyond the scope of the present paper, but in Section~\ref{sec:simpleprocesses} we provide analytical and numerical results for several simple partonic scattering processes, in particular the ones relevant for $Z$- and Higgs-boson production, also in association with a jet. While the SLLs are numerically suppressed in $2\to 0$ and $2 \to 1$ partonic processes for reasons that we will elucidate, we find numerically significant effects for $2\to 2$ processes. In the latter case several color configurations contribute to a given partonic channel, and we show that the interference between different configurations leads to SLL effects depending on the Born-level kinematics. As with any calculation relying on the leading double-logarithmic approximation, our results suffer from large uncertainties due to neglected higher-order terms. An important example of such a higher-order effect is the scale dependence of the strong coupling $\alpha_s(\mu)$. However, we can take this particular effect into account when solving the RG equation order by order and compare with the fixed-coupling results. Before concluding, we discuss in Section~\ref{sec:discussion} the systematics of the expansion and what other single-logarithmic contributions would need to be considered in order to obtain more accurate predictions for physical cross sections.

\section{Factorization of jet cross sections at hadron colliders}
\label{sec:factorization}

In this paper, we develop an effective field-theory based approach for a systematic theoretical description of non-global observables at hadron colliders, such as gap-between-jets cross sections. The starting point is the factorization formula~\cite{Balsiger:2018ezi,Becher:2021zkk}
\begin{align}\label{eq:factorization_formula}
   \sigma_{2\to M}(Q_0) =\int \! d x_1 \int \!d x_2 \, \sum_{m={2+M}}^\infty \!\big\langle \bm{\mathcal{H}}_m(\{\underline{n}\},s,x_1,x_2,\mu) \otimes \bm{\mathcal{W}}_m(\{\underline{n}\},Q_0,x_1,x_2,\mu) \big\rangle \,.
\end{align}
Here $s$ denotes the squared center-of-mass energy, $x_1$ and $x_2$ are the longitudinal momentum fractions carried by the colliding partons, and $Q_0$ is the soft scale associated with the veto imposed on radiation between the beam remnants and the final-state jets. The tuple $\{\underline{n}\}=\{n_1,\dots,n_m\}$ collects light-like 4-vectors aligned with the directions of the initial-state ($i=1,2$) and final-state ($i=3,\dots,m$) particles. The above formula generalizes the analogous result for $e^+e^-$ colliders~\cite{Becher:2015hka,Becher:2016mmh}, which involves hard functions $\bm{\mathcal{H}}_m$ describing the energetic partons inside the jets and soft functions $\bm{\mathcal{S}}_m$ describing the soft emissions off these hard partons. In the hadron-collider case, the hard functions of multiplicity $m$ also contain the two initial-state partons, and the functions $\bm{\mathcal{W}}_m$ describe both the soft emissions off the hard partons and the collinear dynamics associated with the initial state. In~\cite{Becher:2015hka,Becher:2016mmh} we have also considered the case of narrow jets, which requires the resummation of collinear logarithms associated with small jet opening angles. In the present paper, we restrict ourselves to the case of large opening angles for simplicity.

The hard functions $\bm{\mathcal{H}}_m$ describe all possible $m$-particle scattering processes $1+2\to 3+\dots+m$, where $i$ represents the $i$-th particle ($i=q,\bar{q},g$ for colored partons). To keep the notation compact, we do not indicate the different partonic configurations, but it is understood that one must sum not only over different values of $m$, but over all possible channels. The brackets $\langle\spac\dots\rangle$ denote a sum (average) over final-state (initial-state) color and spin indices. The relevant spin/color multiplicity factors are
\begin{align} \label{eq:spin_color_average_factors}
   \mathcal{N}_i = \begin{cases}
    ~ 2N_c & \text{for } i=q,\bar{q} \,, \\
    ~ (d-2)(N_c^2-1) & \text{for } i=g \,,
    \end{cases}
\end{align}
and analogously for color-neutral particles. Here $d=(4-2\epsilon)$ is the dimension of spacetime. Although in our discussion we only consider unpolarized hadron beams, a generalization to fixed helicities would be straightforward. 

The hard functions are obtained after imposing appropriate kinematic constraints, such as cuts on the transverse momenta and rapidities of the leading jets. One then integrates over the phase space of the final-state particles but for fixed directions of the outgoing particles. To define these directions, we choose reference 4-vectors $n_i^\mu=(1,\bm{n}_i)$ in the laboratory frame,\footnote{In~\cite{Becher:2021zkk} we have defined the hard functions in the partonic center-of-mass frame. However, in order to implement experimental cuts, it is more convenient to work in the laboratory frame.} 
so that the associated particle momenta are given by $p_i^\mu=E_i\,n_i^\mu$. For each light-cone vector $n_i^\mu$, we introduce a conjugate vector $\bar n_i^\mu=(1,-\bm{n}_i)$, so that $n_i\cdot \bar{n}_i =2$. The hard functions are then defined as
\begin{align} \label{eq:hard_function_definition}
\begin{aligned}
   \bm{\mathcal{H}}_m %(\{\underline{n}\},\hat{s}) 
   &= \frac{1}{2x_1 x_2\spac s} \prod_{i=3}^m \int\!\frac{dE_i\,E_i^{d-3}}{{\tilde{c}}^\epsilon \,(2\pi)^{2}}\,
    |\mathcal{M}_m(\{\underline{p}\})\rangle \langle\mathcal{M}_m(\{\underline{p}\})| \\[1mm]
   &\quad\times (2\pi)^d\,2\,\delta(\bar{n}_1\cdot p_{\rm tot}-x_1 \sqrt{s})\,\delta(\bar{n}_2\cdot p_{\rm tot}-x_2 \sqrt{s})\,
    \delta^{(d-2)}(p^\perp_{\rm tot})\,\Theta_{\rm hard}\!\left(\left\{\underline{n}\right\}\right) ,
\end{aligned}
\end{align} 
where $p_{\rm tot}$ is the total momentum of the final-state particles and $p^{\perp}_{\rm tot}$ denotes the $(d-2)$ components transverse to the beam directions $n_1$ and $n_2$. The energies of the incoming partons are $E_1 = x_1 \sqrt{s}/2$ and $E_2 = x_2 \sqrt{s}/2$. The angular constraint $\Theta_{\rm hard}\!\left(\left\{\underline{n}\right\}\right)$ ensures that the hard partons cannot enter the gap or veto region. Note that some of the final-state particles can be color neutral. In particular, we will also consider the production of $Z$- or Higgs-bosons in association with $M\ge 0$ jets. We stress that the amplitude in~\eqref{eq:hard_function_definition} is squared in the sense of a density matrix. We use the color/helicity-space formalism~\cite{Catani:1996vz}, in which the color and helicity indices of the amplitude $|\mathcal{M}_m(\{\underline{p}\})\rangle$ and its conjugate are not contracted. 

The symbol $\otimes$ in~\eqref{eq:factorization_formula} indicates an integration over the directions $\{n_3,\dots,n_m\}$ of the final-state particles in the hard scattering process. These integrals must be performed after the hard functions are combined with the low-energy matrix elements $\bm{\mathcal{W}}_m$, which encode the soft and collinear dynamics in the process of interest. Following~\cite{Becher:2021urs}, we have included a factor $\tilde{c}^\epsilon= (e^{\gamma_E}/\pi)^\epsilon$ in the denominators of the energy integrals in the definition of the hard function, where $\gamma_E$ is Euler's constant. The same factor is added to the $(d-2)$-dimensional angular integrals, for which we use the measure
\begin{align} \label{eq:angular_integrals_measure}
   [d\Omega_i] = \tilde{c}^\epsilon\,\frac{d^{d-2}\Omega_i}{2 (2\pi)^{d-3}} \,.
\end{align}
We thus define
\begin{equation} \label{eq:angular_integrals}
\begin{aligned}
    &\bm{\mathcal{H}}_m(\{\underline{n}\},s,x_1,x_2,\mu) \otimes 
     \bm{\mathcal{W}}_m(\{\underline{n}\},Q_0,x_1, x_2,\mu) \\
    \equiv{} & \prod_{i=3}^m \int [d\Omega_i]\,\bm{\mathcal{H}}_m(\{\underline{n}\},s,x_1,x_2,\mu)\,
     \bm{\mathcal{W}}_m(\{\underline{n}\},Q_0,x_1,x_2,\mu) \,.
\end{aligned}
\end{equation}
The factors of $\tilde{c}^\epsilon$ cancel in the combination of the hard functions and angular integrals but avoid a proliferation of $\gamma_E$'s and logarithms of $\pi$ at intermediate stages. 

\begin{figure}
    \centering
    \includegraphics[scale=1]{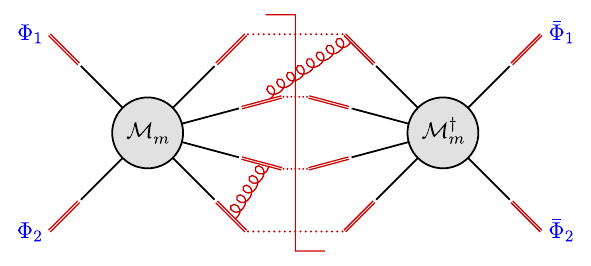} 
    \caption{Pictorial representation of the factorization formula~\eqref{eq:factorization_formula}. In black, a hard function $\bm{\mathcal{H}}_m$ in~\eqref{eq:hard_function_definition} is shown, which is multiplied by soft Wilson lines for each hard parton (red double lines). The color indices of the Wilson lines along the directions of the final-state particles in the amplitude are connected (dotted lines) to the ones sourced by the particles in the conjugate amplitude. The Wilson lines of the initial-state partons connect to the collinear fields (blue), see~\eqref{eq:W_function_definition}. We also included a real and a virtual soft gluon, which are part of the matrix element $\bm{\mathcal{W}}_m$.}
    \label{fig:factorization}
\end{figure}

The low-energy matrix elements $\bm{\mathcal{W}}_m$ involve soft Wilson lines $\bm{S}_i(n_i)$ along the directions of all hard particles in the process (for color-neutral particles, one uses $\bm{S}_i(n_i)=\bm{1}$), and collinear fields for the two incoming partons. They are given by Fourier transforms 
\begin{align}
    \bm{\mathcal{W}}_m(\{\underline{n}\},Q_0,x_1,x_2) = 
    \int_{-\infty}^{\infty} \frac{dt_1}{2\pi}\,e^{- i x_1 t_1 \bar{n}_1\cdot p_1}  \int_{-\infty}^{\infty} \frac{dt_2}{2\pi}\,e^{-i x_2 t_2 \bar{n}_2\cdot p_2} \,\widetilde{\bm{\mathcal{W}}}_m(\{\underline{n}\},Q_0, t_1, t_2) 
\end{align}
of matrix elements of the form
\begin{align} \label{eq:W_function_definition}
\begin{aligned}
   &\widetilde{\bm{\mathcal{W}}}_m(\{\underline{n}\},Q_0, t_1,t_2) \\[3mm]
   ={} & \hspace{0.15cm}\int\limits_{X_s}\hspace{-0.5cm}\sum\, \mathcal{P}^{(1)}_{\bar{\alpha}\alpha}\,\mathcal{P}^{(2)}_{\bar{\beta}\beta}\,  \langle  H_1(p_1) H_2(p_2)  |\, \bar{\Phi}_{1}^{\bar{\alpha}}(t_1 \bar{n}_1) \, \bar{\Phi}_{2}^{\bar{\beta}}(t_2 \bar{n}_2) \, \bm{S}_1^\dagger(n_1) \,  \dots\,  \bm{S}_m^\dagger(n_m)\,  |X_s \rangle  \\
   &\times \langle X_s | \,\bm{S}_1(n_1) \,  \dots\,  \bm{S}_m(n_m)\, \Phi_{1}^{\alpha}(0) \, \Phi_{2}^{\beta}(0) \, |H_1(p_1) H_2(p_2) \rangle \, \theta( Q_0 - E^\perp_{\rm \, out}) \,,
\end{aligned}
\end{align}
where $H_1$ and $H_2$ are the colliding hadrons. In these expressions, the fields $\Phi_{i}$ are the gauge-invariant collinear building blocks~\cite{Bauer:2001yt,Hill:2002vw} in the directions of the two hadrons, as appropriate for a given partonic channel, i.e.\ $\Phi_{i}\in\{\chi_i,\bar\chi_i,\mathcal{A}_{i\perp}\}$ for a quark, anti-quark or gluon, while $\bar{\Phi}_i$ is equal to the corresponding conjugate fields. Note that the argument of the collinear fields indicates the spacetime point at which they are localized, whereas the argument of the soft Wilson lines indicates their direction. All soft Wilson lines are located at the point $x=0$. Since we only consider unpolarized hadron beams, the Dirac and Lorentz indices in~\eqref{eq:W_function_definition} are contracted with the relevant spin sums, i.e.\ 
\begin{align}
\begin{aligned}
   \mathcal{P}^{(i)}_{\bar{\alpha}\alpha }\,\bar{\chi}_i^{\bar{\alpha}}(t\bar{n}_i)\,\chi_i^\alpha(0)
   &= \left(\frac{{\bar{n}\!\!\!/}_{i}}{2}\right)_{\bar\alpha\alpha}\,\bar{\chi}_i^{\bar{\alpha}}(t\bar{n}_i)\,\chi_i^\alpha(0) 
    = \bar{\chi}_i(t\bar{n}_i)\,\frac{{{\bar{n}\!\!\!/}_i}}{2}\,\chi_i(0) \,, \\
   \mathcal{P}^{(i)}_{\bar{\alpha}\alpha}\,\mathcal{A}_{\perp i}^{\bar{\alpha}}(t\bar{n}_i)\,\mathcal{A}_{\perp i}^{\alpha}(0)   
   &= (-g_{\bar\alpha\alpha}) (-i\partial_t)\,\mathcal{A}_{\perp i}^{\bar{\alpha}}(t\bar{n}_i)\,\mathcal{A}_{\perp i}^{\alpha}(0) 
    =  i \partial_t\,\mathcal{A}_{\perp i}^\mu(t\bar{n}_i)\,\mathcal{A}_{\perp i\mu}(0) \,.
\end{aligned}
\end{align}
Therefore, $\bm{\mathcal{W}}_m$ acts as a unity matrix in helicity space. The additional derivative arising in the gluon case ensures that the collinear matrix element corresponds to the usual definition of the gluon PDF. The factorization theorem is depicted in Figure~\ref{fig:factorization}, which also shows how the color indices of the Wilson lines in $\bm{\mathcal{W}}_m$ are connected to the hard functions and the collinear fields (dotted lines). 

In the factorization formula~\eqref{eq:factorization_formula} the soft Wilson lines $\bm{S}_i(n_i)$ in~\eqref{eq:W_function_definition} multiply the amplitudes $|\mathcal{M}_m(\{\underline{p}\})\rangle$ in the hard functions~\eqref{eq:hard_function_definition}, while the conjugate Wilson lines $\bm{S}_i^\dagger(n_i)$ multiply the conjugate amplitude $\langle\mathcal{M}_m(\{\underline{p}\})|$. In~\eqref{eq:W_function_definition}, the jet-veto scale $Q_0$ is defined to be the upper limit on the total transverse momentum $E^\perp_{\rm \, out} = \sum_i |p^\perp_i|$ of the particles outside of the jets, but many other kinematic restrictions could be considered. For example, in order to be less sensitive to the underlying event and pile-up, one can instead define $Q_0$ as the upper limit on the transverse momentum of jets inside the veto region, as was done by the ATLAS collaboration in~\cite{ATLAS:2011yyh,ATLAS:2014lzu}. In the leading-logarithmic approximation, one is not sensitive to the precise definition of the observable, but only to the associated energy scale $Q_0$.

We note that the $n_1$- and $n_2$-collinear fields in~\eqref{eq:W_function_definition} are fields obtained after the soft-collinear decoupling transformation~\cite{Bauer:2001yt}
\begin{equation}
   \Phi_i(t\bar{n}_i)\to\bm{S}_i(n_i)\,\Phi_i(t\bar{n}_i)
\end{equation}
has been applied, which explains the appearance of the soft Wilson lines $\bm{S}_{1,2}$ in~\eqref{eq:W_function_definition}. However, it is important to note that the low-energy matrix elements in~\eqref{eq:W_function_definition} are \emph{not} factorized into soft and $n_i$-collinear fields without interactions among them, because the low-energy effective theory still includes Glauber-gluon interactions between the soft and collinear sectors, which break this factorization. For the case of forward scattering, the Glauber Lagrangian was derived in~\cite{Rothstein:2016bsq}. The non-trivial interactions are associated with the scale $Q_0$, and below this scale we can match onto an effective theory that only involves soft and collinear fields associated with the scale $\Lambda_{\rm QCD}$ of non-perturbative physics. Since the emissions below the scale $Q_0$ are not restricted, we expect that the Glauber interactions will cancel by the same mechanism which is at work for the Drell--Yan process~\cite{Collins:1983ju,Collins:1988ig,Collins:1989gx}. In the absence of these interactions, the collinear low-energy matrix elements reduce to the usual collinear PDFs for quarks and gluons inside the hadron $H_i$, i.e.\
\begin{equation}
   f_{i}(x_i,\mu) = \int_{-\infty}^{\infty} \frac{dt}{2\pi} e^{- i x_i t\bar{n}_i\cdot p_i}\, 
   \langle H_i(p_i)|\,\bar{\Phi}_{i}^{\bar{\alpha}}(t\bar{n}_i)\,\mathcal{P}^{(i)}_{\bar{\alpha}\alpha}\, 
   \Phi_{i}^{\alpha}(0)\,|H_i(p_i)\rangle \,,
\end{equation}
which multiply a matrix element of soft Wilson lines. Above the scale $Q_0$, however, such a factorization no longer holds.

An important ingredient for the resummation of large logarithms is the RG equation for the hard functions, which we write in the form~\cite{Becher:2021zkk}
\begin{equation}\label{eq:hardRG}
    \frac{d}{d\ln\mu}\,\bm{\mathcal{H}}_m(\{\underline{n}\},s,\mu)
    = - \sum_{l=2+M}^{m} \bm{\mathcal{H}}_l(\{\underline{n}\},s,\mu) \, \ast \bm{\Gamma}^H_{lm}(\{\underline{n}\},s,\mu) \,.
\end{equation}
Here and below we omit the momentum-fraction variables for the initial-state partons in the hard functions and anomalous-dimension coefficients, which in the convolution on the right-hand side are combined by
\begin{equation} \label{eq:ast_convolution}
   (f\ast g)(x_i) \equiv \int_0^1\!d\xi_i \, f(\xi_i \spac x_i) \, g(\xi_i) \,.
\end{equation}
There is one such convolution for each initial-state parton. Most of the discussion in this paper concerns the soft part of the anomalous-dimension matrix, which has a trivial dependence on the momentum fractions proportional to $\delta(1-\xi_1)\,\delta(1-\xi_2)$, because soft emissions take away an insignificant fraction of longitudinal momentum. This renders the convolution~\eqref{eq:ast_convolution} trivial. For the discussion of soft effects, we will thus omit the convolution symbol $\ast$ when writing products of anomalous dimensions.

The evolution equation~\eqref{eq:hardRG} exhibits the familiar structure of RG equations in the presence of operator mixing. However, its solution is a highly non-trivial task even in the leading-logarithmic approximation. The reason is that the anomalous-dimension matrix is an operator not only in the high-dimensional color space of the initial- and final-state particles, but also in the infinite space of particle multiplicities. This is a key feature of our approach and reflects the intrinsic complexity of the problem at hand. The evolution equations shows that higher-multiplicity hard functions mix with lower-multiplicity functions under scale evolution. At one-loop order, and written in the space of particle-multiplicities, the anomalous-dimension matrix takes the form
\begin{align} \label{eq:gammaHmat}
   \bm{\Gamma}^H(\{\underline{n}\},s,\mu) = \frac{\alpha_s}{4\pi}\,
    \begin{pmatrix}
    \,\bm{V}_{2+M} & \bm{R}_{2+M\phantom{+1}} & 0 & 0 & \hdots \\
    0 & \bm{V}_{2+M+1} & \bm{R}_{2+M+1} & 0 & \hdots \\
    0 & 0 & \bm{V}_{2+M+2} &  \bm{R}_{2+M+2} & \hdots \\
    0 & 0 & 0 &  \bm{V}_{2+M+3} & \hdots \\
    \vdots & \vdots & \vdots & \vdots & \ddots
    \end{pmatrix}
    + \mathcal{O}(\alpha_s^2) \,,
\end{align}
where $(2+M)$ is the minimal number of particles for an $M$-jet process at a hadron collider. The virtual-correction matrix elements $\bm{V}_{m}$ on the diagonal leave the number of partons unchanged, while the real-emission operators $\bm{R}_{m}$ map a hard function with $m$ partons onto one with $(m+1)$ partons.\footnote{Recall that $\bm{\Gamma}^H$ stands to the right of the hard functions in~\eqref{eq:hardRG}.} 
With each higher order in perturbation theory an additional off-diagonal in the upper right half of the matrix is filled, but the entries below the diagonal remain zero to all orders.

By solving the RG equation~\eqref{eq:hardRG} we can evolve the hard functions from their natural scale $\mu_h\sim Q\sim\sqrt{\hat{s}}$, where $\hat{s}=x_1 x_2 s$ is the partonic center-of-mass energy, down to the scale $\mu_s\sim Q_0$ of the low-energy dynamics. A formal solution is given by the path-ordered exponential
\begin{equation} \label{eq:U}
   \bm{U}(\{\underline{n}\},s,\mu_h,\mu_s) = {\bf P} \exp\left[\, \int_{\mu_s}^{\mu_h} \frac{d\mu}{\mu}\, \bm{\Gamma}^H(\{\underline{n}\},s,\mu) \right] ,
\end{equation}
which is defined by its series expansion
\begin{equation} \label{eq:Uexp}
\begin{aligned}
   \bm{\mathcal{H}}(\mu_h) \ast \bm{U}(\mu_h,\mu_s) 
   &= \bm{\mathcal{H}}(\mu_h) + \int_{\mu_s}^{\mu_h}\!\frac{d\mu_1}{\mu_1}\,
    \bm{\mathcal{H}}(\mu_h) \ast \bm{\Gamma}^H(\mu_1) \\
   &\quad + \int_{\mu_s}^{\mu_h}\!\frac{d\mu_1}{\mu_1}\,
    \int_{\mu_1}^{\mu_h}\!\frac{d\mu_2}{\mu_2}\,
    \bm{\mathcal{H}}(\mu_h) \ast \bm{\Gamma}^H(\mu_2) \ast \bm{\Gamma}^H(\mu_1)
    + \dots \,,
\end{aligned}
\end{equation}
where the anomalous-dimension matrices on the right-hand side are ordered in the direction of decreasing scale values (i.e.\ $\mu_2>\mu_1$ in the second line). In the last two equations we have suppressed the energy and direction arguments of the various functions for simplicity. As the convolution~\eqref{eq:ast_convolution} is not associative, we use the convention that the leftmost product is convoluted first, i.e.\ $f\ast g\ast h \equiv (f\ast g)\ast h$.
It is closely associated with the standard Mellin convolution
\begin{equation}
    (f\star g)(x_i) = \int \!d\xi_i \! \int \!d\xi_i^\prime \, \delta(x_i - \xi_i \spac \xi_i^\prime) \, f(\xi_i) \, g(\xi_i^\prime) \,,
\end{equation}
which arises in the DGLAP evolution equation. The relation reads
\begin{equation}
    (f \ast g) \ast h = f \ast (g \star h) \,.
\end{equation}
It is thus possible to first Mellin convolute the anomalous dimensions in~\eqref{eq:Uexp} before combining them with the hard function.
Below, we will show that in the absence of Glauber phases the collinear part of the anomalous dimension indeed produces the standard DGLAP evolution cf.~\eqref{eq:DGLAP_evolution_GammaC}.

In the following, we first present a detailed derivation of the anomalous dimension $\bm{\Gamma}^H$ at one-loop order (Section~\ref{sec:gamma}). In contrast to the case of $e^+ e^-$ collisions, the anomalous dimension not only contains soft contributions, but also collinear and soft$\,+\,$collinear contributions associated with the initial-state partons. The soft$\,+\,$collinear parts exhibit a logarithmic dependence on the factorization scale $\mu$, which leads to double logarithms upon performing the  scale integrals in~\eqref{eq:Uexp}. This feature is the source of the SLLs. Following our earlier work~\cite{Becher:2021zkk}, we then calculate the leading double-logarithmic terms from~\eqref{eq:factorization_formula} and~\eqref{eq:Uexp} order by order in perturbation theory, by evaluating the relevant color traces (Sections~\ref{sec:lltraces} and~\ref{sec:traces}) and iterated scale integrals (Section~\ref{sec:integrals_resummation}). For this calculation it is sufficient to work with the lowest-order expressions for the low-energy matrix elements and combine them with the expression for the RG-evolved hard functions evaluated at leading double-logarithmic order~\cite{Becher:2021zkk}.
We can thus neglect all quantum corrections at the scale $\mu=Q_0$, so that the functions $\bm{\mathcal{W}}_m$ are given by their tree-level expressions
\begin{equation}\label{eq:Wlead}
   \bm{\mathcal{W}}_m(\{\underline{n}\},Q_0,x_1,x_2,\mu_s) 
   = f_1(x_1,\mu_s)\,f_2(x_2,\mu_s)\,\bm{1} + \mathcal{O}(\alpha_s) \,.
\end{equation}

In the future, it will be very interesting to study the low-energy matrix elements $\bm{\mathcal{W}}_m$ in more detail. The fact that the evolution of the hard functions produces double logarithms, but the low-energy theory naively only knows about a single scale $Q_0$, implies that the low-energy matrix elements must suffer from a collinear anomaly, which produces rapidity logarithms~\cite{Becher:2010tm,Chiu:2012ir}. The presence of rapidity logarithms is characteristic for processes involving Glauber gluons~\cite{Rothstein:2016bsq}, but since the double-logarithmic terms only start at four-loop order, these must have quite an intricate structure in our case, which waits to be explored. A resummation of the rapidity logarithms will be required to extend our results beyond the leading double-logarithmic approximation.

\section{Derivation of the anomalous dimension}\label{sec:gamma}

We will now derive the one-loop anomalous dimension $\bm{\Gamma}^H$ of the hard functions. The corresponding anomalous dimension relevant for the case of $e^+e^-$ collisions was derived in~\cite{Becher:2016mmh,Becher:2021urs}, where in the second reference also the two-loop contribution was obtained. An important complication present in the hadron-collider case we consider here is the occurrence of collinear terms in the anomalous dimension. Indeed, it is these collinear terms which in conjunction with the Glauber phases lead to the SLLs. 

The anomalous dimension $\bm{\Gamma}^H$ is related to infrared divergences in the functions $\bm{\mathcal{H}}_m$ and can be obtained by considering soft and collinear limits of these functions. In the following, we will first consider the soft limit of the hard amplitudes and recapitulate the result obtained for the $e^+e^-$ case. After this, we turn to a detailed analysis of collinear singularities. We show that for final-state partons, the singularities exactly cancel between real and virtual corrections. For the initial-state partons, there are singularities that lead to a collinear anomalous dimension.

\subsection{Soft part of the anomalous dimension}

The gap-between-jets observable at $e^+e^-$ colliders is single logarithmic, and instead of the matrix elements $\bm{\mathcal{W}}_m$ one has matrix elements $\bm{\mathcal{S}}_m$ which only involve soft Wilson lines. The physics of the NGLs is driven by soft emissions and we have extracted the one-loop anomalous dimension by considering the soft limit of the hard functions $\bm{\mathcal{H}}_m$, which leads to the result~\cite{Becher:2016mmh}
\begin{align}\label{eq:oneLoopRG}
\begin{aligned}
    \bm{V}^S_m  &= 2\sum_{(ij)} \int \frac{d\Omega(n_k)}{4\pi} \left(\bm{T}_{i,L}\cdot  \bm{T}_{j,L}+\bm{T}_{i,R}\cdot  \bm{T}_{j,R}\right) \overline{W}_{ij}^k  \\
    &\quad - 2i\pi \,\sum_{(ij)} \left(\bm{T}_{i,L}\cdot  \bm{T}_{j,L} - \bm{T}_{i,R}\cdot  \bm{T}_{j,R}\right) \Pi_{ij} , \\
    \bm{R}^S_m &= -4\sum_{(ij)}\,\bm{T}_{i,L}\circ\bm{T}_{j,R}  \,\overline{W}_{ij}^{k}\,  \Theta_{\rm hard}(n_{k})\,.
\end{aligned}
\end{align}
A detailed derivation of this result, including the imaginary part in the second line, can be found in~\cite{Becher:2021urs}. For $e^+e^-$ collisions, $\bm{V}^S_m$ and $\bm{R}^S_m $ are the full result for the entries of the anomalous-dimension matrix~\eqref{eq:gammaHmat}, but for hadron colliders we need to add the collinear parts computed below. The superscript $S$ indicates that these terms are associated with the purely soft singularities of $\bm{\mathcal{H}}_m$.
 
Let us now explain the notation in~\eqref{eq:oneLoopRG}. The symbol $(ij)$ on the sums runs over all (unordered) pairs of parton indices with $i\ne j$. The quantity $\overline{W}_{ij}^{k}$ describes the angular dependence and is related to the soft dipole
\begin{align} \label{eq:Dipole}
    W_{ij}^{k} =\frac{n_i\cdot n_j}{n_i \cdot n_k \, n_j \cdot n_k}\,,
\end{align}
which is the product of the two eikonals, summed over the spin of the emitted gluon. To restrict the anomalous dimension~\eqref{eq:oneLoopRG} to the purely soft contributions, the collinear limits of the soft dipole $W_{ij}^k$ were subtracted using
\begin{align} \label{eq:subtractedDipole}
    \overline{W}_{ij}^k = W_{ij}^k - \frac{1}{n_i\cdot n_k}\,\delta(n_i- n_k)  - \frac{1}{n_j\cdot n_k}\,\delta(n_j- n_k) \,.
\end{align}
It is understood that the angular delta distribution $\delta(n_i- n_k)$ only acts on the test function, not on the coefficient multiplying it. The hard gluons in the real emission are restricted to lie inside the jet region by the constraint $\Theta_{\rm hard}(n_{k})$, while the virtual corrections are unrestricted. 
We use the color-space formalism, where $\bm{T}_i$ denotes a color generator acting on particle $i$. The color matrices $\bm{T}_{i,L}$ act on the amplitude while $\bm{T}_{j,R}$ multiplies the conjugate, for example
\begin{equation}
(\bm{T}_{1,L}\cdot  \bm{T}_{2,L} + \bm{T}_{3,R}\cdot \bm{T}_{4,R})\, \bm{\mathcal{H}}_m = \bm{T}_{1}\cdot  \bm{T}_{2}\, \bm{\mathcal{H}}_m + \, \bm{\mathcal{H}}_m\, \bm{T}_{3}\cdot \bm{T}_{4} \,.
\end{equation}
The color matrices in the virtual part act on the color indices of the $m$ partons of the amplitude and $\bm{T}_{i}\cdot\bm{T}_{j}=\sum_a \bm{T}_{i}^a\,\bm{T}_{j}^a$. This is the usual color-space notation. The color matrices in the real emission matrix $\bm{R}^S_m$ are different. They take an amplitude with $m$ partons and associated color indices and map it into an amplitude with $(m+1)$ partons, see Figure~\ref{fig:action_GammaBar}. Explicitly, we have
\begin{equation}
\bm{T}_{i,L}\circ \bm{T}_{j,R} \,\bm{\mathcal{H}}_m = \bm{T}_{i}^a\, \bm{\mathcal{H}}_m \,\bm{T}_{j}^{\tilde{a}}\,,
\end{equation}
where $a$ and $\tilde{a}$ are the color indices of the additional emitted gluon in the amplitude and the conjugate amplitude. In contrast to the virtual case, these color indices cannot immediately be contracted because later emissions can attach to the new gluon. 

The terms in the second line of~\eqref{eq:oneLoopRG} are purely imaginary. An imaginary part is present whenever $i$ and $j$ are both incoming or both outgoing partons, the prefactor is $\Pi_{ij}=1$ in these cases and zero otherwise. The imaginary part can be simplified using color conservation $\sum_i  \bm{T}_{i}=0$. For concreteness, consider the process $1+ 2 \to 3+ \dots + m$. We then have
\begin{align}\label{eq:GlauberSimp}
\sum_{(ij)} \bm{T}_{i}\cdot  \bm{T}_{j} \,\Pi_{ij} &= 2\, \bm{T}_1\cdot \bm{T}_2 + \sum_{i=3}^m \bm{T}_i \cdot (-\bm{T}_1-\bm{T}_2 - \bm{T}_i) \nonumber\\
&= 2\,\bm{T}_1\cdot \bm{T}_2 +  (\bm{T}_1+\bm{T}_2)\cdot  (\bm{T}_1+\bm{T}_2) - \sum_{i =3}^m C_i  \\[-1mm]
& = 4\, \bm{T}_1\cdot \bm{T}_2 +C_1+C_2 - \sum_{i =3}^m C_i\,, \nonumber
\end{align}
where $\bm{T}_i \cdot \bm{T}_i = C_i\, \bm{1}$ is the quadratic Casimir of the representation associated with leg~$i$, and $C_i$ evaluates to $C_F$ for (anti-)quark legs and to $C_A$ for gluons.
The constant imaginary part arises both from the generators $\bm{T}_{i,L}$ acting on the amplitude and the generators  $\bm{T}_{i,R}$ acting on the conjugate amplitude. These terms cancel in the anomalous dimension.
In case where one or both incoming particles are color-neutral the term  $\bm{T}_1\cdot \bm{T}_2$ is not present and the Coulomb phase never contributes to the cross section. The phase terms completely vanish and can be dropped from the anomalous-dimension matrix as we did in our previous paper~\cite{Becher:2016mmh}.  A non-trivial phase can arise if the initial state carries color, as is the case for the partonic amplitudes relevant for hadronic collisions. Note that, after using color conservation to rewrite the sum of the final-state Glauber phases in terms of the initial-state color generators, the coefficient of the $\bm{T}_1\cdot \bm{T}_2$ term has doubled. In order to account for the final-state phases, the Lagrangian of~\cite{Rothstein:2016bsq} needs to be adapted to our problem. For example, an immediate consequence of~\eqref{eq:GlauberSimp} is that the coefficient of the Glauber terms must be twice as large as in the case of forward scattering.
 
In $e^+e^-$ collisions the soft$\,+\,$collinear parts of the anomalous dimension cancel between the real and virtual entries and one could work in terms of the unsubtracted dipoles $W_{ij}^k$ as explained in~\cite{Becher:2021urs}.  For hadron collider processes, due to the presence of Glauber phases, the soft$\,+\,$collinear singularities associated with the initial state will not cancel and lead to SLLs.  Furthermore, we will also need a purely collinear anomalous dimension that corresponds, up to the color structure, to the usual DGLAP evolution of the PDFs. To extract the collinear pieces of the anomalous dimension, we will now first consider collinear singularities in the virtual corrections and then collinear limits of the hard function.
  
\subsection{Singularities in virtual corrections}

The soft and collinear divergences of massless scattering amplitudes $|\mathcal{M}_m(\{\underline{p}\}) \rangle$ are well known~\cite{Becher:2009cu, Gardi:2009qi, Becher:2009qa,Dixon:2009ur,Ahrens:2012qz,Becher:2019avh}. The amplitudes can be renormalized multiplicatively through
\begin{equation}
    |\mathcal{M}_m(\{\underline{p}\},\mu) \rangle = \lim_{\epsilon\to 0} \bm{Z}^{-1}(\{\underline{p}\},\mu,\epsilon)\, |\mathcal{M}_m(\{\underline{p}\},\epsilon) \rangle\,.
\end{equation}
The renormalization factor can be obtained from an anomalous-dimension matrix, which up to two-loop order takes the form~\cite{Becher:2009cu}
\begin{equation}\label{eq:GammaAmp}
\bm{\Gamma}^\mathcal{M}(\{\underline{p}\},\mu) 
   = \sum_{(ij)}\,\frac{\bm{T}_i\cdot\bm{T}_j}{2}\,\gamma_{\rm cusp}(\alpha_s)\,\ln\frac{\mu^2}{-s_{ij}} 
    + \sum_i\,\gamma^i(\alpha_s)\,\bm{1}\,,
\end{equation}
where $s_{ij}= 2\sigma_{ij}\, p_i\cdot p_j+ i0$ with $\sigma_{ij}=2\,\Pi_{ij}-1$. The hard functions $\bm{\mathcal{H}}_m$ are given by squared amplitudes with particles along fixed directions, so that this anomalous dimension is relevant. However, according to the definition~\eqref{eq:hard_function_definition} the hard functions are integrated over the energies of the outgoing partons in the presence of the phase space constraints. Since the collinear part of the anomalous dimension depends logarithmically on the energies through the cusp logarithms, the result~\eqref{eq:GammaAmp} does not immediately translate into a result for the anomalous dimension of the hard functions. 

We will now prove that the collinear pieces of the anomalous dimension~\eqref{eq:GammaAmp} associated with final-state partons cancel against collinear singularities of real-emission corrections present in hard functions with additional collinear legs. This cancellation can be shown to take place before the energy integrals are carried out. To simplify the notation for our discussion, we write the hard functions in the form
\begin{equation}\label{eq:hard_function_definitionHat}
\begin{aligned}
   \bm{\mathcal{H}}_m(\{\underline{n}\},s,x_1,x_2) 
   &= \int \!\! d{\mathcal{E}}_m \, \widetilde{\bm{\mathcal{H}}}_m(\{\underline{p}\})\,,
\end{aligned} 
\end{equation}
where the ``unintegrated'' hard functions
\begin{equation}
 \widetilde{\bm{\mathcal{H}}}_m(\{\underline{p}\})  = |\mathcal{M}_m(\{\underline{p}\})\rangle \langle\mathcal{M}_m(\{\underline{p}\})| 
\end{equation}
are simply given by the squared amplitude, and $\int\! d{\mathcal{E}}_m$ collects the final-state parton energy integrals together with the momentum-conservation and phase-space constraints on hard radiation, see~\eqref{eq:hard_function_definition}.

To discuss the collinear singularities, we rewrite the logarithmic part of the anomalous dimension in~\eqref{eq:GammaAmp} in the form
\begin{equation}
\ln\frac{\mu^2}{-s_{ij}} = \ln\frac{2}{n_i\cdot n_j} + \ln\frac{\mu}{2E_i} + \ln\frac{\mu}{2E_j} +i \pi \,\Pi_{ij} \,.
\end{equation}
The energy-dependent parts only depend on a single parton and can be simplified using color conservation
\begin{equation}\label{eq:GammaAmpSep}
\begin{aligned}
    \bm{\Gamma}^\mathcal{M}(\{\underline{p}\},\mu) 
    & = \sum_{(ij)}\,\frac{\bm{T}_i\cdot\bm{T}_j}{2}\,\gamma_{\rm cusp}(\alpha_s)\, \left(\ln\frac{2}{n_i\cdot n_j} +i \pi \,\Pi_{ij} \right) \\  
    & \quad+ \sum_{i} \left(- C_i \,\gamma_{\rm cusp}(\alpha_s)\, \ln\frac{ \mu}{2 E_i} +\gamma^i(\alpha_s) \right)\bm{1} \\[1mm]
    &= \bm{\Gamma}^\mathcal{M}_s + \sum_{i} \Gamma^\mathcal{M}_{c,i}\, \bm{1} \,, 
\end{aligned}
\end{equation}
where $\Gamma^\mathcal{M}_{c,i}$ contains the collinear as well as the soft$\,+\,$collinear singularities.
Using the relation
\begin{equation}\label{eq:virtualint}
    \int \frac{d\Omega(n_k)}{4\pi}\, \overline{W}_{ij}^k = \ln\frac{n_i\cdot n_j}{2} \,,
\end{equation}
where the integral is related to the ones defined in~\eqref{eq:angular_integrals_measure} by
\begin{equation}
    \int \frac{d\Omega(n_k)}{4\pi} = \int [d\Omega_k] + \mathcal{O}(\epsilon)\,,
\end{equation}
we see that the soft anomalous dimension $\bm{\Gamma}^\mathcal{M}_s$ gives rise to the collinearly subtracted version of the virtual part $\bm{V}^S_m$ of the one-loop anomalous dimension~\eqref{eq:oneLoopRG}. The treatment of the collinear part of the anomalous dimension is more subtle because it depends on the energy, which is integrated over for final-state partons. 

For later use, let us write out the divergences associated with the collinear anomalous dimension $\Gamma^\mathcal{M}_{c,i}$ at one loop. We have
\begin{equation}\label{eq:collinear_singularity_virtual}
   \bm{\mathcal{H}}_m(\{\underline{n}\},s) 
   = \sum_i \int \! d{\mathcal{E}}_m \, \frac{\alpha_s}{4\pi} \left[-C_i\spac \gamma_0^{\rm cusp} \left(\frac{1}{2\epsilon^2} + \frac{1}{\epsilon} \ln\frac{ \mu}{2 E_i} \right) +  \frac{\gamma^i_0}{\epsilon} \right] \widetilde{\bm{\mathcal{H}}}_m(\{\underline{p}\}) +\dots\,,
\end{equation}
where the ellipsis denotes terms that are free of collinear singularities at one loop. The factor of two compared to the result~\eqref{eq:GammaAmp} arises because we get divergences from both the amplitude and its conjugate.

\subsection{Collinear limits of hard functions}
 
 To analyze the limits where two of the partons in $\bm{\mathcal{H}}_{m+1}$ become collinear, we make use of splitting amplitude factorization
\begin{equation}\label{eq:coll}
   |\mathcal{M}_{m+1}(\{p_1,p_2,p_3,\dots,p_{m+1}\})\rangle 
   = \mbox{\bf Sp}(\{p_1,p_2\})\,
   |\mathcal{M}_m(\{P,p_3,\dots, p_{m+1}\})\rangle + \dots \,
\end{equation}
in the region where two partons become collinear with $p_1 \approx z P$ and $p_2 \approx(1-z) P$ and $P^2\to 0$. We write the collinear limit of $1$ and $2$ for notational convenience, but in the application to the hard function, we will need to consider two cases: (i) both collinear partons are in the final state and (ii) one parton is in the initial state, one in the final state.

\begin{figure}
    \centering
    \includegraphics[scale=1]{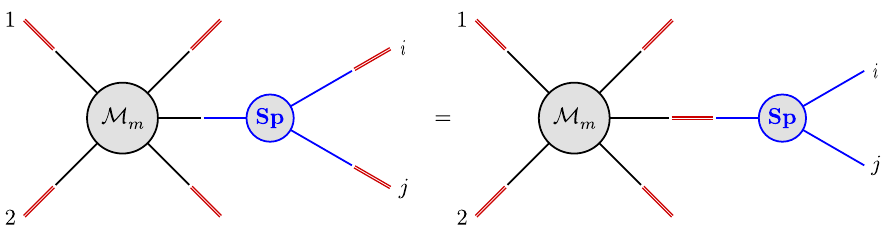}
    \caption{In the limit where partons $i$ and $j$ become collinear, the hard functions factorize. The soft Wilson lines (red double lines) associated with  $i$ and $j$ combine into a single Wilson line for the parent parton.}
    \label{fig:finalColl}
\end{figure}

In our factorization theorem, the hard functions are multiplied by Wilson lines along the directions of the hard partons. In particular, the soft function for the two collinear partons contains Wilson lines for partons $1$ and $2$ along the common direction $n_P$. Since the generators $1$ and $2$ commute, the Wilson lines combine into a single Wilson line with color $(\bm{T}_1^a+\bm{T}_2^a)$,
\begin{equation}\label{eq:WilsonComb}
\bm{S}_1(n_1)\,\bm{S}_2(n_2 ) = \bm{S}_1(n_P)\, \bm{S}_2(n_P) = \bm{S}_{1+2}(n_P)\,.
\end{equation} 
In~\cite{Becher:2021urs} it was shown that this implies 
\begin{align}\label{eq:WilsonSplit}
\begin{aligned}
   &\bm{S}_{1+2}(n_P)\,\mbox{\bf Sp}(\{p_1,p_2\})\,
    |\mathcal{M}_{m}(\{P, p_3, \dots, p_{m+1}\})\rangle \\[2mm]
   ={}& \mbox{\bf Sp}(\{p_1,p_2\})\,\bm{S}_{P}(n_P)\,
    |\mathcal{M}_{m}(\{P, p_3, \dots, p_{m+1}\})\rangle \,.
\end{aligned}
\end{align}
This relation is illustrated in Figure~\ref{fig:finalColl} and is an operator version of the usual QCD coherence, which states that the soft emissions from two collinear partons are the same as the collinear emissions from the parent parton. It follows immediately from charge conservation, which implies that the color state of the partons after the decay corresponds to the color state of the parent parton (see e.g.\ \cite{Catani:2003vu,Becher:2009qa})
\begin{equation}\label{eq:splitting_functions_final_definitionColor}
    \left(\bm{T}_1^a+\bm{T}_2^a\right) \mbox{\bf Sp}(\{p_1,p_2\}) = \mbox{\bf Sp}(\{p_1,p_2\})\, \bm{T}_P^a\,.
\end{equation}

Let us now first consider case (i) with a hard function $\bm{\mathcal{H}}_{m+1}$ in a kinematical situation where two final-state partons $i$ and $j$ become collinear. Parameterizing the energies as $E_i=z E_P$ and $E_j=(1-z) E_P$ and using that the measurement function is collinear safe, we find 
\begin{equation}
    \int \! d{\mathcal{E}}_{m+1} =  \int_0^1 dz \left[ z(1-z) \right]^{d-3}  \int \! d{\mathcal{E}}_{m} \, \frac{E_P^{d-2}}{\tilde{c}^\epsilon (2\pi)^{2}}\,.
\end{equation}
Note the presence of the additional factor of $E_P^{d-2}$. 

The above equations imply that in the limit where $i$ and $j$ become collinear, the product of hard and soft-collinear functions before the angular integrations take the form
\begin{equation}\label{eq:HW_collinear_limit_final}
\begin{aligned}
    &\int \! d{\mathcal{E}}_{m+1}  \,\big\langle \widetilde{\bm{\mathcal{H}}}_{m+1}(\{\underline{p}\})   \,\,\bm{\mathcal{W}}_{m+1}(\{\underline{n}\},Q_0) \big\rangle \,
    \underset{i\parallel j}{\longrightarrow} \int_0^1 dz 
    \left[ z(1-z) \right]^{d-3} \\
    &\times \int \! d{\mathcal{E}}_{m} \, \frac{E_P^{d-2}}{\tilde{c}^\epsilon (2\pi)^{2}} \,\big\langle \mbox{\bf Sp}(\{p_i,p_j\})\, \widetilde{\bm{\mathcal{H}}}_m(\{\underline{p}\})  \,\,\bm{\mathcal{W}}_{m}(\{\underline{n}\},Q_0) \,\mbox{\bf Sp}^\dagger(\{p_i,p_j\})\big\rangle\,.
\end{aligned}
\end{equation}
Recall that, as explained after~\eqref{eq:W_function_definition}, the quantity $\bm{\mathcal{W}}$ acts on both the amplitude and the conjugate amplitude in the hard function, i.e.\ half of the Wilson lines in $\bm{\mathcal{W}}$ act on the left-hand side of $\widetilde{\bm{\mathcal{H}}}$. 

Explicit expressions for the tree-level splitting amplitudes can be found in (12)\,--\,(15) of~\cite{Catani:2011st}. We use the fact that the product $\widetilde{\bm{\mathcal{H}}}_m\,\bm{\mathcal{W}}_{m}$ is independent of the colors and spins of partons $i$ and $j$ to carry out the associated sums in $\langle\dots\rangle$. Then it is possible to rewrite the result as a sum over color and spin of the parent parton $P$
\begin{align} \label{eq:splitting_functions_final_definition}
    \big\langle \mbox{\bf Sp}(\{p_i,p_j\})\,\widetilde{\bm{\mathcal{H}}}_m\,\bm{\mathcal{W}}_{m}\, \mbox{\bf Sp}^\dagger(\{p_i,p_j\}) \big\rangle &=4\pi \alpha_s \tilde{\mu}^{2\epsilon}\frac{2}{s_{ij}}\,\mathcal{P}_{i+j \leftarrow P}(z) \big\langle \widetilde{\bm{\mathcal{H}}}_m\,\bm{\mathcal{W}}_{m} \big\rangle  \,,
\end{align}
where $s_{ij} = 2 E_P^2 \,z(1-z)\, n_i\cdot n_j$. The scale $\tilde{\mu}$ is related to the scale $\mu$ in the $\overline{\rm MS}$ scale through $\tilde{\mu}^2 = e^{\gamma_E}/(4\pi) \mu^2$. We find the spin averages of the squared splitting amplitudes, which are commonly referred to as splitting functions $\mathcal{P}_{i+j\leftarrow P}$. Note that the product~\eqref{eq:HW_collinear_limit_final} will be integrated over $(m+1)$ directions. When one integrates over the direction $n_j$, one encounters a collinear singularity
\begin{equation}\label{eq:angle}
\int\! [d\Omega_j] \,\frac{1}{n_i\cdot n_j} = -\frac{1}{2\epsilon} +\mathcal{O}(\epsilon) = -\frac{1}{2\epsilon} \int\! [d\Omega_j] \, \delta(n_i-n_j) +\mathcal{O}(\epsilon)\,.
\end{equation}
We should thus view the splitting function as a distribution that produces a singularity when integrating over the direction $n_j$. The remaining angular integration over the direction of parton $i$ can then be reinterpreted as the integration over the direction of the parent parton so that we are then left with the integral over the $m$ directions relevant for the parent hard function $\bm{\mathcal{H}}_{m}$. 

To proceed we now perform the integrals over $z$. In the integrand, we have the spin-averaged splitting functions, which can be found in eqs.~(14)\,--\,(17) of~\cite{Catani:1999ss}, and are given by
\begin{equation}\label{eq:splitting_functions_final_explicit}
\begin{aligned}
    \mathcal{P}_{q+g \leftarrow q}(z) &= \mathcal{P}_{\bar{q}+g \leftarrow \bar{q}}(z)  = C_F \left[ \frac{1+z^2}{1-z} - \epsilon (1-z) \right] ,
    \\
    \mathcal{P}_{g+q \leftarrow q}(z)  &= \mathcal{P}_{g+\bar{q} \leftarrow \bar{q}}(z) = \mathcal{P}_{q+g \leftarrow q}(1-z)  \,,
    \\[1mm]
    \mathcal{P}_{q+\bar{q} \leftarrow g}(z) & = \mathcal{P}_{\bar{q}+q \leftarrow g}(z)  = T_F \left[1 - \frac{2 z(1-z)}{1-\epsilon}\right] ,
    \\
    \mathcal{P}_{g+ g \leftarrow g}(z)  &= 2C_A \left[\frac{z}{1-z}+ \frac{1-z}{z} + z(1-z)\right].
\end{aligned}
\end{equation}
Combining~\eqref{eq:HW_collinear_limit_final} and~\eqref{eq:splitting_functions_final_definition} yields the integrals
\begin{align}\label{eq:splitting_functions_final_integrated}
\int_0^1 dz \left[ z(1-z) \right]^{-2\epsilon} \,\frac{1}{2} \left(\mathcal{P}_{q+g \leftarrow q}(z) +\mathcal{P}_{g+q \leftarrow q}(z)\right)  &=C_F\left(-\frac{1}{\epsilon} - \frac32\right) +\mathcal{O}(\epsilon) \,, 
\end{align}
and
\begin{equation}\label{eq:gluesplit}
\begin{aligned}
    &\int_0^1 dz \left[ z(1-z) \right]^{-2\epsilon} \,\frac{1}{2} \bigg( \mathcal{P}_{g+g \leftarrow g}(z) +  \sum_q\left[ \mathcal{P}_{q+\bar{q} \leftarrow g}(z)+  \mathcal{P}_{\bar{q}+q \leftarrow g}(z)\right]\bigg) \\
    ={} & C_A\left(-\frac{1}{\epsilon} - \frac{11}{6}\right) +  \frac{4}{6}\,T_F \, n_f +\mathcal{O}(\epsilon)\,,
\end{aligned}
\end{equation}
where we have averaged over the splittings  $q+g \leftarrow q$ and $g+q \leftarrow q$ which are both part of the same $(m+1)$-parton configuration. The same is true for $q+\bar{q} \leftarrow g$ and $\bar{q}+q \leftarrow g$. Since the integral is symmetric under $z\to 1-z$ and the two contributions map onto each other under this transformation, we could instead also simply only consider one of the two channels. For $g+g \leftarrow g$ the factor $1/2$ on the left-hand side of~\eqref{eq:gluesplit} ensures that we do not over-count identical particles.

Adding the integral over directions and putting things together, we find the following result for the collinear contribution associated with a parent parton $P$ splitting into collinear partons $i$ and $j$: 
\begin{equation}\label{eq:collinear_singularity_real_final}
\begin{aligned}
   &\int \! d{\mathcal{E}}_{m+1}  \,\big\langle \widetilde{\bm{\mathcal{H}}}_{m+1}(\{\underline{p}\}) \otimes \bm{\mathcal{W}}_{m+1}(\{\underline{n}\},Q_0) \big\rangle \\
    \underset{i \parallel j}{\longrightarrow} &\int \! d{\mathcal{E}}_{m}\, \frac{\alpha_s}{4\pi} \left[ C_P \,\gamma_0^{\rm cusp}\,\left(\frac{1}{2\epsilon^2} + \frac{1}{\epsilon}\ln\frac{ \mu}{2 E_P}\right) - \frac{\gamma^P_0}{\epsilon} \right] \,\big\langle \widetilde{\bm{\mathcal{H}}}_m(\{\underline{p}\}) \otimes \bm{\mathcal{W}}_{m}(\{\underline{n}\},Q_0) \big\rangle \,.
\end{aligned}
\end{equation}
Note that we integrate over $(m+1)$ angles on the left-hand side and over $m$ angles on the right-hand side. The extra angular integration has been performed to get rid of the angular $\delta$-distribution and sets $n_j \to n_i$ which is identified with the parent parton direction. Doing so, we observe that the result on the right-hand side is equal and opposite to the one associated with virtual collinear singularities shown in~\eqref{eq:collinear_singularity_virtual} and after we sum over the splitting of all parent partons $P$, i.e.\ over all the $(m-2)$ final-state partons in the hard function $\bm{\mathcal{H}}_m$, we find that the collinear singularities associated with final-state partons exactly cancel.

Next, let us consider case (ii) where a final-state parton becomes collinear to an initial-state parton. We can infer this splitting from time-like result~\eqref{eq:splitting_functions_final_definition} by crossing one of the final-state momenta to the initial state. For concreteness, let us study the case where leg~$1$ becomes collinear to leg~$j$ by crossing $p_i \to - p_1$. We also need to take into account the difference in kinematics. The final-state collinear splittings describe the process $P \to p_i + p_j$, where in the collinear limit $p_i = z\, P$. For the initial-state splitting, we instead consider $p_i \to P + p_j$ with $P=\xi \,p_1$ in the collinear limit, which implies that we should substitute $z \to 1/\xi$ in the time-like splitting functions in~\eqref{eq:splitting_functions_final_definition}.\footnote{At higher orders, careful analytic continuation is needed to correctly reproduce the complex phases in the amplitude when performing the crossing $p_i \to - p_1$ ~\cite{Catani:2011st}. We work at tree level.} The result is that for a space-like splitting $p_1 \to P + p_j$ the trace in~\eqref{eq:HW_collinear_limit_final} evaluates to
\begin{align} \label{eq:splitting_functions_initial_definition}
    \big\langle \mbox{\bf Sp}(\{p_1,p_j\})\,\widetilde{\bm{\mathcal{H}}}_m\,\bm{\mathcal{W}}_{m}\, \mbox{\bf Sp}^\dagger(\{p_1,p_j\}) \big\rangle &=4\pi \alpha_s \tilde{\mu}^{2\epsilon} \frac{2}{(-s_{1j})} \frac{1}{\xi}\,\mathcal{P}_{1 \rightarrow P}(\xi) \big\langle \widetilde{\bm{\mathcal{H}}}_m\,\bm{\mathcal{W}}_{m} \big\rangle \,,
\end{align} 
where $\mathcal{P}_{1 \rightarrow P}(\xi)$ are the unregularized DGLAP splitting functions. In contrast to the time-like case, it is customary to only indicate the incoming parton $1$ and the parton $P$ entering the hard scattering. The radiated collinear parton $j$ can be inferred from fermion flavor conservation. The extra factor of $1/\xi$ in~\eqref{eq:splitting_functions_initial_definition} compared to the time-like case~\eqref{eq:splitting_functions_final_definition} will correct the flux factor in the cross section to one relevant for the scattering of the incoming parton $P$ since $s_{P2} = \xi s_{12}$. 

Performing the crossing carefully and taking into account that the average factors~\eqref{eq:spin_color_average_factors} change between the left- and right-hand side of~\eqref{eq:splitting_functions_initial_definition} if parton $P$ and $1$ are different, one finds that $\mathcal{P}_{1 \rightarrow P}(\xi) = \mathcal{P}_{ P + j \leftarrow 1 }(\xi)$, i.e. the one-loop functional form of the DGLAP kernels is identical to the one which arises in the time-like splitting~\eqref{eq:splitting_functions_final_explicit} for the given partonic channel. 

Let us now analyze the integration measure of the hard function in the collinear limit. We parameterize the energy of the final-state collinear particle as $E_j = E_1 \spac (1-\xi)$ and rewrite the hard function as a hard function for the process with initial-state parton $P$ instead $1$ and without the final-state parton $j$. The denominator in~\eqref{eq:splitting_functions_initial_definition} is $s_{1j} = -2 E_1^2\, n_1\cdot n_j\, (1-\xi)$. Inserting these expressions, one would naively rewrite the contribution from $(m+1)$ partons as
\begin{equation}\label{eq:HW_collinear_limit_initial}
\begin{aligned}
    &\int \! d{\mathcal{E}}_{m+1}  \,\big\langle \widetilde{\bm{\mathcal{H}}}_{m+1}(\{\underline{p}\}) \, \bm{\mathcal{W}}_{m+1}(\{\underline{n}\},Q_0) \big\rangle \\
    \underset{1 \parallel j}{\longrightarrow} & \,\frac{\alpha_s}{\pi} \left(\frac{\mu}{2 E_1}\right)^{2\epsilon}  \int_0^1 d\xi \, (1-\xi)^{-2\epsilon} \, \mathcal{P}_{1 \to P}(\xi) \,\frac{1}{n_1\cdot n_j} \int \! d{\mathcal{E}}_{m}\, 
     \,\big\langle \widetilde{\bm{\mathcal{H}}}_m(\{\underline{p}\})\, \bm{\mathcal{W}}_{m}(\{\underline{n}\},Q_0) \big\rangle \,.
\end{aligned}
\end{equation}
To obtain the collinear divergence, we first extract the divergence of the angular integral using~\eqref{eq:angle}. The splitting kernels $\mathcal{P}_{q\to q}$ and $\mathcal{P}_{g\to g}$ also have a soft divergence when $\xi \to 1$. This divergence is regularized by the $d$-dimensional energy integral which yields a factor of $(1-\xi)^{-2\epsilon}$ in the integrand. We can isolate the divergence using the relation
\begin{equation}\label{eq:PlusDistribution}
    (1-\xi)^{-1-2\epsilon} = -\frac{1}{2\epsilon} \delta(1-\xi) + \left[\frac{1}{1-\xi}\right]_+  + \mathcal{O}(\epsilon)\,.
\end{equation}
Doing so leads to the following result for the collinearly divergent part 
\begin{align} \label{eq:collinear_singularity_real_initial_INCORRECT}
\begin{aligned}
    &\int \! d{\mathcal{E}}_{m+1} \, \big\langle \widetilde{\bm{\mathcal{H}}}_{m+1}(\{\underline{p}\}) \, \bm{\mathcal{W}}_{m+1}(\{\underline{n}\},Q_0,x_1,x_2) \big\rangle \\
    \underset{1 \parallel j}{\longrightarrow} &\,  \int_0^1 \!d\xi \, \frac{\alpha_s}{4\pi}\left[ C_1 \spac \gamma_0^{\rm cusp} \,\delta(1-\xi)\,\left( \frac{1}{2\epsilon^2} + \frac{1}{\epsilon} \ln\frac{\mu}{2E_1} \right)\delta_{1P}  - \frac{2}{\epsilon}\,\overline{\mathcal{P}}_{1 \to P}(\xi) \right]  \delta(n_1-n_j)\\
    &\times\int \! d{\mathcal{E}}_{m} \, \big\langle \widetilde{\bm{\mathcal{H}}}_m(\{\underline{p}\})\, \bm{\mathcal{W}}_{m}(\{\underline{n}\},Q_0,x_1,x_2) \big\rangle \,,
\end{aligned}
\end{align}
where $\overline{\mathcal{P}}_{1 \to P}(\xi) $ are the splitting function from which the soft singularity has been subtracted using~\eqref{eq:PlusDistribution}. After adding the collinearly singular terms~\eqref{eq:collinear_singularity_virtual} arising in the virtual part, we would find that the soft$\,+\,$collinear pieces would cancel and we would recover the standard $\overline{\text{MS}}$ DGLAP kernels.

\begin{figure}
    \centering
    \includegraphics[scale=1]{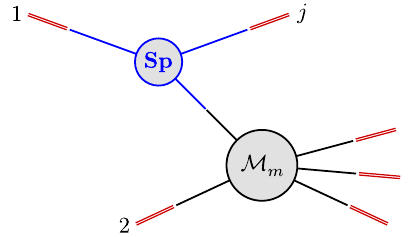}
    \caption{In the limit where the initial-state parton 1 and the final-state parton $j$ become collinear, the hard functions factorize, but in contrast to the final-state collinear limit depicted in Figure~\ref{fig:finalColl}, the soft Wilson lines associated with 1 and $j$ do not combine.}
    \label{fig:initialColl}
\end{figure}

However, in the above derivation there is a subtle mistake. In our argument based on crossing, we have implicitly used relations~\eqref{eq:WilsonComb} and~\eqref{eq:WilsonSplit} to combine Wilson lines and simplify the color structure, but in the space-like limit $1\parallel j$ the Wilson line $\bm{S}_1$ is associated with an incoming particle, while $\bm{S}_j$ describes emissions from an outgoing line, see Figure~\ref{fig:initialColl}. These two Wilson lines do not combine into a single outgoing Wilson line even if their direction $n_P$ is the same. In terms of Feynman diagrams the difference between the two Wilson lines is the sign of the $i 0$ prescription in the associated light-cone direction, which is of course directly related to the Glauber effects~\eqref{eq:GlauberSimp} we investigate in this paper. Since we are unable to use~\eqref{eq:WilsonSplit}, the color generators associated with the emission cannot be commuted through the $\bm{\mathcal{W}}_{m+1}$ function, which remains in the original $(m+1)$-parton space. Let us denote the color matrix associated with the splitting amplitude $\mbox{\bf Sp}(p_1,p_j)$ by $\bm{\mathcal{C}}_{1\to P}$. Since the color structure of the splitting amplitudes cannot be simplified with~\eqref{eq:WilsonSplit}, the corrected version of~\eqref{eq:collinear_singularity_real_initial_INCORRECT} then reads
\begin{align}\label{eq:collinear_singularity_real_initial} 
\begin{aligned}
    & \int \! d{\mathcal{E}}_{m+1} \, \big\langle \widetilde{\bm{\mathcal{H}}}_{m+1}(\{\underline{p}\}) \, \bm{\mathcal{W}}_{m+1}(\{\underline{n}\},Q_0,x_1,x_2) \big\rangle \\
    \underset{1 \parallel j}{\longrightarrow} & \, \int_0^1 \!d\xi \, \frac{\alpha_s}{4\pi}\left[ C_1 \spac \gamma_0^{\rm cusp} \,\delta(1-\xi)\,\left( \frac{1}{2\epsilon^2} + \frac{1}{\epsilon} \ln\frac{\mu}{2E_1} \right)\delta_{1P}  - \frac{2}{\epsilon}\,\overline{\mathcal{P}}_{1 \to P}(\xi) \right]  \delta(n_1-n_j)\\
    &\times \int \! d{\mathcal{E}}_{m} \, \big\langle \bm{\mathcal{C}}^{\phantom{\dagger}}_{1\to P}\, \widetilde{\bm{\mathcal{H}}}_m(\{\underline{\hat{p}}\})  \,\bm{\mathcal{C}}^\dagger_{1\to P} \, \bm{\mathcal{W}}_{m+1}(\{\underline{n}\},Q_0,x_1,x_2) \big\rangle \,.
\end{aligned}
\end{align}
The matrix $\bm{\mathcal{C}}_{1\to P}$ connects the colors of the three partons involved in the splitting and maps from the $m$-parton space with momenta $\{\underline{\hat{p}}\} = \{ P, p_2, \dots, p_{j-1},p_{j+1},\dots, p_{m+1}\}$  before the splitting to the $(m+1)$-parton space with directions $\{\underline{n}\} = \{ n_1, n_2, \dots, n_{m+1}\}$ after the splitting. We have normalized these matrices to unity for trivial $\bm{\mathcal{W}}_{m+1}=\bm{1}$ 
\begin{equation}
   \bm{\mathcal{C}}^\dagger_{1\to P}\, \bm{\mathcal{C}}^{\phantom{\dagger}}_{1\to P} =  \bm{1}\,.
\end{equation}
For the $q\to q$ or $g\to g$ splittings, the matrix $\bm{\mathcal{C}}_{1\to P}$ describes the emission of an additional collinear gluon, which can be described in the color-space formalism. With our normalization, we have
\begin{equation}
    \bm{\mathcal{C}}^{\phantom{\dagger}}_{1\to P}\, \bm{\mathcal{H}}_m\, \bm{\mathcal{C}}^\dagger_{1\to P}  =  \bm{\mathcal{H}}_m \,\frac{1}{C_P} \, \bm{T}_{P,L}\circ\bm{T}_{P,R}\,   .
\end{equation}
The subscripts $L,R$ indicate on which side the color generator multiplies the hard function. For the soft terms proportional to $\delta(1-\xi)$, the normalization factor $C_P$ simply cancels the Casimir in the prefactor. 

\begin{figure}[t!]
    \centering
    \includegraphics[scale=1]{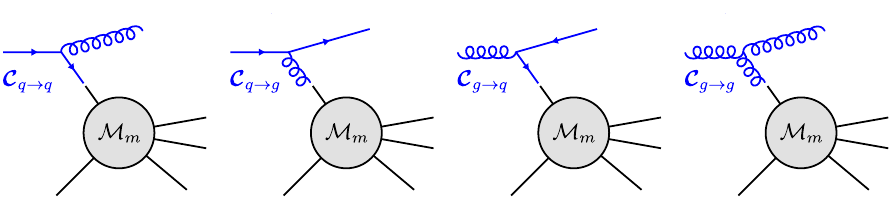} 
    \caption{Color structures $\bm{\mathcal{C}}_{1\to P}$ for different collinear splittings. In the first three cases, the color structure is given by the color generator associated with the quark-gluon vertex, appropriately contracted with the hard function. In the last case, the color structure is given by the $SU(N_c)$ structure constant. The first and last example can be written as $\bm{T}_1^a$ in the color space formalism, where $a$ is the color index of the emitted collinear gluon.}
    \label{fig:collinear_splittings_initial}
\end{figure}

After this discussion, we can now present the result for the full anomalous dimension, including both the soft part and the collinear pieces associated with the initial-state collinear singularities. At the one-loop order, we split the anomalous dimension into a soft part and a sum of purely collinear terms
\begin{equation}\label{eq:gammaH}
\bm{\Gamma}^H(\xi_1,\xi_2) = \delta(1-\xi_1) \,\delta(1-\xi_2)\, \bm{\Gamma}^S+\bm{\Gamma}_1^C(\xi_1)\,\delta(1-\xi_2) + \delta(1-\xi_1)\, \bm{\Gamma}_2^C(\xi_2) \, .
\end{equation}
To separate the soft$\,+\,$collinear parts from the purely collinear ones, we introduce a reference scale $\mu_h \sim \sqrt{\hat{s}}$ and split
\begin{equation}
    \ln\frac{\mu}{2E_i} = \ln\frac{\mu}{\mu_h} + \ln\frac{\mu_h}{2E_i}\, 
\end{equation}
for $i=1,2$. The large logarithms $\ln\frac{\mu}{\mu_h}$ are included with the soft anomalous dimension $\bm{\Gamma}^S\equiv\bm{\Gamma}^S(\mu_h,\mu)$ and the remaining $\mathcal{O}(1)$ terms are included in $\bm{\Gamma}_i^C$. In the partonic center-of-mass frame $2E_1=2E_2=\sqrt{\hat{s}}$ so that the extra term is absent for the choice $\mu_h \sim\sqrt{\hat{s}}$ which we adopted in our previous paper~\cite{Becher:2021zkk}. In the laboratory frame, we have instead $2E_1=x_1\sqrt{s}$ and $2E_2=x_2\sqrt{s}$. To obtain the cross section, the hard functions $\bm{\mathcal{H}}_{m}$ and the soft-collinear functions $\bm{\mathcal{W}}_{m+1}$ are integrated over the momentum fractions. The above anomalous dimension multiplies these functions in the sense of Mellin convolutions over $\xi_1$ and $\xi_2$. Since the soft part has trivial dependence on the momentum fractions, we have suppressed this dependence in~\cite{Becher:2021zkk}.

The real and virtual pieces of the purely collinear part $\bm{\Gamma}_i^C$ of the one-loop anomalous-dimension matrix~\eqref{eq:gammaHmat} are given by
\begin{equation}\label{eq:collineargammas}
\begin{aligned}
\bm{V}_{i}^C(\xi_i)&= -2\spac \left(\gamma_0^i -C_i\spac \gamma_0^{\rm cusp} \ln\frac{\mu_h}{2E_i} \right)\,\delta(1-\xi_i)\,,\\
\bm{R}_i^C(\xi_i) &= 2
\left(2\spac \overline{\mathcal{P}}_{i \to P}(\xi_i) - C_i \spac \gamma_0^{\rm cusp}\spac \delta_{iP} \spac \ln\frac{\mu_h}{2E_i} \,\delta(1-\xi_i)\right) \bm{\mathcal{C}}^{\phantom{\dagger}}_{i\to P,L}\, \bm{\mathcal{C}}^\dagger_{i\to P,R} \spac \delta(n_k-n_i)\,,  
\end{aligned}
\end{equation}
where the $L$ and $R$ subscript indicate how the color matrices act on the hard function. Note that the collinear real-emission operator has different channels $1\to P$. For example, $\bm{R}_q^C$ acts on hard functions with initial state $P$, which could be a quark or gluon, and produces a new hard function with initial-state quark. With the default choice $\mu_h \sim \sqrt{\hat{s}}$ the logarithms in the collinear anomalous dimension operators evaluate (modulo a sign) to the rapidity difference between the lab and the partonic center-of-mass frame. After convolution with the PDFs, this is an order one logarithm. Furthermore, for channels in which the two initial-state partons transform in the same color representation, the logarithms immediately cancel for the default scale choice, as is evident from~\eqref{eq:GammaAmpSep}.

It is interesting that the hard evolution is not driven by the usual DGLAP kernels but by a real part that has a non-trivial color structure and a color-diagonal virtual part. Only if the color structure of the soft-collinear functions $\bm{\mathcal{W}}_{m+1}$ is such that the color structure of the real emissions trivializes, the two parts will combine into the standard $\overline{\text{MS}}$ kernels. For the same reason also the soft$\,+\,$collinear pieces do not cancel out, which will lead to double-logarithmic terms in the evolution, the SLLs. 

\begin{figure}[t!]
    \centering
    \includegraphics[scale=1]{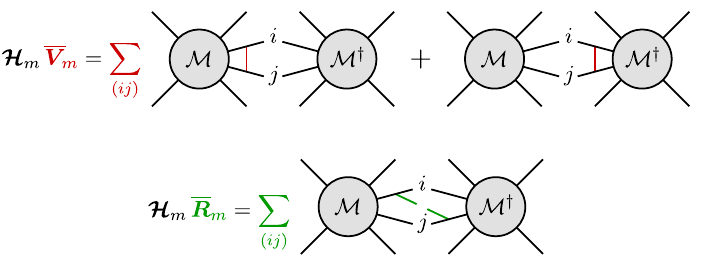}
    \caption{Action of the real-emission operator $\bm{\overline{R}}_m$ and the virtual piece $\overline{\bm{V}}\!_m$ on a hard function $\bm{\mathcal{H}}_m$.  Due to the emitted gluon (green line), the product $\bm{\mathcal{H}}_m\,\bm{\overline{R}}_m$ defines a hard function with $(m+1)$ external legs.}
    \label{fig:action_GammaBar}
\end{figure}

\begin{figure}[t!]
    \centering
    \includegraphics[scale=1]{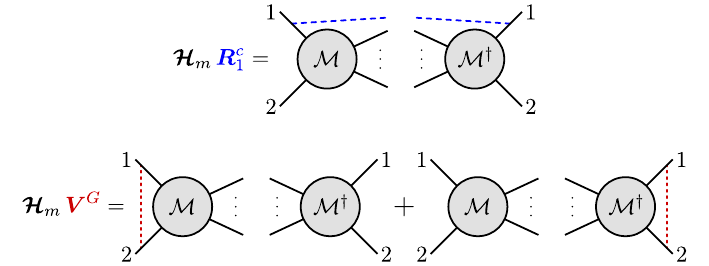} 
    \caption{Action of the cusp operator $\bm{R}^c_1$ and the virtual piece $\bm{V}^G$ on a hard function $\bm{\mathcal{H}}_m$. The operator $\bm{R}^c_1$ adds an additional final-state leg (dashed blue line) along the direction of the incoming parton $1$.}
    \label{fig:action_GammaC_VG}
\end{figure}

The soft piece $ \bm{\Gamma}^S$ of the anomalous dimension can be split into the following parts~\cite{Becher:2021zkk}
\begin{equation}\label{eq:cuspterms}
\begin{aligned}
   \bm{V}^S_m &= \overline{\bm{V}}\!_m + \bm{V}^G + \sum_{i=1,2} \bm{V}^c_i\,
    \ln\frac{\mu^2}{\mu_h^2} \,, \\
   \bm{R}^S_m &= \overline{\bm{R}}_m + \sum_{i=1,2} \bm{R}^c_i\,
    \ln\frac{\mu^2}{\mu_h^2} \,.
\end{aligned}
\end{equation}
The entries $\overline{\bm{R}}_m$ and $\overline{\bm{V}}\!_m$ are the collinearly subtracted real and virtual corrections, and $\bm{V}^G$ contains the Glauber phases. We have shown that final-state collinear singularities cancel between the real and virtual pieces so that we are left with the initial-state soft-collinear terms $\bm{R}^c_i$ and $\bm{V}^c_i$. Let us explicitly list the three ingredients.
The wide-angle soft emissions read
\begin{align}
\begin{aligned}
       \overline{\bm{V}}\!_m 
   &= 2\spac\sum_{(ij)}\,\big( \bm{T}_{i,L}\cdot\bm{T}_{j,L} 
    + \bm{T}_{i,R}\cdot\bm{T}_{j,R} \big)
    \int\frac{d\Omega(n_k)}{4\pi}\,\spac\overline{W}_{ij}^k \,, \\
     \overline{\bm{R}}_m 
   &= - 4\spac\sum_{(ij)}\,\bm{T}_{i,L}\circ\bm{T}_{j,R}\,\spac
    \overline{W}^{k}_{ij}\,\Theta_{\rm hard}(n_{k}) \,.
 \end{aligned}
\end{align} 
The angular integral in the virtual terms $\overline{\bm{V}}\!_m$ could be carried out using~\eqref{eq:virtualint}, but it is convenient to keep it to make real-virtual cancellations manifest. The real emission piece, on the other hand, generates a new parton along direction $n_{k}$ and the corresponding angular integration can only be carried out at the end.
The Glauber terms are given by
 \begin{align}  
   \bm{V}^G
   & = - 2 \spac i\pi \spac \gamma_0^{\rm cusp}\,\big( \bm{T}_{1,L}\cdot\bm{T}_{2,L}
    - \bm{T}_{1,R}\cdot\bm{T}_{2,R} \big) \,,
\end{align}
and the coefficients of the cusp logarithms are
\begin{equation}
\begin{aligned}
  \bm{V}^c_i  &= \gamma_0^{\rm cusp}\spac C_i\,\bm{1} \,, 
  \\
  \bm{R}^c_i 
   &= - \gamma_0^{\rm cusp} \spac\bm{T}_{i,L}\circ\bm{T}_{i,R}\,\delta(n_{k}-n_i) \,.
\end{aligned}
\end{equation}
The action of the different parts of the soft anomalous dimension on the hard function $\bm{\mathcal{H}}_m$ is depicted in Figures~\ref{fig:action_GammaBar} and~\ref{fig:action_GammaC_VG}.

\section{Color traces for the leading double-logarithmic terms}
\label{sec:lltraces}

To extract the logarithmically-enhanced terms, we will now compute the  evolution~\eqref{eq:Uexp} order by order. To do so, we will start with the lowest multiplicity hard function for the given process and multiply by powers of $\bm{\Gamma}^H$. We will first evaluate the color structure of these products and then perform the relevant $\mu$-integrals. In evaluating the products, we keep in mind the simple multiplicity matrix structure, namely that the real-emission contributions ($\overline{\bm{R}}_m$, $\bm{R}^c_i$ and $\bm{R}_i^C$) add an extra leg to a given hard function, while the virtual pieces ($\overline{\bm{V}}\!_m$, $\bm{V}^c_i$, $\bm{V}^G$ and $\bm{V}_i^C$) keep the number of legs the same. To streamline the notation, we will no longer write out the multiplicity indices on the anomalous dimensions and the hard and soft functions. In multiplicity space, the hard functions are vectors which we indicate by the notation $\bm{\mathcal{H}}$. The Born-level hard function only contains a lowest-order entry, $\bm{\mathcal{H}}_{2\to M}\equiv(\bm{\mathcal{H}}_{2+M},0,0,\dots)$. We also combine the real and virtual pieces of the soft anomalous dimension into the matrix notation
\begin{align}\label{eq:gammaSimp}
   \bm{\Gamma}^c
   &= \sum_{i=1,2}\, \gamma_0^{\rm cusp}\spac  \big[ C_i\,\bm{1} - \bm{T}_{i,L}\circ\bm{T}_{i,R}\,\delta(n_k-n_i) \big] \,,
   \nonumber\\[1mm]
   \bm{V}^G
   &= -2i\pi \spac \gamma_0^{\rm cusp}\spac \left( \bm{T}_{1,L}\cdot\bm{T}_{2,L} - \bm{T}_{1,R}\cdot\bm{T}_{2,R} \right) ,
   \\[2.5mm]\nonumber
   \overline{\bm{\Gamma}} 
   &= 2 \sum_{(ij)} \left( \bm{T}_{i,L}\cdot\bm{T}_{j,L} + \bm{T}_{i,R}\cdot\bm{T}_{j,R} \right)
    \int\frac{d\Omega(n_k)}{4\pi}\,\overline{W}_{ij}^k 
    - 4 \sum_{(ij)} \bm{T}_{i,L}\circ\bm{T}_{j,R}\,\overline{W}_{ij}^{k}\,\Theta_{\rm hard}(n_{k}) \,.
\end{align}
As in~\eqref{eq:hardRG} and~\eqref{eq:Uexp}, these are matrices in multiplicity space that multiply the hard function from the right and the order of the matrices determines the order in which they act on the hard function. At the same time, they contain color matrices that can act on the amplitude or the conjugate amplitude in each step, i.e.\ multiply the color indices of the hard function on the left or on the right. The vector $n_k$ in~\eqref{eq:gammaSimp} corresponds to the direction of the emitted gluon. Each emission generates a new vector and in a product of anomalous dimensions we will label  the vectors with an index $n_{k_\ell}$ with $\ell=0, 1, \dots $, where $\ell=0$ is the last emission, $\ell=1$ the second to last, and so on. 

Three properties of the different components of the anomalous dimension~\eqref{eq:gammaSimp} greatly simplify our calculations. Color coherence, the fact that the sum of the soft emissions off two collinear partons has the same effect as a single soft emission off the parent parton, implies that
\begin{equation}\label{eq:coherence}
   \bm{\mathcal{H}}\,\bm{\Gamma}^c\,\bm{\overline{\Gamma}} 
   = \bm{\mathcal{H}}\,\bm{\overline{\Gamma}} \,\bm{\Gamma}^c \,,
\end{equation}
in other words they commute when multiplying a hard function $\bm{\mathcal{H}}$
\begin{equation}
    [\bm{\Gamma}^c ,\bm{\overline{\Gamma}} ] = 0.
\end{equation}
To derive this relation, we note that the contributions $\overline{\bm{R}}_m$ and $\overline{\bm{V}}\!_m$ only depend on the sum of colors if two partons $i$ and $j$ become collinear, e.g. 
\begin{equation}\label{eq:colorcomb}
\bm{T}_{i,L}\cdot\bm{T}_{k,R} \overline{W}^{q}_{ik} +\bm{T}_{j,L}\cdot\bm{T}_{k,R} \overline{W}^{q}_{jk} = (\bm{T}_{i,L}+\bm{T}_{j,L})\cdot\bm{T}_{k,R} \overline{W}^{q}_{ik}\,,
\end{equation}
because the associated dipoles are identical if the particles are collinear. Then one uses the property~\eqref{eq:splitting_functions_final_definitionColor} to transform the sum of the color generators of the collinear partons into the color generator of the parent parton. Next, the cyclicity of the trace ensures that
\begin{equation}\label{eq:collsafety} 
   \big\langle \bm{\mathcal{H}}\,\bm{\Gamma}^c\otimes\bm{1} \big\rangle = 0 \,, \qquad
   \big\langle \bm{\mathcal{H}}\,\bm{V}^G\otimes\bm{1} \big\rangle = 0 \,.
\end{equation}
The first relation is a consequence of collinear safety: the singularity associated with a collinear real emission cancels against the one in the associated virtual correction. The second equation describes a cancellation of complex phases between the amplitude and its conjugate. The three properties hold for an arbitrary hard function $\bm{\mathcal{H}}$ obtained, for example, from the lowest-order hard function $\bm{\mathcal{H}}_{2\to M}$ after applying the one-loop anomalous dimension several times. 

To get the leading SLLs at a given order, we want to maximize the number of insertions $\bm{\Gamma}^c$, but the properties~\eqref{eq:collsafety} imply that we need a factor of $\bm{V}^G\spac \bm{\overline{\Gamma}}$ at the end of the evolution, otherwise the $\bm{\Gamma}^c$ immediately vanish. The insertion of this Glauber phase breaks color coherence~\cite{Forshaw:2006fk,Catani:2011st,Forshaw:2012bi,Schwartz:2017nmr}. To get a real, non-vanishing contribution to the cross section a second insertion of $\bm{V}^G$ is needed on top of this, while all remaining insertions can be due to double-logarithmic $\bm{\Gamma}^c$ terms. Based on these considerations we concluded in~\cite{Becher:2021zkk} that the leading SLLs arise from the color traces
\begin{equation}\label{eq:colortraces}
   C_{rn} = \big\langle \bm{\mathcal{H}}_{2\to M} \left( \bm{\Gamma}^c \right)^r
    \bm{V}^G \left( \bm{\Gamma}^c \right)^{n-r} \bm{V}^G\,\overline{\bm{\Gamma}} 
    \otimes\bm{1} \big\rangle \,,
\end{equation}
where $0\le r\le n$. The additional $p\le n$ collinear gluons, which can be emitted from the $n$ insertions of $\bm{\Gamma}^c$, will be labeled by indices $k_1,\dots,k_p$ (reading the insertions of $\bm{\Gamma}^c$ from right to left) and the final wide-angle gluon emission will have direction vector $n_{k_0}$. Our normalization of the hard functions is such that their trace is equal to the contribution of the given partonic channel to the Born-level cross section, i.e.
\begin{equation}\label{eq:trace1}
   \big\langle \bm{\mathcal{H}}_{2\to M} \otimes\bm{1} \big\rangle = \hat{\sigma}_{2 \to M} \,.
\end{equation}
Note that for a given $M$, there are several partonic channels
$1+2\to 3+\dots+(2+M)$ contributing. As explained earlier, we suppress the channel indices to keep the notation compact.

The symbol $\circ$ in the real-emission terms of the $\bm{\Gamma}^c$ and $\overline{\bm{\Gamma}}$ generates an additional gluon. There are two cases, where the associated color sum can be evaluated. First of all, if the rest of the color structure does not act on the color index of the additional gluon, it can immediately be summed over
\begin{equation}\label{eq:circdef0}
   \bm{T}_{i,L}\circ\bm{T}_{i,R} 
   = \bm{T}_{i,L}^a\,\bm{T}_{i,R}^a \,. 
\end{equation}
Secondly, if the remaining color structure only contains a single color generator $\bm{T}_{k_\ell}^c$ associated with the emitted gluon $k_\ell$, we can insert the explicit form of the generator and sum
\begin{equation} \label{eq:circdef}
   \bm{T}_{i,L}\circ\bm{T}_{i,R}\,\bm{T}_{k_\ell}^c
   = \bm{T}_{i,L}^a\,\bm{T}_{i,R}^b \left( -i f_{cba} \right) .
\end{equation}
Here and below, a sum over repeated color indices is implied.

The symbol $\otimes$ in~\eqref{eq:colortraces} includes, in particular, integrations over the directions $n_{k_\ell}$ of the emitted collinear gluons, which simply has the effect of replacing $\delta(n_k-n_i)\to 1$ in the expression for $\bm{\Gamma}^c$. It also includes an integration over the direction $n_{k_0}$, which has the effect of adding an integral $\int\frac{d\Omega(n_{k_0})}{4\pi}$ in front of the second term in $\overline{\bm{\Gamma}}$. The trivial consequences of these angular integrations are a result of the fact that the low-energy matrix elements $\bm{\mathcal{W}}_m$ are proportional to the trivial color structure $\bm{1}$ in lowest order, see~\eqref{eq:Wlead}.

So far, we have only considered the different pieces of the soft anomalous dimension, but let us also briefly discuss the purely collinear part. Combining real and virtual as in the soft case 
\begin{equation} \label{eq:collinear_part_anomalous_dimension}
\begin{aligned}
\bm{\Gamma}_i^C(\xi_i) &= \frac{\alpha_s}{4\pi} \left[2  \left(2\spac\overline{\mathcal{P}}_{i \to P}(\xi_i) - C_i \gamma_0^{\rm cusp} \ln\frac{\mu_h}{2E_i} \delta(1-\xi_i)\, \delta_{iP}\right) \delta(n_k-n_i)\, \bm{\mathcal{C}}^{\phantom{\dagger}}_{i\to P}\, \bm{\mathcal{C}}^\dagger_{i\to P} \right. \\ 
& \hspace{1.15cm} \left. - 2\spac \left(\gamma_0^i-  C_i \gamma_0^{\rm cusp} \ln\frac{\mu_h}{2E_i} \right) \delta(1-\xi_i)\, \delta_{iP}\right] ,
\end{aligned}
\end{equation}
we find that also the purely collinear anomalous dimension commutes with the wide-angle emissions
\begin{equation}\label{eq:coherenceColl}
   \bm{\mathcal{H}}\,\bm{\Gamma}_i^C(\xi_i)\,\bm{\overline{\Gamma}} 
   = \bm{\mathcal{H}}\,\bm{\overline{\Gamma}} \,\bm{\Gamma}_i^C(\xi_i) \,.
\end{equation}
As for~\eqref{eq:coherence}, this property follows from~\eqref{eq:colorcomb} and~\eqref{eq:splitting_functions_final_definitionColor}.
Furthermore, when inserted in the last step one can perform the color sum over the emitted parton, after which the real and virtual parts combine into the usual DGLAP kernels
\begin{equation} \label{eq:DGLAP_evolution_GammaC}
   \big\langle (\bm{\mathcal{H}} \ast \bm{\Gamma}_i^C)(x_i)\otimes\bm{1} \big\rangle\,  f_i(x_i) =   \big\langle \bm{\mathcal{H}}(x_i)\otimes\bm{1} \big\rangle \, \frac{\alpha_s}{\pi} \, (\mathcal{P}_{i\to P} \star f_i)(x_i) \,.
\end{equation}
Here the soft-collinear cusp pieces have cancelled out between the real and virtual terms. As mentioned above, this will no longer be the case beyond the leading order due to Glauber phases in the low-energy matrix elements. Compared to the contributions $C_{rn}$, contributions involving the collinear anomalous dimension involve fewer powers of logarithms at a given order and we will not discuss them further in this paper, but it is interesting that in the presence of Glauber phases the collinear evolution becomes more complicated than the one associated with the DGLAP equations.

\section{Iterated scale integrals and resummation}
\label{sec:integrals_resummation}

Expanding the path-ordered exponential in~\eqref{eq:U} one generates the ordered product~\eqref{eq:Uexp} of integrals over the anomalous dimension. The leading double-logarithmic terms in this series result from the iteration of the soft anomalous dimension $\bm{\Gamma}^S$ in~\eqref{eq:gammaH}, which in turn consists of the three elements given in~\eqref{eq:gammaSimp}. For these terms the Mellin convolutions are trivial, and we obtain the iterated integral
\begin{equation}\label{eq:muints}
   \int_{\mu_s}^{\mu_h}\!\frac{d\mu_1}{\mu_1} \int_{\mu_1}^{\mu_h}\!\frac{d\mu_2}{\mu_2}\,\,\dots 
   \int_{\mu_{n-1}}^{\mu_h}\!\frac{d\mu_n}{\mu_n}\,\bm{\Gamma}^S(\mu_h,\mu_n)\,\dots\,\bm{\Gamma}^S(\mu_h,\mu_2)\,
    \bm{\Gamma}^S(\mu_h,\mu_1) \,.
\end{equation}
To obtain the series of SLLs one picks out the different components of the anomalous dimension ($\bm{\Gamma}^c$, $\bm{V}^G$ or $\overline{\bm{\Gamma}}$) corresponding to the color traces in~\eqref{eq:colortraces}. This leads to the iterated integral 
\begin{equation}\label{eq:iteratedInts}
\begin{aligned}
   I_{rn}(\mu_h,\mu_s) 
   &= \int_{\mu_s}^{\mu_h}\!\frac{d\mu_1}{\mu_1}\,\frac{\alpha_s(\mu_1)}{4\pi} 
    \int_{\mu_1}^{\mu_h}\!\frac{d\mu_2}{\mu_2}\,\frac{\alpha_s(\mu_2)}{4\pi} \\
   &\quad\times \int_{\mu_2}^{\mu_h}\!\frac{d\mu_3}{\mu_3}\,\frac{\alpha_s(\mu_3)}{4\pi}\,\ln\frac{\mu_3^2}{\mu_h^2}
    \,\,\dots \int_{\mu_{n-r+1}}^{\mu_h}\!\frac{d\mu_{n-r+2}}{\mu_{n-r+2}}\,\frac{\alpha_s(\mu_{n-r+2})}{4\pi}\,
    \ln\frac{\mu_{n-r+2}^2}{\mu_h^2} \\
   &\quad\times \int_{\mu_{n-r+2}}^{\mu_h}\!\frac{d\mu_{n-r+3}}{\mu_{n-r+3}}\,\frac{\alpha_s(\mu_{n-r+3})}{4\pi} \\
   &\quad\times \int_{\mu_{n-r+3}}^{\mu_h}\!\frac{d\mu_{n-r+4}}{\mu_{n-r+4}}\,\frac{\alpha_s(\mu_{n-r+4})}{4\pi}\, 
    \ln\frac{\mu_{n-r+4}^2}{\mu_h^2}\,\,\dots 
    \int_{\mu_{n+2}}^{\mu_h}\!\frac{d\mu_{n+3}}{\mu_{n+3}}\,\frac{\alpha_s(\mu_{n+3})}{4\pi}\,
    \ln\frac{\mu_{n+3}^2}{\mu_h^2} \,.
\end{aligned}
\end{equation}
The integrals in the second and fourth line of this expression result from the $(n-r)$ and $r$ insertions of $\bm{\Gamma}^c$ in~\eqref{eq:colortraces}, respectively. 

From~\eqref{eq:factorization_formula},~\eqref{eq:Uexp} and~\eqref{eq:Wlead}, we find that in leading double-logarithmic approximation the cross section for a $2\to M$ jet process can be written in the form 
\begin{equation}
   \sigma_{2\to M}^{\rm SLL}(Q_0) 
   = \sum_{i \in \{q,\bar q,g\}} \int\!dx_1 \int\!dx_2\,f_{1}(x_1,\mu_s)\,f_{2}(x_2,\mu_s)\, 
    \sum_{n=0}^\infty \sum_{r=0}^n\,I_{rn}(\mu_h,\mu_s)\,C_{rn} \,,
\end{equation}
with $\mu_h\sim\sqrt{\hat s}$ and $\mu_s\sim Q_0$. Note that the color traces depend on the partonic channels contributing to the Born-level $2\to M$ scattering process, $C_{rn}\equiv C_{rn}^{1+2\to 3+\dots+(2+M)}$, but this dependence is implicit in our notation. The formula for the cross section includes a sum over all contributing partonic channels, and $i=1,2,\dots (2+M)$ refers to any of the partons involved in the process.

\subsection{Evaluation of the iterated integrals}

In the strict leading double-logarithmic approximation, the running of the coupling can be neglected in~\eqref{eq:iteratedInts}, because this is a single-logarithmic effect. Then the above expression simplifies to~\cite{Becher:2021zkk}
\begin{equation}\label{eq:norunning}
   I_{rn}(\mu_h,\mu_s) \Big|_{\text{no running}}
   = \left( \frac{\alpha_s(\bar\mu)}{4\pi} \right)^{n+3} 
    \frac{\left(-4\right)^n n!}{(2n+3)!}\,\frac{(2r)!}{4^r \left(r!\right)^2}\,
    \ln^{2n+3}\!\bigg(\frac{\mu_h}{\mu_s}\bigg) \,.   
\end{equation}
We will see that the scale ambiguity under variations of $\bar\mu$ is significant because the coupling $\alpha_s(\bar\mu)$ enters with a large power. To estimate the corresponding uncertainty, we will vary the reference scale $\bar\mu$ in the interval between $\mu_s$ and $\mu_h$.

The ambiguity in the choice of $\bar\mu$ can be avoided if we evaluate the ordered integrals including the scale dependence of the running QCD coupling. At leading order, we use
\begin{equation}
\begin{aligned}
   \int_{\mu}^{\mu_h}\!\frac{d\nu}{\nu}\,\frac{\alpha_s(\nu)}{4\pi} 
   &= \frac{1}{2\beta_0}\,\int_{x_\mu}^1\!\frac{dx}{x} \,, \\
   \int_{\mu}^{\mu_h}\!\frac{d\nu}{\nu}\,\frac{\alpha_s(\nu)}{4\pi}\,\ln\frac{\nu^2}{\mu_h^2} 
   &= - \frac{1}{2\beta_0}\,\frac{4\pi}{\beta_0\spac\alpha_s(\mu_h)}\,
    \int_{x_\mu}^1\!\frac{dx}{x}\,(1-x) \,,
\end{aligned}
\end{equation}
where $x_\mu=\alpha_s(\mu_h)/\alpha_s(\mu)$, and $\beta_0$ is the one-loop coefficient of the QCD $\beta$-function. We then obtain
\begin{equation}\label{eq:Intsresult}
\begin{aligned}
   I_{rn}(\mu_h,\mu_s) 
   &= \left( \frac{1}{2\beta_0} \right)^{n+3} 
    \left[ \frac{-4\pi}{\beta_0\spac\alpha_s(\mu_h)} \right]^n
    \int_{x_s}^1\!\frac{dx_1}{x_1} \int_{x_1}^1\!\frac{dx_2}{x_2} \\
   &\quad\times \int_{x_2}^1\!\frac{dx_3}{x_3}\,(1-x_3)\,\,\dots 
    \int_{x_{n-r+1}}^1\!\frac{dx_{n-r+2}}{x_{n-r+2}}\,(1-x_{n-r+2}) 
    \int_{x_{n-r+2}}^1\!\frac{dx_{n-r+3}}{x_{n-r+3}} \\
   &\quad\times \int_{x_{n-r+3}}^1\!\frac{dx_{n-r+4}}{x_{n-r+4}}\,(1-x_{n-r+4})\,\,\dots 
    \int_{x_{n+2}}^1\!\frac{dx_{n+3}}{x_{n+3}}\,(1-x_{n+3}) \\
   &\equiv \left( \frac{1}{2\beta_0} \right)^{n+3} 
    \left[ \frac{-4\pi}{\beta_0\spac\alpha_s(\mu_h)} \right]^n h_{nr}(x_s) \,,  
\end{aligned}
\end{equation}
where $x_s=\alpha_s(\mu_h)/\alpha_s(\mu_s)<1$. These iterated integrals generate simple functions $h_{nr}(x_s)$ containing logarithms and polynomials. For the calculation of the super-leading terms up to five-loop order one needs
\begin{align}\label{eq:hints}
   h_{00}(x) 
   &= - \frac{\ln^3\!x}{6} \,, \nonumber\\
   h_{10}(x) 
   &= \frac{\ln^4\!x}{24} - \frac{\ln^2\!x}{2} - (2+x) \ln x - 3 + 3x \,, \nonumber\\
   h_{11}(x) 
   &= \frac{\ln^4\!x}{24} + \frac{\ln^3\!x}{6} + \frac{\ln^2\!x}{2} + \ln x + 1 - x \,, \nonumber\\
   h_{20}(x) 
   &= - \frac{\ln^5\!x}{120} + \frac{\ln^3\!x}{6} + \left( \frac54 + x \right) \frac{\ln^2\!x}{2} 
    + \left( \frac12 - 2x - \frac{x^2}{8} \right) \ln x - \frac{21}{16} + x + \frac{5x^2}{16} \,, \nonumber\\
   h_{21}(x) 
   &= - \frac{\ln^5\!x}{120} - \frac{\ln^4\!x}{24} - \frac{\ln^3\!x}{6} 
    + \left( - \frac32 + x \right) \frac{\ln^2\!x}{2} - \left( \frac{11}{4} + 2x \right) \ln x 
    - \frac{39}{8} + 5x - \frac{x^2}{8} \,, \nonumber\\
   h_{22}(x) 
   &= - \frac{\ln^5\!x}{120} - \frac{\ln^4\!x}{24} - \frac{\ln^3\!x}{12} 
    + \frac{\ln^2\!x}{8} + \left( \frac98 + x \right) \ln x 
    + \frac{33}{16} - 2x - \frac{x^2}{16} \,.
\end{align}
The approximate result~\eqref{eq:norunning} can be recovered using the leading-order expression
\begin{equation}
   x_s = \frac{\alpha_s(\mu_h)}{\alpha_s(\mu_s)}
   \approx 1 - \frac{\beta_0\spac\alpha_s}{2\pi}\,\ln\frac{\mu_h}{\mu_s}
\end{equation}
and expanding the result to $\mathcal{O}(\alpha_s^{2n+3})$. We will find that the numerical results obtained using the expression~\eqref{eq:Intsresult} are close to the results obtained from the simpler form~\eqref{eq:norunning} with the intermediate scale choice $\bar\mu=\sqrt{\mu_h\spac\mu_s}$. The simpler form is, however, more convenient for performing the all-order resummation of the leading double logarithms.

It is straightforward to include the effects of the two-loop cusp anomalous dimension and two-loop $\beta$-function in this analysis, which eventually would be necessary to extend our calculation to a systematic analysis in leading order of RG-improved perturbation theory. This is accomplished by replacing 
\begin{equation} \label{eq:inclusion_2-loop_cusp_run}
   \int_y^1\!\frac{dx}{x}\,(1-x)
   \to \int_y^1\!\frac{dx}{x}\,(1-x)
    \left\{ 1 + \frac{\alpha_s(\mu_h)}{4\pi} \left[
    \left( \frac{\gamma_1^{\rm cusp}}{\gamma_0^{\rm cusp}} - \frac{\beta_1}{\beta_0} \right) \frac{1}{x}
    + \frac{\beta_1}{\beta_0}\,\frac{\ln x}{1-x} \right] + \dots \right\}
\end{equation}
for all ``cusp terms'' generating double logarithms in~\eqref{eq:Intsresult}, and
\begin{equation}\label{eq:inclusion_2-loop_cusp_run_Glauber}
   \int_y^1\!\frac{dx}{x}
   \to \int_y^1\!\frac{dx}{x} \left[ 1 + \frac{\alpha_s(\mu_h)}{4\pi} 
    \left( \frac{\gamma_1^{\rm cusp}}{\gamma_0^{\rm cusp}} - \frac{\beta_1}{\beta_0} \right) \frac{1}{x}
    + \dots \right]
\end{equation}
for the two Glauber terms. In the approximation where one works with a fixed coupling $\alpha_s(\bar\mu)$, as in~\eqref{eq:norunning}, one would obtain 
\begin{equation}
   I_{rn}(\mu_h,\mu_s) \Big|_{\text{no running}}
   = \left( \frac{\alpha_s(\bar\mu)}{4\pi} \right)^{n+3} 
    \left[ 1 + \frac{\gamma_1^{\rm cusp}}{\gamma_0^{\rm cusp}}\,\frac{\alpha_s(\bar\mu)}{4\pi} \right]^{n+2} 
    \frac{\left(-4\right)^n n!}{(2n+3)!}\,\frac{(2r)!}{4^r \left(r!\right)^2}\,
    \ln^{2n+3}\!\bigg(\frac{\mu_h}{\mu_s}\bigg) \spac .   
\end{equation}
To approximately take this effect into account in our numerical results presented in Section~\ref{sec:simpleprocesses}, we will simply replace 
\begin{equation} \label{eq:inclusion_2-loop_cusp}
    \alpha_s(\bar{\mu}) \to \left[ 1 + \frac{\gamma_1^{\rm cusp}}{\gamma_0^{\rm cusp}}\,\frac{\alpha_s(\bar\mu)}{4\pi} \right] \alpha_s(\bar{\mu})
\end{equation}
in the fixed-order results. For the results obtained using a running coupling, we will multiply the integrals $I_{rn}$ in~\eqref{eq:Intsresult} with the same factor to the $(3+n)$-th power evaluated at $\bar{\mu}=\sqrt{Q\spac Q_0}$, to avoid reevaluating the integrals in~\eqref{eq:Intsresult} according to~\eqref{eq:inclusion_2-loop_cusp_run}. Numerically this has the effect of increasing the running coupling by about six percent. 

\subsection{Resummation and asymptotic behavior of the super-leading logarithms}

We will derive an exact, closed-form expression for the color traces $C_{rn}$ for an arbitrary $2\to M$ process in Section~\ref{sec:traces}. While the resulting expressions for specific partonic channels can be lengthy, we find that in all cases the dependence on $r$ and $n$ can be factorized in the general form
\begin{equation}
   C_{rn} = \left( \gamma_0^{\rm cusp}\spac N_c \right)^n \bigg[ k_0\,\delta_{r0} + \sum_{i=1}^6\,k_i\,v_i^r \bigg] \,,
\end{equation}
with $\gamma_0^{\rm cusp}=4$ and process-dependent coefficients $k_i$ and parameters
\begin{equation} \label{eq:def_eigenvalues_v}
    v_1 = \frac{1}{2} \,, \qquad v_2 = 1 \,, \qquad v_{3,4} = \frac{3 N_c\pm 2}{2 N_c} \,, \qquad v_{5,6} = \frac{2\left( N_c\pm 1\right)}{N_c} \,,
\end{equation}
where $v_3$ and $v_5$ correspond to the plus signs. These $v_i$ arise as eigenvalues of $\bm{\Gamma}^c$ acting on the space of color structures, see Section~\ref{sec:traces}. Neglecting the running of the coupling, as is formally permitted at strict double-logarithmic accuracy, we derived in~\eqref{eq:norunning} a simple expression for the integrals $I_{rn}$. Using this expression and the power-like dependence of $C_{rn}$ on $r$, we find that the SLL contribution to the partonic cross section is given by the double sum
\begin{equation}\label{eq:resummedsigma}
 \hat{\sigma}_{2\to M}^{\rm SLL} = \sum_{n=0}^\infty \sum_{r=0}^n\,I_{rn}(\mu_h,\mu_s)\,C_{rn}
   = \left( \frac{\alpha_s(\bar\mu)}{4\pi} \right)^3 \frac16\,\ln^3\!\bigg(\frac{\mu_h}{\mu_s}\bigg)\,
    \bigg[ k_0\,\Sigma_0(w) + \sum_{i=1}^6\,k_i\,\Sigma(v_i,w) \bigg] \,,
\end{equation}
where 
\begin{equation}\label{eq:wdef}
    w = \frac{N_c\spac\alpha_s(\bar\mu)}{\pi}\,\ln^2\!\bigg(\frac{\mu_h}{\mu_s}\bigg)
\end{equation}
encodes the double-logarithmic dependence, and all relevant sums can be expressed in terms of the functions
\begin{equation}\label{eq:sum0}
   \Sigma_0(w) 
   = \sum_{n=0}^\infty\,\frac{\left(-4\right)^n 3!\,n!}{(2n+3)!}\,w^n
   = {}_2F_2\!\left(1,1;2,\frac{5}{2};-w\right) ,
\end{equation}
and
\begin{equation}
   \Sigma(v,w) = \sum_{n=0}^\infty \sum_{r=0}^n \frac{\left(-4\right)^n 3!\,n!}{(2n+3)!}\,\frac{(2r)!}{4^r \left(r!\right)^2}\,v^r\spac w^n \,,
\end{equation}
which satisfy $\Sigma_0(w)=\Sigma(0,w)$ and are normalized such that
\begin{equation}
   \Sigma_0(0) = \Sigma(v,0) = 1 \,.
\end{equation}
For small values $w\ll 1$, one finds the Taylor series
\begin{equation}\label{eq:Sigmaseries}
   \Sigma(v,w) = 1 - \frac{w}{5} \left( 1 + \frac{v}{2} \right) + \frac{4w^2}{105} \left( 1 + \frac{v}{2} + \frac{3v^2}{8} \right) + \dots \,.
\end{equation}
Notice that the expansion of each $\Sigma(v,w)$ function in powers of $w$ generates an alternating-sign perturbative series. This important fact can be traced back to the sign of the cusp logarithm in the soft anomalous dimension~\eqref{eq:cuspterms}.

Setting $n=m+r$, we can extend the sum over $r$ to infinity. Rewriting the factorials as Pochhammer symbols $(x)_n=\Gamma(x+n)/\Gamma(x)$, we obtain the representation
\begin{equation}\label{eq:Sigma}
   \Sigma(v,w) = \sum_{m=0}^\infty \sum_{r=0}^\infty \frac{\big(1\big)_{m+r} \big(1\big)_m \big(\frac{1}{2}\big)_r}{\big(2\big)_{m+r} \big(\frac{5}{2}\big)_{m+r}}\,\frac{(-w)^{m}\,(-v\spac w)^r}{m!\,r!} \,,
\end{equation}
which shows that the sum is a Kamp\'e de F\'eriet function\footnote{We thank B.~Ananthanarayan and Souvik Bera for pointing this out.}
\begin{equation}\label{eq:Feriet}
    \Sigma(v,w) = {}^{1+1} F_{2+0}\Big(\,
%    \begin{matrix} \phantom{1,\spac} 1:1,\frac{1}{2} \,; \\ 2,\frac{5}{2}: \emptyset,  \emptyset \,; \end{matrix}\,, -w, -v\spac w\Big) \,.
    \begin{matrix} \phantom{1,\spac} 1:1,\frac{1}{2} \,; \\ 2,\frac{5}{2}: \phantom{\emptyset, \emptyset} \,; \end{matrix} -w, -v\spac w\Big) \,.
\end{equation}
The arguments in the upper line indicate the Pochhammer symbols in the numerator, and the lower line corresponds to the ones in the denominator. A useful integral representation of $\Sigma(v,w)$ can be obtained by first performing the sum over $r$ in~\eqref{eq:Sigma} in terms of a hypergeometric function, then using an integral representation for this function, and finally performing the sum over $m$ in terms of the error function $\text{erf}(y)$. We find
\begin{equation}\label{eq:intref}
   \Sigma(v,w) = \int_0^1\!dx\,\frac{3}{4\sqrt{x}} \left[ \frac{2}{y^2(x)} - \frac{\sqrt{\pi}\,\text{erf}\left(y(x)\right)}{y^3(x)} \right] ,
\end{equation}
with $y(x)=\sqrt{w\spac\big(1+(v-1)\spac x\big)}$. 
For the special case $v=1$, it follows that
\begin{equation}\label{eq:int1}
   \Sigma_1(w) \equiv \Sigma(1,w) 
   = \frac{3}{w} - \frac{3\sqrt{\pi}\,\text{erf}\left(\sqrt{w}\right)}{2\spac w^{3/2}} \,.
\end{equation}
It is possible to carry out the integral in~\eqref{eq:intref} and obtain a representation in terms of the Owen $T$-function, which has an implementation in {\sc Mathematica}. We obtain
\begin{equation}
\begin{aligned}
   \Sigma(v,w) 
   &= \frac{3}{2z\sqrt{w}} \left[ 4\pi\,T\bigg(\sqrt{2}\,z,\frac{\sqrt{w}}{z}\bigg)
    - \frac{\sqrt{\pi}\,z\,\text{erf}\left(\sqrt{v\spac w}\right)}{\sqrt{v}\spac w} 
    + \frac{\sqrt{\pi}\,e^{-w}\,\text{erf}(z)}{\sqrt{w}} \right. \\ 
   &\hspace{1.8cm} \left. + \pi\,\text{erf}\left(\sqrt{w}\right) \text{erf}(z) + 2 \arccos\left(\frac{1}{\sqrt{v}}\right) - \pi \right] ,
\end{aligned}
\end{equation}
with $z=\sqrt{(v-1+i\varepsilon)\spac w}$, where the $i\varepsilon$ prescription is needed for the analytic continuation to $v<1$. 

\begin{figure}
    \centering
    \includegraphics[scale=1]{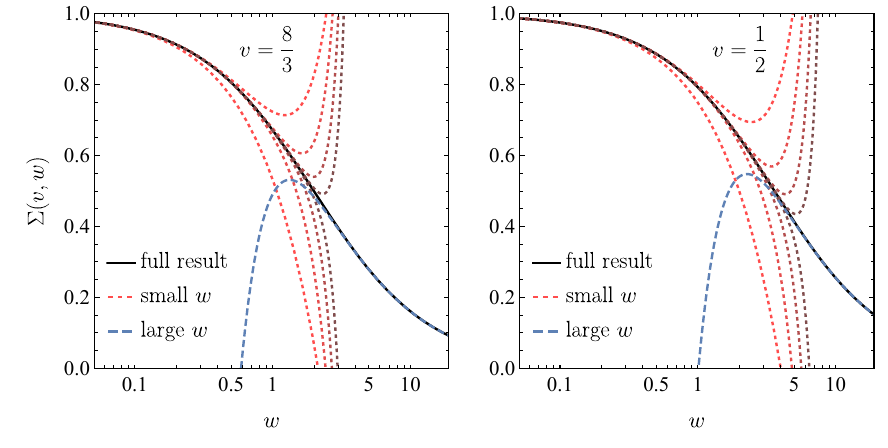}
    \caption{Plot of the function $\Sigma(v,w)$ for the largest parameter $v_5$ and the smallest non-zero value $v_1$. The full result is shown as a solid line. The red dotted lines show the perturbative expansion up to the eighth order in $w$. The blue dashed line is the large-$w$ asymptotics shown in~\eqref{eq:sigma_large_w}.}
    \label{fig:Feriet_asymptotics}
\end{figure}

It will be useful to derive the asymptotic behavior of the function $\Sigma(v,w)$ in the limit $w\to\infty$. For $\Sigma_0(w)$ we find from~\eqref{eq:sum0}
\begin{equation}\label{eq:sig0limit}
   \Sigma_0(w) = \frac{3}{2w}\,\Big( \ln(4w) + \gamma_E - 2 \Big) + \frac{3}{4 w^2} + \mathcal{O}(w^{-3}) \,.
\end{equation}
For the general case $v\ne 0$, we can derive the asymptotic behavior from~\eqref{eq:intref}, finding 
\begin{equation}\label{eq:sigma_large_w}
   \Sigma(v,w) = \frac{3\arctan\left(\sqrt{v-1}\right)}{\sqrt{v-1}\,w} - \frac{3\sqrt{\pi}}{2\sqrt{v}\,w^{3/2}} + \mathcal{O}(w^{-2}) \,.
\end{equation}
Note that the limits $v\to 0$ and $w\to \infty$ do not commute, and hence one does not recover~\eqref{eq:sig0limit} from~\eqref{eq:sigma_large_w}. It is interesting to contrast the asymptotic form of $\Sigma(v,w)$ and $\Sigma_0(w)$ to the standard double-logarithmic behavior, which in the variable $w$ translates to $e^{-c w}$, where the coefficient $c$ of the double logarithm depends on the process under consideration. The standard behavior leads to the exponential Sudakov suppression. The suppression is much weaker for the SLLs. We will come back to this after we analyze the color traces for a few simple processes.

\begin{figure}
    \centering
    \includegraphics[scale=1]{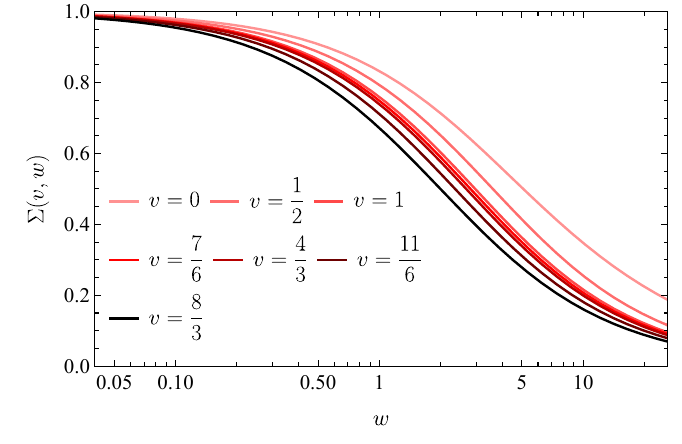}
    \caption{Behavior of the functions $\Sigma(v,w)$ for different values of $v$ corresponding to the eigenvalues in~\eqref{eq:def_eigenvalues_v}. Darker colors correspond to larger values of $v$.}
    \label{fig:Feriet_all}
\end{figure}

The functional form of $\Sigma(v,w)$ for two different values of $v$ is illustrated in Figure~\ref{fig:Feriet_asymptotics}, where we also show the perturbative expansion up to the eighth order in $w$ (dotted lines) and the asymptotic form~\eqref{eq:sigma_large_w} (dashed line). Note that in the phenomenologically interesting region $w\gtrsim 1$ the convergence of the Taylor series~\eqref{eq:Sigmaseries} is slow. In Figure~\ref{fig:Feriet_all} we show the functions $\Sigma(v,w)$ for all relevant eigenvalues $v_i$. We observe that the shape is fairly universal. As discussed in Section~\ref{sec:simpleprocesses}, this induces  cancellations that strongly reduce the super-leading effects in $2\to 0 $ and $2\to 1$ processes, for which the results can be expressed in terms of differences of $\Sigma(v_i,w)$ functions belonging to different eigenvalues.

\section{Evaluation of the color traces}
\label{sec:traces}

The second relation in~\eqref{eq:collsafety} implies that we can replace the last two color operators under the color trace in~\eqref{eq:colortraces} by their commutator $[\bm{V}^G,\bm{\overline{\Gamma}}]$. Introducing the abbreviation 
\begin{equation}\label{eq:Hinitial}
   \bm{\mathcal{H}} 
   = \bm{\mathcal{H}}_{2\to M} \left( \bm{\Gamma}^c \right)^r \bm{V}^G 
    \left( \bm{\Gamma}^c \right)^{n-r} ,
\end{equation}
we find after a straightforward calculation
\begin{align}
   \bm{\mathcal{H}}\,[\bm{V}^G,\bm{\overline{\Gamma}}]
   &= - 16\pi f_{abc} \sum_{i,j} \left( \delta_{i1}-\delta_{i2} \right) \nonumber\\[-1mm]
   &\times \bigg\{ \Big[\! \left( \bm{T}_1^a\,\bm{T}_2^b\,\bm{T}_j^c 
    + \bm{T}_j^c\,\bm{T}_1^a\,\bm{T}_2^b \right) \bm{\mathcal{H}}
    + \bm{\mathcal{H}} \left( \bm{T}_1^a\,\bm{T}_2^b\,\bm{T}_j^c 
    + \bm{T}_j^c\,\bm{T}_1^a\,\bm{T}_2^b \right) \!\Big] 
    \int\frac{d\Omega(n_{k_0})}{4\pi}\,\overline{W}_{ij}^{k_0} \nonumber\\
   &\qquad - 2 \left( \bm{T}_1^a\,\bm{T}_2^b\,\bm{\mathcal{H}}\,\bm{T}_j^c 
    + \bm{T}_j^c\,\bm{\mathcal{H}}\,\bm{T}_1^a\,\bm{T}_2^b \right) 
    \overline{W}_{ij}^{k_0}\,\Theta_{\rm hard}(n_{k_0}) \bigg\} \,.
\end{align}
Note that one of the two indices in the sum over $i$ and $j$ in $\bm{\overline{\Gamma}}$ must be equal to 1 or 2, corresponding to an attachment of the emitted soft gluon (with index $k_0$) on one of the initial-state partons, while the second index can be arbitrary. When the above result is inserted under the color trace in~\eqref{eq:colortraces}, we can use the cyclicity of the trace to move all color generators to the right-hand side of $\bm{\mathcal{H}}$, and the symbol $\otimes$ implies that we must integrate over the direction $n_{k_0}$ of the emitted gluon. We obtain 
\begin{equation}
\begin{aligned}
   \big\langle \bm{\mathcal{H}}\,\bm{V}^G\,\overline{\bm{\Gamma}}\otimes\bm{1} \big\rangle 
   &= - 32\pi f_{abc} \sum_j \big\langle \bm{\mathcal{H}}
    \left( \bm{T}_1^a\,\bm{T}_2^b\,\bm{T}_j^c + \bm{T}_j^c\,\bm{T}_1^a\,\bm{T}_2^b \right) \big\rangle \\
   &\quad\times \int\frac{d\Omega(n_{k_0})}{4\pi} \left( \overline{W}_{1j}^{k_0} - \overline{W}_{2j}^{k_0} \right)
    \Theta_{\rm veto}(n_{k_0}) \,,
\end{aligned}
\end{equation}
where $\Theta_{\rm veto}(n_{k_0})\equiv 1-\Theta_{\rm hard}(n_{k_0})$, and we have used the fact that $\overline{W}_{ii}^k=0$ by definition. The right-hand side still contains the integrals over the directions of the up to $(M+n)$ remaining partons in $\bm{\mathcal{H}}$, but for brevity we omit the symbol $\otimes \,\bm{1}$ here and below. It is not difficult to show that the color trace on the right-hand side of this relation vanishes if $j=1,2$ refers to one of the initial-state partons. We thus obtain
\begin{equation}\label{eq:step1}
   \big\langle \bm{\mathcal{H}}\,\bm{V}^G\,\overline{\bm{\Gamma}}\otimes\bm{1} \big\rangle 
   = - 64\pi f_{abc} \sum_{j>2} J_j\,\big\langle \bm{\mathcal{H}}\,\bm{T}_1^a\,\bm{T}_2^b\,\bm{T}_j^c \big\rangle \,,
\end{equation}
where we have defined
\begin{equation}\label{eq:Jints}
   J_j \equiv \int\frac{d\Omega(n_k)}{4\pi} \left( W_{1j}^k - W_{2j}^k \right) \Theta_{\rm veto}(n_k) \,.
\end{equation}
The fact that the angular integration is now restricted to the region outside the jets allows us to replace the subtracted dipoles $\overline{W}_{ij}^k$ defined in~\eqref{eq:subtractedDipole} with the original, unsubtracted ones. Already at this stage, we observe that all information about the phase-space restrictions on the direction of the emitted gluon $k_0$ is contained in the angular integrals $J_j$. If the gluon is emitted from one of the hard final-state partons present in the Born process, then $n_j$ is equal to the direction of that parton. If instead the gluon is radiated off one of the gluons collinear to the beam, which can be obtained from any of the insertions of $\bm{\Gamma}^c$ in the structure $\bm{\mathcal{H}}$ defined in~\eqref{eq:Hinitial}, then its direction $n_j$ is equal to $n_1$ or $n_2$. In this case, we encounter the integral 
\begin{equation}\label{eq:J12def}
   J_{12}\equiv J_2 = - J_1 = \int\frac{d\Omega(n_k)}{4\pi}\,W_{12}^k\,\Theta_{\rm veto}(n_k) \,. 
\end{equation}
Overall, there are thus $(M+1)$ independent kinematic structures $J_j$ for a $2\to M$ jet process. 

Given the result~\eqref{eq:step1}, we will now successively evaluate the effects of the various insertions of $\bm{\Gamma}^c$ and $\bm{V}^G$ contained in the original structure $\bm{\mathcal{H}}$ in~\eqref{eq:Hinitial}, working from right to left.

\subsection[First set of insertions of \texorpdfstring{$\Gamma^c$}{Gamma c}]{First set of insertions of $\bm{\Gamma}^c$}

We first evaluate the action of the right-most factor of $\bm{\Gamma}^c$ in the hard function $\bm{\mathcal{H}}$ in~\eqref{eq:Hinitial} on the result shown above, assuming that $(n-r)\ge 1$ (otherwise this step is skipped). We obtain
\begin{equation}\label{eq:firstGc}
   \big\langle \bm{\mathcal{H}}\,\bm{\Gamma}^c\,\bm{V}^G\,\overline{\bm{\Gamma}}\otimes\bm{1} \big\rangle
   = - 256\pi f_{abc} \sum_{j>2} J_j \sum_{i=1,2}
    \big\langle \bm{\mathcal{H}}\,C_i\,\bm{T}_1^a\,\bm{T}_2^b\,\bm{T}_j^c 
    - \bm{\mathcal{H}}\,\bm{T}_{i,L}\circ\bm{T}_{i,R}\,\bm{T}_1^a\,\bm{T}_2^b\,\bm{T}_j^c \big\rangle \,,
\end{equation}
where now
\begin{equation}\label{eq:Hnext}
   \bm{\mathcal{H}} 
   = \bm{\mathcal{H}}_{2\to M} \left( \bm{\Gamma}^c \right)^r \bm{V}^G 
    \left( \bm{\Gamma}^c \right)^{n-r-1}
\end{equation}
contains one insertion of $\bm{\Gamma}^c$ less than before. For the term involving the $\circ$ symbol in~\eqref{eq:firstGc} we need to distinguish the two cases where parton $j$ coincides with the collinear gluon $k_1$ emitted from the explicit factor $\bm{\Gamma}^c$, or where it is one of the remaining partons. Using the definition of the $\circ$ symbol shown in~\eqref{eq:circdef}, we find
\begin{equation}\label{eq:twoterms}
   \sum_{j>2} J_j \big\langle \bm{\mathcal{H}}\,\bm{T}_{i,L}\circ\bm{T}_{i,R}\,\bm{T}_1^a\,\bm{T}_2^b\,\bm{T}_j^c \big\rangle 
   = {\sum_{j>2}}^{\,\prime} J_j \big\langle \bm{\mathcal{H}}\,\bm{T}_i^A\,\bm{T}_1^a\,\bm{T}_2^b\,\bm{T}_j^c\,\bm{T}_i^A \big\rangle 
    -if_{cBA}\,J_{k_1} \big\langle \bm{\mathcal{H}}\,\bm{T}_i^B\,\bm{T}_1^a\,\bm{T}_2^b\,\bm{T}_i^A \big\rangle \spac,
\end{equation}
where the prime on the sum in the first term means that $j\ne k_1$. We find that the last term on the right-hand side vanishes after contraction with $f_{abc}$ for each $i=1,2$ separately. Physically, this means that the (virtual or real) gluon emitted by the insertion of $\overline{\bm{\Gamma}}$ does \emph{not} attach to the collinear gluon emitted by the insertion of $\bm{\Gamma}^c$. To arrive at this result, we have used the identity 
\begin{equation}\label{eq:3fidentity}
   f_{aAB} f_{bBC} f_{c\spac CA} = \frac{N_c}{2}\,f_{abc} \,.
\end{equation}
Considering the first term in~\eqref{eq:twoterms}, our strategy is to move the color generator $\bm{T}_i^A$ that sits next to $\bm{\mathcal{H}}$ all the way to the right, where it multiplies the second insertion of $\bm{T}_i^A$ to produce a factor $C_i$, yielding a contribution which cancels the first term on the right-hand side of~\eqref{eq:firstGc}. The only leftover contributions are those from the commutator terms, and we obtain
\begin{equation}
   \big\langle \bm{\mathcal{H}}\,\bm{\Gamma}^c\,\bm{V}^G\,\overline{\bm{\Gamma}}\otimes\bm{1} \big\rangle
   = - 64\pi\,(4 N_c)\spac f_{abc} {\sum_{j>2}}^{\,\prime}\spac J_j\, 
    \big\langle \bm{\mathcal{H}}\,\bm{T}_1^a\,\bm{T}_2^b\,\bm{T}_j^c \big\rangle \,.
\end{equation}
Surprisingly, we find that the insertion of $\bm{\Gamma}^c$ has the effect of reproducing the previous structure~\eqref{eq:step1} up to an overall color factor $(4N_c)$, combined with the restriction that the sum over $j$ no longer contains the collinear gluon $k_1$.

We can repeat this argument for the remaining $(n-r-1)$ insertions of $\bm{\Gamma}^c$ on the right-hand side of~\eqref{eq:Hnext}, finding
\begin{equation}\label{eq:step2}
   \big\langle \bm{\mathcal{H}} \left( \bm{\Gamma}^c \right)^{n-r} 
    \bm{V}^G\,\overline{\bm{\Gamma}} \otimes \bm{1} \big\rangle 
   = - 64\pi \left( 4 N_c \right)^{n-r} f_{abc}  {\sum_{j>2}}^{\,\prime} \spac J_j\,
    \big\langle \bm{\mathcal{H}}\,\bm{T}_1^a\,\bm{T}_2^b\,\bm{T}_j^c \big\rangle \,,
\end{equation}
where now
\begin{equation}\label{eq:newHstructure}
   \bm{\mathcal{H}}
   = \bm{\mathcal{H}}_{2\to M} \left( \bm{\Gamma}^c \right)^r \bm{V}^G \,.
\end{equation}
The prime on the sum over $j$ now means that this sum does not include any of the collinear gluons emitted from the $(n-r)$ insertions of $\bm{\Gamma}^c$. Importantly, however, the sum \emph{does} include the collinear emissions of the remaining $r$ insertions of $\bm{\Gamma}^c$ still contained in the structure $\bm{\mathcal{H}}$ in~\eqref{eq:newHstructure}.

\subsection{Insertion of the second Glauber phase}

A more complicated structure arises when one applies the second insertion of $\bm{V}^G$ to expression~\eqref{eq:step2}. After some algebra making use of the group-theory identity~\eqref{eq:3fidentity}, we find
\begin{equation}\label{eq:step3}
\begin{aligned}
   & \big\langle \bm{\mathcal{H}}\,\bm{V}^G \left( \bm{\Gamma}^c \right)^{n-r} 
    \bm{V}^G\,\overline{\bm{\Gamma}} \otimes \bm{1} \big\rangle \\[1mm]
   ={} & - 256\pi^2 \left( 4 N_c \right)^{n-r} f_{abe}\spac f_{cde} {\sum_{j>2}}^{\,\prime}
   \spac J_j\,
    \big\langle \bm{\mathcal{H}} \left( \bm{T}_2^a\,\{ \bm{T}_1^b,\bm{T}_1^c \}\,\bm{T}_j^d
    - \bm{T}_1^a\,\{ \bm{T}_2^b,\bm{T}_2^c \}\,\bm{T}_j^d \right) \!\big\rangle \,,
\end{aligned}
\end{equation}
where now
\begin{equation}\label{eq:Hremainder}
   \bm{\mathcal{H}} = \bm{\mathcal{H}}_{2\to M} \left( \bm{\Gamma}^c \right)^r .
\end{equation}
The result involves anti-commutators of color generators, which in general cannot be simplified using the Lie algebra of $SU(N_c)$. 

As a side remark, let us mention that if particles~1 and 2 transform in the fundamental or anti-fundamental representation of $SU(N_c)$, one can use the relation (for $i=1,2$)
\begin{equation}\label{eq:fundrep}
   \{ \bm{T}_i^a,\bm{T}_i^b \}
   = \frac{1}{N_c}\,\delta_{ab}\,\bm{1} + \sigma_i\,d_{abc}\,\bm{T}_i^c \,,
\end{equation}
where $\sigma_i=1$ for an initial-state anti-quark and $\sigma_i=-1$ for an initial-state quark, to eliminate the anti-commutators. The $d$-symbol appearing on the right-hand side is totally symmetric in its indices. The result~\eqref{eq:step3} can then be simplified by means of the identity
\begin{equation}
   f_{abe}\spac f_{cde}\!\left( \bm{T}_2^a\,\{ \bm{T}_1^b,\bm{T}_1^c \}\,\bm{T}_j^d
    - \bm{T}_1^a\,\{ \bm{T}_2^b,\bm{T}_2^c \}\,\bm{T}_j^d \right) 
   = (\bm{T}_1-\bm{T}_2)\cdot\bm{T}_j - \frac{N_c}{2} \left( \sigma_1 - \sigma_2 \right) 
    d_{abc}\,\bm{T}_1^a\spac\bm{T}_2^b\spac\bm{T}_j^c \,.
\end{equation}
Based on this observation, we have shown in~\cite{Becher:2021zkk} that in this case, the color traces evaluate to (only for particles~1 and 2 in the (anti-)fundamental representation)
\begin{align}\label{eq:oursimpleresult}
  C_{rn} &= - 2^{8-r} \pi^2 \left( 4 N_c \right)^n 
    \bigg\{ \sum_{j=3}^{M+2}\spac J_j\spac\big\langle \bm{\mathcal{H}}_{2\to M}
    \Big[ (\bm{T}_1-\bm{T}_2)\cdot\bm{T}_j - 2^r\spac\frac{N_c}{2} \left( \sigma_1 - \sigma_2 \right) 
    d_{abc}\,\bm{T}_1^a\spac\bm{T}_2^b\spac\bm{T}_j^c \Big] \big\rangle \nonumber\\
   &\hspace{3.68cm} - 2 \left( 1 - \delta_{r0} \right) J_{12}\, 
    \big\langle \bm{\mathcal{H}}_{2\to M}\,\big[ C_F\,\bm{1} + \left( 2^r - 1 \right) \bm{T}_1\cdot\bm{T}_2 \big]
    \spac\big\rangle \bigg\} \,.
\end{align}
In the following, we will generalize this simple result to the general case, where particles~1 or 2 (or both) do not transform in the (anti-)fundamental representation.

\subsection[Remaining insertions of \texorpdfstring{${\Gamma}^c$}{Gamma c}]{Remaining insertions of $\bm{\Gamma}^c$}
\label{subsec:difficult}

When one applies the right-most factor of $\bm{\Gamma}^c$ in~\eqref{eq:Hremainder} to the color structure shown in~\eqref{eq:step3}, additional structures are generated. In the following discussion, we assume that $r\ge 1$, otherwise this step is skipped. As in~\eqref{eq:twoterms}, we need to distinguish between the terms in the sum where $j$ is equal to the collinear gluon emitted by the new insertion of $\bm{\Gamma}^c$, and the remaining terms where $j$ labels a different parton. Contrary to the discussion following relation~\eqref{eq:twoterms}, we find that the contribution where $j=k_{n-r+1}$ refers to the new collinear gluon does \emph{not} vanish. Instead, it produces a term involving the angular integral $J_{12}$ defined in~\eqref{eq:J12def}, because the collinear gluon moves in the direction of either particle 1 or particle 2. After some algebra, we arrive at the result
\begin{align}\label{eq:step4}
   & \big\langle \bm{\mathcal{H}}\,\bm{\Gamma}^c\,\bm{V}^G \left( \bm{\Gamma}^c \right)^{n-r} 
    \bm{V}^G\,\overline{\bm{\Gamma}}\otimes\bm{1} \big\rangle \nonumber\\   
   ={} & - 256\pi^2 \left( 4 N_c \right)^{n-r} f_{abe}\spac f_{cde}\,
    \Bigg\{   {\sum_{j>2}}^{\,\prime}\spac J_j\,
    \Big[\spac 6 N_c\,\big\langle \bm{\mathcal{H}} \left( \bm{T}_2^a\,\{ \bm{T}_1^b,\bm{T}_1^c \}\,\bm{T}_j^d
    - \bm{T}_1^a\,\{ \bm{T}_2^b,\bm{T}_2^c \}\,\bm{T}_j^d \right) \!\big\rangle \nonumber\\
   &\hspace{4.7cm} - 4 f_{Bbg}\spac f_{Ccg}\, 
    \big\langle \bm{\mathcal{H}} \left( \bm{T}_2^a\,\{ \bm{T}_1^B,\bm{T}_1^C \}\,\bm{T}_j^d
    - \bm{T}_1^a\,\{ \bm{T}_2^B,\bm{T}_2^C \}\,\bm{T}_j^d \right) \!\big\rangle \Big] \nonumber\\
   & + 4 J_{12}\,\bigg[\spac \frac{N_c}{2}\,
    \big\langle \bm{\mathcal{H}}\,\bm{T}_2^a\,\{ \bm{T}_1^b,\bm{T}_1^c \}\,(\bm{T}_1^d - \bm{T}_2^d) \big\rangle \nonumber\\
   &\hspace{1.4cm} + f_{Aag}\spac f_{Ddg}\,
    \big\langle \bm{\mathcal{H}}\,\bm{T}_2^A\,\{ \bm{T}_1^b,\bm{T}_1^c \}\,\bm{T}_2^D \big\rangle 
    - f_{Bbg}\spac f_{Ddg}\,
    \big\langle \bm{\mathcal{H}}\,\bm{T}_2^a\,\{ \bm{T}_1^B,\bm{T}_1^c \}\,\bm{T}_1^D \big\rangle \nonumber\\
    &\hspace{1.4cm} - f_{Ccg}\spac f_{Ddg}\,
    \big\langle \bm{\mathcal{H}}\,\bm{T}_2^a\,\{ \bm{T}_1^b,\bm{T}_1^C \}\,\bm{T}_1^D \big\rangle
    + (1\leftrightarrow 2) \bigg] \Bigg\} \,.
\end{align}
The primed sum over $j$ now includes only the final-state partons contained in the residual structure 
\begin{equation}
   \bm{\mathcal{H}} = \bm{\mathcal{H}}_{2\to M} \left( \bm{\Gamma}^c \right)^{r-1} .
\end{equation}
Note the important fact that in each term on the right-hand side the number of color generators is still four, as in~\eqref{eq:step3}. In the terms under the sum over $j$, the original color structure is reproduced times a factor $6N_c$, and a second, analogous structure is generated, which involves four rather than two $f$-symbols. The contribution proportional to $J_{12}$ has a more complicated form since it contains color structures in which three generators act on the same initial-state parton.

In the result~\eqref{eq:step4} we encounter products of four $f$-symbols, which can be simplified using the notation
\begin{equation}
   \left( F^a \right)_{bc} = -i f_{abc}
\end{equation}
for the generators in the adjoint representation of $SU(N_c)$ and noting the trace relations~\cite{Haber:2019sgz}
\begin{equation}\label{eq:traces1}
\begin{aligned}
   \mbox{Tr}\,\big( F^a F^b \big) 
   &= N_c\,\delta_{ab} \,, \\
   \mbox{Tr}\,\big( F^a F^b F^c \big) 
   &= \frac{N_c}{2}\,if_{abc} \,, \\
   \mbox{Tr}\,\big( F^a F^b F^c F^d \big) 
   &= \delta_{ad}\,\delta_{bc} + \frac12 \left( \delta_{ab}\,\delta_{cd} + \delta_{ac}\,\delta_{bd} \right)
    + \frac{N_c}{4} \left( f_{ade}\spac f_{bce} + d_{ade}\spac d_{bce} \right) ,
\end{aligned}
\end{equation}
the second of which is equivalent to~\eqref{eq:3fidentity}. Next, we use the symmetry properties of the anti-commutators and define the symbols
\begin{equation}\label{eq:F2F4}
\begin{aligned}
   F_{abcd}^{(2)} 
   &= \frac12 \left( f_{abe}\spac f_{cde} + f_{ace}\spac f_{bde} \right) , \\
   F_{abcd}^{(4)} 
   &= \frac12\,\mbox{Tr}\,\big( F^a \{ F^b, F^c\}\spac F^d \big) 
    = \delta_{ad}\,\delta_{bc} + \frac12 \left( \delta_{ab}\,\delta_{cd} + \delta_{ac}\,\delta_{bd} \right)
    + \frac{N_c}{4}\,d_{ade}\spac d_{bce} \,,
\end{aligned}
\end{equation}
both of which are symmetric in the index pairs $(b,c)$ and $(a,d)$. The result~\eqref{eq:step4} can then be recast in the form
\begin{equation}\label{eq:step4b}
\begin{aligned}
   & \big\langle \bm{\mathcal{H}}\,\bm{\Gamma}^c\,\bm{V}^G \left( \bm{\Gamma}^c \right)^{n-r} 
    \bm{V}^G\,\overline{\bm{\Gamma}} \otimes \bm{1} \big\rangle \\   
   ={} & - 256\pi^2 \left( 4 N_c \right)^{n-r}\! \bigg\{  {\sum_{j>2}}^{\,\prime} \spac J_j
    \left[ 6 N_c\spac F_{abcd}^{(2)} + 4 F_{abcd}^{(4)} \right]
    \big\langle \bm{\mathcal{H}} \left( \bm{T}_2^a\,\{ \bm{T}_1^b,\bm{T}_1^c \}\,\bm{T}_j^d
    - (1\leftrightarrow 2) \right) \!\big\rangle \\
   &\hspace{3.1cm} - 4 J_{12} \left[ \frac{N_c}{2}\,F_{abcd}^{(2)} + F_{abcd}^{(4)} \right]\! 
    \big\langle \bm{\mathcal{H}} \left( \bm{T}_2^a\,\{ \bm{T}_1^b,\bm{T}_1^c \}\,(\bm{T}_1^d + \bm{T}_2^d) 
    + (1\leftrightarrow 2) \right) \!\big\rangle\! \bigg\} \spac.
\end{aligned}
\end{equation}
Note that the structure $F_{abcd}^{(4)}$ can be generated from $F_{abcd}^{(2)}$ using the relation
\begin{equation}
   F_{abcd}^{(4)} = - f_{Bbe}\spac f_{Cce}\,F_{aBCd}^{(2)} \,.
\end{equation}

Let us now explore what happens when we pull out additional insertions of $\bm{\Gamma}^c$ from the structure $\bm{\mathcal{H}}$ and apply them to the color structures in~\eqref{eq:step4b}. We will first discuss the structures under the sum over $j$ and then focus on the new structures multiplying $J_{12}$. Naively, one would expect that at least in these latter structures the number of generators acting on particle~1 or 2 is increased by one under each insertion of $\bm{\Gamma}^c$. Fortunately, we will find that group theory is kind here and this is actually not the case.

\subsubsection*{\boldmath Color traces under the sum over $j$}

Considering the terms under the sum over $j$, we observe that the trace over color structures is the same as in~\eqref{eq:step3}. The effect of the insertion of $\bm{\Gamma}^c$ is simply to replace the coefficient according to
\begin{equation}\label{eq:F2replace}
  F_{abcd}^{(2)} ~\to~ 6 N_c\spac F_{abcd}^{(2)} + 4 F_{abcd}^{(4)} \,.
\end{equation}
This pattern repeats itself when we apply additional insertions of $\bm{\Gamma}^c$. In the next step, we find the replacement rule
\begin{equation}
\begin{aligned}
  6 N_c\spac F_{abcd}^{(2)} + 4 F_{abcd}^{(4)} ~\to~
  &\, 6 N_c \left[ 6 N_c\spac F_{abcd}^{(2)} + 4 F_{abcd}^{(4)} \right]
   + 4 \left[ 6 N_c\spac F_{abcd}^{(4)} + 4 F_{abcd}^{(6)} \right] \\[1mm]
  &= 36 N_c^2\spac F_{abcd}^{(2)} + 48 N_c\spac F_{abcd}^{(4)} + 16 F_{abcd}^{(6)} \,,
\end{aligned}
\end{equation}
where 
\begin{equation}
   F_{abcd}^{(6)} = - f_{Bbe}\spac f_{Cce}\,F_{aBCd}^{(4)} \,.
\end{equation}
This would seem to generate increasingly complicated tensor structures, but using the explicit form of $F_{abcd}^{(4)}$ in~\eqref{eq:F2F4} we find that this is, in fact, not the case. Instead, we obtain
\begin{equation}\label{eq:F6reduce}
   F_{abcd}^{(6)} = F_{abcd}^{(2)} - N_c\,\delta_{ad}\,\delta_{bc} - \frac{N_c^2}{8}\,d_{ade}\spac d_{bce} \,.
\end{equation}
To arrive at this result, we have defined the matrices
\begin{equation}
   \left( D^a \right)_{bc} = d_{abc}
\end{equation}
and used the trace relation~\cite{Haber:2019sgz}
\begin{equation}\label{eq:traces2}
   \mbox{Tr}\,\big( F^a F^b D^c \big) = \frac{N_c}{2}\,d_{abc} \,.
\end{equation}
Generalizing relation~\eqref{eq:F6reduce} to higher orders leads to 
\begin{equation}
   F_{abcd}^{(4+2n)} = F_{abcd}^{(2n)} + \left( -N_c \right)^n \delta_{ad}\,\delta_{bc} 
    - \frac12 \left( \frac{N_c}{2} \right)^{n+1} d_{ade}\spac d_{bce}
\end{equation}
for all $n\in\mathbb{N}$. It follows that any symbol $F_{abcd}^{(2n)}$ for $n\ge 3$ can be reduced to the two symbols in~\eqref{eq:F2F4} plus terms proportional to $\delta_{ad}\,\delta_{bc}$ and $d_{ade}\spac d_{bce}$. In other words, only four color tensors are generated by successive applications of $\bm{\Gamma}^c$, namely
\begin{equation}
   f_{abe}\spac f_{cde} \,, \qquad
   d_{ade}\spac d_{bce} \,, \qquad
   \delta_{ab}\,\delta_{cd} \,, \qquad
   \delta_{ad}\,\delta_{bc} \,.
\end{equation}
There is no need to symmetrize the first and the third structure in the index pair $(b,c)$ because the color trace
\begin{equation}
   \big\langle \bm{\mathcal{H}} \left( \bm{T}_2^a\,\{ \bm{T}_1^b,\bm{T}_1^c \}\,\bm{T}_j^d
    - (1\leftrightarrow 2) \right) \!\big\rangle
\end{equation}
with which these structures are contracted already has this symmetry.

At this point, we arrive at the result
\begin{equation}\label{eq:prelim}
    C_{rn} = - 256\pi^2 \left( 4 N_c \right)^{n-r} 
    \Bigg[ \sum_{j=3}^{M+2}\spac J_j\,\sum_{i=1}^4\,c_i^{(r)}\,
    \big\langle \bm{\mathcal{H}}_{2\to M}\,\bm{O}_i^{(j)} \big\rangle 
    + \text{terms proportional to $J_{12}$} \Bigg] \,,
\end{equation}
where the basis operators are defined as
\begin{equation}\label{eq:Oibasis}
\begin{aligned}
   \bm{O}_1^{(j)} 
   &= f_{abe}\spac f_{cde}\,\bm{T}_2^a\,\{ \bm{T}_1^b,\bm{T}_1^c \}\,\bm{T}_j^d - (1\leftrightarrow 2) \,, \\
   \bm{O}_2^{(j)} 
   &= d_{ade}\spac d_{bce}\,\bm{T}_2^a\,\{ \bm{T}_1^b,\bm{T}_1^c \}\,\bm{T}_j^d - (1\leftrightarrow 2) \,, \\
   \bm{O}_3^{(j)} 
   &= \bm{T}_2^a\,\{ \bm{T}_1^a,\bm{T}_1^b \}\,\bm{T}_j^b - (1\leftrightarrow 2) \,, \\
   \bm{O}_4^{(j)} 
   &= 2 C_1\,\bm{T}_2\cdot\bm{T}_j - 2 C_2\,\bm{T}_1\cdot\bm{T}_j \,.
\end{aligned}
\end{equation}
They are antisymmetric in the parton indices 1 and 2. From~\eqref{eq:step3} it follows that for the special case where $r=0$ we have 
\begin{equation}\label{eq:cinitial}
   c_i^{(0)} = \delta_{i1} \,.
\end{equation}
Applying $s$ insertions of $\bm{\Gamma}^c$ we generate the right-hand side of~\eqref{eq:prelim} with coefficients $c_i^{(s)}$. (We also generate terms proportional to $J_{12}$, which will be discussed below.) Applying $\bm{\Gamma}^c$ one more time, the four structures change to 
\begin{equation}\label{eq:recO}
\begin{aligned}
   \bm{O}_1^{(j)} 
   &\to 6 N_c\,\bm{O}_1^{(j)} + N_c\,\bm{O}_2^{(j)} + 4\spac\bm{O}_3^{(j)} + 4\spac\bm{O}_4^{(j)} \,, \\
   \bm{O}_2^{(j)} 
   &\to 4 N_c\,\bm{O}_2^{(j)} \,, \\
   \bm{O}_3^{(j)} 
   &\to 4\spac\bm{O}_1^{(j)} + 6 N_c\,\bm{O}_3^{(j)} \,, \\
   \bm{O}_4^{(j)} 
   &\to 2 N_c\,\bm{O}_4^{(j)} \,.
\end{aligned}
\end{equation}
The first relation follows from~\eqref{eq:F2replace}, and the remaining relations are readily derived by repeating the derivation of~\eqref{eq:step4} from~\eqref{eq:step3}, after replacing the overall color tensor $f_{abe}\spac f_{cde}$ with $d_{ade}\spac d_{bce}$, $\delta_{ab}\,\delta_{cd}$, and $\delta_{ad}\,\delta_{bc}$, respectively, and making use of the trace relations in~\eqref{eq:traces1} and~\eqref{eq:traces2}. The above replacement rules lead to the recurrence relations 
\begin{equation}
\begin{aligned}
   c_1^{(s+1)} &= 6 N_c\,c_1^{(s)} + 4\spac c_3^{(s)} \,, \\
   c_2^{(s+1)} &= N_c\,c_1^{(s)} + 4 N_c\,c_2^{(s)} \,, \\
   c_3^{(s+1)} &= 4\spac c_1^{(s)} + 6 N_c\,c_3^{(s)} \,, \\
   c_4^{(s+1)} &= 4\spac c_1^{(s)} + 2 N_c\,c_4^{(s)} \,.
\end{aligned}
\end{equation}
Solving this set of equations with the initial conditions in~\eqref{eq:cinitial}, we find
\begin{equation}\label{eq:cisolu}
\begin{aligned}
   c_1^{(r)} &= 2^{r-1}\spac\big[ \left( 3 N_c+2 \right)^r + \left( 3 N_c-2 \right)^r \big] \,, \\
   c_2^{(r)} &= 2^{r-2} N_c 
    \left[ \frac{\left( 3 N_c+2 \right)^r}{N_c+2} + \frac{\left( 3 N_c-2 \right)^r}{N_c-2} 
    - \frac{\left( 2 N_c \right)^{r+1}}{N_c^2-4} \right] , \\
   c_3^{(r)} &= 2^{r-1}\spac\big[ \left( 3 N_c+2 \right)^r - \left( 3 N_c-2 \right)^r \big] \,, \\
   c_4^{(r)} &= 2^{r-1} 
    \left[ \frac{\left( 3 N_c+2 \right)^r}{N_c+1} + \frac{\left( 3 N_c-2 \right)^r}{N_c-1} 
    - \frac{2\spac N_c^{r+1}}{N_c^2-1} \right] .
\end{aligned}
\end{equation}

\begin{figure}
    \centering
    \includegraphics[scale=1]{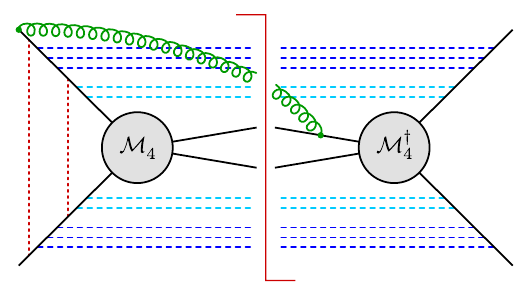}
    \caption{Representative diagram depicting a contribution to the color trace $C_{rn}$ relevant for $M=2$ jet production. The soft wide-angle gluon emission mediated by $\overline{\bm{\Gamma}}$ is shown in green, Glauber exchanges described by $\bm{V}^G$ are drawn as red dotted lines, and collinear gluon emissions governed by $\bm{\Gamma}^c$ are represented by the dashed blue lines. The diagram shows a contribution to $C_{4,10}$ which involves $r=4$ emissions (light-blue lines) before the first Glauber exchange and $(n-r)=6$ emissions between the two Glauber exchanges (dark-blue lines). The same color trace also gets contributions involving Glauber exchanges in the conjugate amplitude and wide-angle soft emissions from other legs of the Born-level hard function.}
    \label{fig:traceExampleO}
\end{figure}

Figure~\ref{fig:traceExampleO} shows a representative example of the contributions from the operators $\bm{O}_i^{(j)}$ in~\eqref{eq:prelim} for the case of a $2\to 2$ hard-scattering process. The wide-angle soft gluon emitted into the gap between the jets must be attached to one of the initial-state partons in the Born amplitude and one of the final-state partons in the conjugate Born amplitude, or vice versa. In the evolution from the hard scale to lower scales (inside to out in the figure), the attachment to the initial-state parton must happen after the second Glauber exchange has taken place. Only in this case a non-zero SLL contribution is obtained.

\subsubsection*{\boldmath Terms proportional to $J_{12}$}

We now return to the terms shown in the last line of~\eqref{eq:step4b}, which using~\eqref{eq:F2F4} can be expressed in terms of traces of $\bm{\mathcal{H}}$ with the four color structures
\begin{align}\label{eq:descendants}
   f_{abe}\spac f_{cde}\,\bm{T}_2^a\,\{ \bm{T}_1^b,\bm{T}_1^c \}\,\big( \bm{T}_1^d + \bm{T}_2^d \big)
    + (1\leftrightarrow 2) 
   &= f_{abe}\spac f_{cde}\,\{ \bm{T}_1^b,\bm{T}_1^c \}\,\{ \bm{T}_2^a,\bm{T}_2^d \} 
    - \frac{N_c^2}{2}\,\bm{T}_1\cdot\bm{T}_2 \,, \nonumber\\
   d_{ade}\spac d_{bce}\,\bm{T}_2^a\,\{ \bm{T}_1^b,\bm{T}_1^c \}\,\big( \bm{T}_1^d + \bm{T}_2^d \big)
    + (1\leftrightarrow 2) 
   &= d_{ade}\spac d_{bce}\,\{ \bm{T}_1^b,\bm{T}_1^c \}\,\{ \bm{T}_2^a,\bm{T}_2^d \} 
    + \frac{N_c^2-4}{3}\,\bm{T}_1\cdot\bm{T}_2 \nonumber\\
   &\quad + 2\spac d_{ade}\spac d_{bce} \left[ \bm{T}_2^a\,\big( \bm{T}_1^b\spac\bm{T}_1^c\spac\bm{T}_1^d \big)_+
    + (1\leftrightarrow 2) \right] , \nonumber\\[1mm]
   \bm{T}_2^a\,\{ \bm{T}_1^a,\bm{T}_1^b \}\,\big( \bm{T}_1^b + \bm{T}_2^b \big) + (1\leftrightarrow 2) 
   &= \{ \bm{T}_1^a,\bm{T}_1^b \}\,\{ \bm{T}_2^a,\bm{T}_2^b \} \nonumber\\[1mm]
   &\quad + \left( 2\spac C_1 + 2\spac C_2 - N_c \right) \bm{T}_1\cdot\bm{T}_2 \,, \nonumber\\[1mm]
   C_1\,\bm{T}_2\cdot \left( \bm{T}_1 + \bm{T}_2 \right) + (1\leftrightarrow 2) 
   &= \left( C_1 + C_2 \right) \bm{T}_1\cdot\bm{T}_2 + 2\spac C_1\spac C_2\,\bm{1} \,. 
\end{align}
On the right-hand side of these equations we have introduced symmetrized products of color generators whenever more than one generator acts on the same parton. In the second relation
\begin{equation}
   \left( \bm{T}^{a_1} \ldots \bm{T}^{a_n} \right)_+
   \equiv \frac{1}{n!} \sum_{\sigma\in S_n} \bm{T}^{a_{\sigma(1)}} \ldots \bm{T}^{a_{\sigma(n)}}
\end{equation}
denotes the symmetrized product of $n$ color generators, where the sum is over all permutations of $\{1,2,\dots,n\}$. To derive the first and second relations we have used the second trace identity in~\eqref{eq:traces1} and the relation~\cite{Haber:2019sgz}
\begin{equation}\label{eq:trace3}
   \mbox{Tr}\,\big( F^a F^b D^c D^d \big) 
   = \frac12 \left( \delta_{ab}\,\delta_{cd} - \delta_{ac}\,\delta_{bd} \right)
    + \frac{N_c^2-8}{4 N_c}\,f_{ade}\spac f_{bce} + \frac{N_c}{4}\,d_{ade}\spac d_{bce} \,.
\end{equation}

The above structures are the ``descendants'' of the operator $\bm{O}_1^{(j)}$, which is the only operator present in~\eqref{eq:step3}. After additional insertions of $\bm{\Gamma}^c$, one also generates the operators $\bm{O}_i^{(j)}$ with $i=2,3,4$ in~\eqref{eq:prelim}. Repeating the analysis for these structures, we find that $\bm{O}_2^{(j)}$ has no descendants, $\bm{O}_3^{(j)}$ gives rise to the first and third structures in~\eqref{eq:descendants}, and $\bm{O}_4^{(j)}$ leads to the fourth structure. In the analysis for $\bm{O}_2^{(j)}$ we need the trace relation~\eqref{eq:trace3}.

It remains to work out what happens if we apply $\bm{\Gamma}^c$ to these four structures themselves, which happens as soon as we act with another insertion of $\bm{\Gamma}^c$ on the structures shown in the last line of~\eqref{eq:step4b}. In essence, this maps 
\begin{equation}
   \bm{S}_i \to 4 \sum_{i=1,2} \left( C_i\,\bm{S}_i - \bm{T}_i^A\spac\bm{S}_i\,\bm{T}_i^A \right)
\end{equation}
for each of the four structures in~\eqref{eq:descendants}. After some lengthy algebra, we find that the set of linearly independent color structures must be generalized to 
\begin{equation}\label{eq:Sibasis}
\begin{aligned}
   \bm{S}_1 &= f_{abe}\spac f_{cde}\,\{ \bm{T}_1^b,\bm{T}_1^c \}\,\{ \bm{T}_2^a,\bm{T}_2^d \} \,, \\[1mm]
   \bm{S}_2 &= d_{ade}\spac d_{bce}\,\{ \bm{T}_1^b,\bm{T}_1^c \}\,\{ \bm{T}_2^a,\bm{T}_2^d \} \,, \\
   \bm{S}_3 &= d_{ade}\spac d_{bce} \left[ \bm{T}_2^a \,\big( \bm{T}_1^b\spac\bm{T}_1^c\spac\bm{T}_1^d \big)_+
    + (1\leftrightarrow 2) \right] , \\
   \bm{S}_4 &= \{ \bm{T}_1^a,\bm{T}_1^b \}\,\{ \bm{T}_2^a,\bm{T}_2^b \} \,, \\[1mm]
   \bm{S}_5 &= \bm{T}_1\cdot\bm{T}_2 \,, \\[1mm]
   \bm{S}_6 &= \bm{1} \,,
\end{aligned}
\end{equation}
which are symmetric in the parton indices 1 and 2. In other words, the linear combinations of the different structures in each line of~\eqref{eq:descendants} are broken up in their substructures. With this generalization, we obtain the mappings
\begin{equation}\label{eq:recS}
\begin{aligned}
   \bm{S}_1 &\to 8 N_c\,\bm{S}_1 + 2 N_c\,\bm{S}_2 + 8\spac\bm{S}_4 + 32\spac C_1\spac C_2\,\bm{S}_6 \,, \\
   \bm{S}_2 &\to 4 N_c\,\bm{S}_2 \,, \\
   \bm{S}_3 &\to 4 N_c\,\bm{S}_3 \,, \\
   \bm{S}_4 &\to 8\spac\bm{S}_1 + 8 N_c\,\bm{S}_4 \,, \\
   \bm{S}_5 &\to 4 N_c\,\bm{S}_5 \,, \\
   \bm{S}_6 &\to 0 \,,
\end{aligned}
\end{equation}
as well as
\begin{equation}\label{eq:recOS}
\begin{aligned}
   \bm{O}_1^{(j)} &\to N_c \left( 2\spac\bm{S}_1 + \bm{S}_2 + 2\spac\bm{S}_3 \right) + 4\spac\bm{S}_4 \\
   &\quad + 16\,\bigg[ C_1 + C_2 - \frac{N_c\left(N_c^2+8\right)}{24} \bigg]\,\bm{S}_5 
    + 16\spac C_1\spac C_2\,\bm{S}_6 \,, \\
   \bm{O}_2^{(j)} &\to 0 \,, \\
   \bm{O}_3^{(j)} &\to 4\spac\bm{S}_1 + 2 N_c\,\bm{S}_4 + 4 N_c \left( C_1 + C_2 - N_c \right) \bm{S}_5 \,, \\
   \bm{O}_4^{(j)} &\to - 4 N_c \left( C_1 + C_2 \right) \bm{S}_5 - 8 N_c\spac C_1\spac C_2\,\bm{S}_6 \,.
\end{aligned}
\end{equation}
Therefore, most remarkably, the basis $\{ \bm{S}_i \}$ closes under repeated application of $\bm{\Gamma}^c$. 

\subsubsection*{Master formula for the color traces}

\begin{figure}
    \centering
    \includegraphics[scale=1]{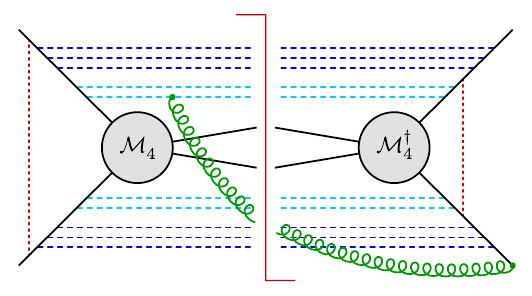}
    \caption{Representative diagram depicting a contribution to the color trace $C_{rn}$, in which the wide-angle soft gluon connects to a collinear gluon emitted from one of the initial-state partons. The meaning of the colors is the same as in Figure~\ref{fig:traceExampleO}.}
    \label{fig:traceExampleS}
\end{figure}

At this point, we obtain the final result 
\begin{equation}\label{eq:MasterFormula}
   C_{rn} = - 256\pi^2 \left( 4 N_c \right)^{n-r} 
    \Bigg[ \sum_{j=3}^{M+2}\spac J_j\,\sum_{i=1}^4\,c_i^{(r)}\,
    \big\langle \bm{\mathcal{H}}_{2\to M}\,\bm{O}_i^{(j)} \big\rangle 
    - J_{12}\,\sum_{i=1}^6\,d_i^{(r)}\,\big\langle \bm{\mathcal{H}}_{2\to M}\,\bm{S}_i \big\rangle \Bigg] \,,
\end{equation}
where the basis operators have been defined in~\eqref{eq:Oibasis} and~\eqref{eq:Sibasis}. It follows from~\eqref{eq:step3} that the coefficients $d_i^{(r)}$ vanish for $r=0$. We find that these coefficients obey the recurrence relations
\begin{equation}
\begin{aligned}
   d_1^{(s+1)} &= 2 N_c\,c_1^{(s)} + 4\spac c_3^{(s)} + 8 N_c\,d_1^{(s)} + 8\spac d_4^{(s)} \,, \\
   d_2^{(s+1)} &= N_c\,c_1^{(s)} + 2 N_c\,d_1^{(s)} + 4 N_c\,d_2^{(s)} \,, \\
   d_3^{(s+1)} &= 2 N_c\,c_1^{(s)} + 4 N_c\,d_3^{(s)} \,, \\
   d_4^{(s+1)} &= 4\spac c_1^{(s)} + 2 N_c\,c_3^{(s)} + 8\spac d_1^{(s)} + 8 N_c\,d_4^{(s)} \,, \\
   d_5^{(s+1)} &= 4 \left( C_1+C_2 \right) \left[ 4\spac c_1^{(s)} + N_c\,c_3^{(s)} - N_c\,c_4^{(s)} \right] \\
   &\quad - \frac{2N_c\,(N_c^2+8)}{3}\,c_1^{(s)} - 4 N_c^2\,c_3^{(s)} + 4 N_c\,d_5^{(s)} \,, \\
   d_6^{(s+1)} &= 8\spac C_1\spac C_2 \left[ 2\spac c_1^{(s)} - N_c\,c_4^{(s)} + 4\spac d_1^{(s)} \right] .
\end{aligned}
\end{equation}
Taking into account the expressions for the coefficients $c_i^{(s)}$ obtained in~\eqref{eq:cisolu}, we find that the solutions to these relations are
\begin{equation}\label{eq:disolu}
\begin{aligned}
   d_1^{(r)} &= 2^{3r-1}\spac\big[ \left( N_c+1 \right)^r + \left( N_c-1 \right)^r \big] 
    - 2^{r-1}\spac\big[ \left( 3 N_c+2 \right)^r + \left( 3 N_c-2 \right)^r \big] \,, \\[1mm]
   d_2^{(r)} &= 2^{3r-2} N_c \left[ \frac{\left( N_c+1 \right)^r}{N_c+2} + \frac{\left( N_c-1 \right)^r}{N_c-2} \right] 
    - 2^{r-2} N_c \left[ \frac{\left( 3 N_c+2 \right)^r}{N_c+2} + \frac{\left( 3 N_c-2 \right)^r}{N_c-2} \right] , \\
   d_3^{(r)} &= 2^{r-1} N_c \left[ \frac{\left( 3 N_c+2 \right)^r}{N_c+2} 
    + \frac{\left( 3 N_c-2 \right)^r}{N_c-2} - \frac{\left(2 N_c\right)^{r+1}}{N_c^2-4} \right] , \\
   d_4^{(r)} &= 2^{3r-1}\spac\big[ \left( N_c+1 \right)^r - \left( N_c-1 \right)^r \big] 
    - 2^{r-1}\spac\big[ \left( 3 N_c+2 \right)^r - \left( 3 N_c-2 \right)^r \big] \,, \\[1mm]
   d_5^{(r)} &= 2^r \left( C_1+C_2 \right) \left[ \frac{N_c+2}{N_c+1} \left( 3 N_c+2 \right)^r 
    - \frac{N_c-2}{N_c-1} \left( 3 N_c-2 \right)^r - \frac{2 N_c^{r+1}}{N_c^2-1} \right] \\
   &\quad - \frac{2^{r-1} N_c}{3}\,\big[ (N_c+4) \left( 3 N_c+2 \right)^r + (N_c-4) \left( 3 N_c-2 \right)^r 
    - \left( 2 N_c \right)^{r+1} \big] \,, \\
   d_6^{(r)} &= 2^{3r+1}\spac C_1\spac C_2\,\big[ \left( N_c+1 \right)^{r-1} + \left( N_c-1 \right)^{r-1} \big]
    \left( 1 - \delta_{r0} \right) \\
   &\quad - 2^{r+1}\spac C_1\spac C_2 \left[ \frac{\left( 3 N_c+2 \right)^r}{N_c+1}  
    + \frac{\left( 3 N_c-2 \right)^r}{N_c-1} - \frac{2 N_c^{r+1}}{N_c^2-1} \right] .
\end{aligned}
\end{equation}

Figure~\ref{fig:traceExampleS} shows a representative example of the contributions from the operators $\bm{S}_i$ in~\eqref{eq:MasterFormula} for the case of a $2\to 2$ hard-scattering process. The wide-angle soft gluon emitted into the gap between the jets must be attached to one of the final-state partons in the Born amplitude and one of the collinear gluons emitted from one of the partons 1 or 2 in the conjugate Born amplitude, or vice versa. The attachment to the initial-state parton must happen after the second Glauber exchange has taken place, while the attachment to the collinear gluon must be on a gluon emitted before the first Glauber exchange. Only in this case a non-zero SLL contribution is obtained.

It is straightforward to express the results for the coefficients $c_i^{(r)}$ and $d_i^{(r)}$ in terms of the eigenvalues $v_i$ defined in~\eqref{eq:def_eigenvalues_v}. We obtain 
\begin{equation}\label{eq:cisoluvi}
\begin{aligned}
   c_1^{(r)} &= (4N_c)^r \left[\frac{1}{2}\,v_3^r + \frac{1}{2}\,v_4^r\right] ,\\
   c_2^{(r)} &= (4N_c)^r \left[-\frac{N_c^2}{2(N_c^2-4)}\,v_2^r + \frac{N_c}{4(N_c+2)}\,v_3^r + \frac{N_c}{4(N_c-2)}\,v_4^r\right] ,\\
   c_3^{(r)} &= (4N_c)^r \left[\frac{1}{2}\,v_3^r - \frac{1}{2}\,v_4^r\right] ,\\
   c_4^{(r)} &= (4N_c)^r \left[-\frac{N_c}{N_c^2-1}\,v_1^r + \frac{1}{2(N_c+1)}\,v_3^r + \frac{1}{2(N_c-1)}\,v_4^r\right] ,
\end{aligned}
\end{equation}
and
\begin{align}\label{eq:disoluvi}
   d_1^{(r)} &= (4N_c)^r \left[-\frac{1}{2}\,v_3^r - \frac{1}{2}\,v_4^r + \frac{1}{2}\,v_5^r + \frac{1}{2}\,v_6^r\right] , \nonumber\\
   d_2^{(r)} &= (4N_c)^r \left[-\frac{N_c}{4(N_c+2)}\,v_3^r - \frac{N_c}{4(N_c-2)}\,v_4^r + \frac{N_c}{4(N_c+2)}\,v_5^r 
    + \frac{N_c}{4(N_c-2)}\,v_6^r\right] ,  \nonumber\\
   d_3^{(r)} &= (4N_c)^r \left[-\frac{N_c^2}{N_c^2-4}\,v_2^r + \frac{N_c}{2(N_c+2)}\,v_3^r + \frac{N_c}{2(N_c-2)}\,v_4^r\right] ,  \nonumber\\
   d_4^{(r)} &= (4N_c)^r \left[-\frac{1}{2}\,v_3^r + \frac{1}{2}\,v_4^r + \frac{1}{2}\,v_5^r - \frac{1}{2}\,v_6^r\right] , \\
   d_5^{(r)} &= (4N_c)^r \left[\frac{N_c^2}{3}\,v_2^r - \frac{N_c}{6}(N_c+4)\,v_3^r - \frac{N_c}{6}(N_c-4)\,v_4^r\right]  \nonumber\\
   &\quad + (4N_c)^r \, (C_1+C_2) \left[ -\frac{2N_c}{N_c^2-1}\,v_1^r + \frac{N_c+2}{N_c+1}\,v_3^r - \frac{N_c-2}{N_c-1}\,v_4^r\right] ,  \nonumber\\ 
   d_6^{(r)} &= (4N_c)^r \, C_1C_2 \bigg[-\frac{4N_c}{N_c^2-1}\,\delta_{r0} + \frac{4N_c}{N_c^2-1}\,v_1^r - \frac{2}{N_c+1}\,v_3^r \nonumber\\
   &\hspace{2.77cm} - \frac{2}{N_c-1}\,v_4^r + \frac{2}{N_c+1}\,v_5^r + \frac{2}{N_c-1}\,v_6^r\bigg] \,.  \nonumber
\end{align}

\subsection*{Final result in the diagonal basis}

Armed with the results~\eqref{eq:cisoluvi} and~\eqref{eq:disoluvi}, we can now diagonalize the recursion matrix and write the result for the color traces in the form
\begin{equation}\label{eq:finalmaster}
   C_{rn} 
   = - 16 \left( \gamma_0^{\rm cusp}\spac\pi \right)^2 \left( \gamma_0^{\rm cusp}\spac N_c \right)^n \, 
    \sum_{i=0}^6\,v_i^r\,\big\langle \bm{\mathcal{H}}_{2\to M}\,\bm{Q}_i \big\rangle \,,
\end{equation}
where the eigenvalues $v_i$ are defined in~\eqref{eq:def_eigenvalues_v} and we included $\delta_{r0}=0^r$ via $v_0=0$ in the sum.
The corresponding eigenoperators are given by 
\begin{equation}\label{eq:Qidef}
\begin{aligned}
   \bm{Q}_0 &= J_{12}\,\left[ \frac{4N_c}{N_c^2-1}\,C_1\spac C_2\,\bm{S}_6 \right] , \\
   \bm{Q}_1 &= \sum_{j=3}^{M+2}\spac J_j\,\left[ - \frac{N_c}{N_c^2-1}\,\bm{O}_4^{(j)} \right]
    + J_{12} \left[ \frac{2N_c}{N_c^2-1}\,(C_1+C_2)\,\bm{S}_5 
    - \frac{4N_c}{N_c^2-1}\,C_1\spac C_2\,\bm{S}_6 \right] , \\
   \bm{Q}_2 &= \sum_{j=3}^{M+2}\spac J_j\,\left[ - \frac{N_c^2}{2(N_c^2-4)}\,\bm{O}_2^{(j)} \right] 
    + J_{12} \left[ \frac{N_c^2}{N_c^2-4}\,\bm{S}_3 - \frac{N_c^2}{3}\,\bm{S}_5 \right] ,\\
   \bm{Q}_{3,4} &= \sum_{j=3}^{M+2}\spac J_j\,\left[ \frac12\,\bm{O}_1^{(j)} 
    + \frac{N_c}{4\spac(N_c\pm 2)}\,\bm{O}_2^{(j)} \pm \frac12\,\bm{O}_3^{(j)} 
    + \frac{1}{2\spac (N_c\pm 1)}\,\bm{O}_4^{(j)} \right] \\
   &\quad + J_{12}\,\bigg[ \frac12\,\bm{S}_1 + \frac{N_c}{4\spac(N_c\pm 2)}\,\bm{S}_2 
    - \frac{N_c}{2\spac(N_c\pm 2)}\,\bm{S}_3 \pm \frac12\,\bm{S}_4 \\
   &\hspace{1.55cm} + \left(\frac{N_c\spac(N_c\pm 4)}{6} \mp (C_1+C_2)\,\frac{N_c\pm 2}{N_c\pm 1}\right) \bm{S}_5
    + \frac{2\spac C_1\spac C_2}{N_c\pm 1}\,\bm{S}_6 \bigg] \,, \\
   \bm{Q}_{5,6} &= - J_{12} \left[ \frac12\,\bm{S}_1 + \frac{N_c}{4\spac(N_c\pm 2)}\,\bm{S}_2 
    \pm \frac12\,\bm{S}_4 + \frac{2\spac C_1\spac C_2}{N_c\pm 1}\,\bm{S}_6 \right] , 
\end{aligned}
\end{equation}
where the subscripts 3 and 5 (4 and 6) refer to the upper (lower) signs. Note that in our final formula the ten color traces which appeared in~\eqref{eq:MasterFormula} have been reduced to only seven independent color structures $\bm{Q}_i$. Along with the operator definitions~\eqref{eq:recO} and~\eqref{eq:Sibasis}, this represents our final result for the coefficient functions $C_{rn}$ for particles transforming in arbitrary representations of $SU(N_c)$. The above result is equivalent to formula~\eqref{eq:MasterFormula} in which the coefficients $C_{rn}$ were expressed directly in terms of the operators $\bm{O}_i$ and $\bm{S}_i$.

% The master formula~\eqref{eq:MasterFormula} along with the operator definitions~\eqref{eq:recO} and~\eqref{eq:Sibasis} and the expressions~\eqref{eq:cisolu} and~\eqref{eq:disolu} for the coefficient functions, or alternatively equation~\eqref{eq:finalmaster} along with the operator definitions~\eqref{eq:Qidef} and the eigenvalues $v_i$ in~\eqref{eq:def_eigenvalues_v}, represent our final solutions for the color structures $C_{rn}$ for particles transforming in arbitrary representations of $SU(N_c)$. 

In the derivation of these results, the number $M$ of final-state particles has been kept arbitrary. While in the early literature SLLs have only  been discussed in the context of $2\to 2$ hard-scattering processes~\cite{Forshaw:2006fk,Forshaw:2008cq,Keates:2009dn}, our formulas can also be applied to the important cases where there is a single final-state jet ($M=1$) or even no final-state jet ($M=0$)~\cite{Becher:2021zkk}. In the latter case, the sums over $j$ in~\eqref{eq:MasterFormula} and~\eqref{eq:Qidef} are absent, but the terms proportional to $J_{12}$ remain. The argument that one needs (at least) two colored particles in the initial state to get a non-trivial effect from Glauber phases can also be applied to the final-state particles, which explains why $2\to 2$ processes have been the focus of previous studies of SLLs. However, the collinear gluons emitted with each insertion of $\bm{\Gamma}^c$ (after the second insertion of the Glauber phase $\bm{V}^G$) can provide the additional hard final-state partons in the cases where $M<2$. For $M=1$ we need one such emission, and hence the SLLs will first arise at four-loop order. For $M=0$ we need two collinear emissions, such that the SLLs appear starting at five-loop order. For the simple processes with $M=0,1$ the color traces can be simplified. This will be discussed in Section~\ref{sec:simpleprocesses}.

\subsection{\boldmath Simplification for QCD with quarks and gluons}
\label{subsec:QCD}

In QCD, all fields transform either in the (anti-)fundamental or the adjoint representation of the gauge group. In this case, the master formula~\eqref{eq:MasterFormula} can be simplified because for such a particle we can simplify
\begin{equation}\label{eq:simplifythis}
   d_{ade}\spac d_{bce}\,\{ \bm{T}_i^b,\bm{T}_i^c \}\,\bm{T}_i^d
   = R_i\,\bm{T}_i^a \,,
\end{equation}
where for quarks and gluons
\begin{equation}\label{eq:Rivalues}
   R_q = R_{\bar q} = \frac{\left(N_c^2-4\right)^2}{2 N_c^2} \,, \qquad
   R_g = \frac{N_c^2-4}{2} \,.
\end{equation}
For the fundamental representation, this relation follows from~\eqref{eq:fundrep}. For the adjoint representation, it follows from the fact that $\bm{T}^a=F^a$, and hence
\begin{equation}
   d_{bce}\,\{ \bm{T}^b,\bm{T}^c \} 
   = 2 d_{bce}\,F^b\spac F^c = N_c\,D^e
\end{equation}
by virtue of relation~\eqref{eq:traces2}. This leads to
\begin{equation}
   d_{ade}\spac d_{bce}\,\{ \bm{T}_i^b,\bm{T}_i^c \}\,\bm{T}_i^d
   = N_c\,d_{ade}\spac D^e F^d 
   = \frac{N_c^2-4}{2}\,F^a \,,
\end{equation}
where we have used the relation~\cite{Haber:2019sgz}
\begin{equation}
   \mbox{Tr}\,\big( D^a D^b F^c \big) = \frac{N_c^2-4}{2 N_c}\,if_{abc} \,.
\end{equation}

The structure appearing in~\eqref{eq:simplifythis} can be recast in the form
\begin{equation}
   d_{ade}\spac d_{bce}\,\{ \bm{T}_i^b,\bm{T}_i^c \}\,\bm{T}_i^d
   = 2\spac d_{ade}\spac d_{bce}\,\big( \bm{T}_i^b\spac\bm{T}_i^c\spac\bm{T}_i^d \big)_+ 
    + \frac{N_c^2-4}{6}\,\bm{T}_i^a \,.
\end{equation}
From this relation, it follows that for particles transforming in the (anti-)fundamental or the adjoint representation of $SU(N_c)$ we can eliminate the structure $\bm{S}_3$ in~\eqref{eq:Sibasis} in favor of simpler structures. We obtain
\begin{equation}
   \bm{S}_3 = \left[ \frac{R_1+R_2}{2} - \frac{N_c^2-4}{6} \right] \bm{S}_5 \,.
\end{equation}
This eliminates the most complicated color structure $\bm{S}_3$ from the master formula~\eqref{eq:MasterFormula} and changes the coefficient of the structure $\bm{S}_5$ to
\begin{equation}
   d_5^{(r)}\to \bar d_5^{(r)}
   \equiv d_5^{(r)} + d_3^{(r)} \left[ \frac{R_1+R_2}{2} - \frac{N_c^2-4}{6} \right] \,.
\end{equation}
Using the solutions in~\eqref{eq:disolu}, we find
\begin{equation}\label{eq:bard5}
\begin{aligned}
   \bar d_5^{(r)} 
   &= 2^r \left( C_1+C_2 \right) \left[ \frac{N_c+2}{N_c+1} \left( 3 N_c+2 \right)^r 
    - \frac{N_c-2}{N_c-1} \left( 3 N_c-2 \right)^r - \frac{2 N_c^{r+1}}{N_c^2-1} \right] \\
   &\quad + 2^r\,\frac{N_c}{4} \left( R_1+R_2 \right) \left[ \frac{\left( 3 N_c+2 \right)^r}{N_c+2} 
    + \frac{\left( 3 N_c-2 \right)^r}{N_c-2} - \frac{\left(2 N_c\right)^{r+1}}{N_c^2-4} \right] \\
   &\quad - 2^r\,\frac{N_c}{4}\,\big[ (N_c+2) \left( 3 N_c+2 \right)^r + (N_c-2) \left( 3 N_c-2 \right)^r 
    - \left( 2 N_c \right)^{r+1} \big] \,.
\end{aligned}
\end{equation}

\subsection{Initial-state partons in the fundamental representation}

The general results~\eqref{eq:MasterFormula} and~\eqref{eq:finalmaster} simplify drastically if particles~1 and 2 transform in the (anti-)fundamental representation of $SU(N_c)$, because we can then use the relation~\eqref{eq:fundrep} to reduce any symmetric product of color generators to structures involving at most one generator for each parton. We will now discuss these simplifications in detail. For the basis color operators $\bm{O}_i^{(j)}$ defined in~\eqref{eq:Oibasis}, we obtain
\begin{equation}
\begin{aligned}
   \bm{O}_1^{(j)} 
   &= \left( \bm{T}_1 - \bm{T}_2 \right)\cdot\bm{T}_j 
    - \frac{N_c}{2} \left( \sigma_1 - \sigma_2 \right) d_{abc}\,\bm{T}_1^a\spac\bm{T}_2^b\spac\bm{T}_j^c \,, \\
   \bm{O}_2^{(j)} 
   &= \frac{N_c^2-4}{N_c} \left( \sigma_1 - \sigma_2 \right) d_{abc}\,\bm{T}_1^a\spac\bm{T}_2^b\spac\bm{T}_j^c \,, \\
   \bm{O}_3^{(j)} 
   &= - \frac{1}{N_c} \left( \bm{T}_1 - \bm{T}_2 \right)\cdot\bm{T}_j 
    + \left( \sigma_1 - \sigma_2 \right) d_{abc}\,\bm{T}_1^a\spac\bm{T}_2^b\spac\bm{T}_j^c \,, \\
   \bm{O}_4^{(j)} 
   &= - \frac{N_c^2-1}{N_c} \left( \bm{T}_1 - \bm{T}_2 \right)\cdot\bm{T}_j \,.
\end{aligned}
\end{equation}
Moreover, for the non-trivial basis color operators $\bm{S}_i$ in~\eqref{eq:Sibasis} we get
\begin{equation}
\begin{aligned}
   \bm{S}_1 &= - \frac{N_c^2-1}{N_c}\,\bm{1} - \frac{N_c^2-4}{2}\,\sigma_1\spac\sigma_2\,\bm{T}_1\cdot\bm{T}_2 \,, \\
   \bm{S}_2 &= \left( \frac{N_c^2-4}{N_c} \right)^2 \sigma_1\spac\sigma_2\,\bm{T}_1\cdot\bm{T}_2 \,, \\
   \bm{S}_3 &= \frac{(N_c^2-4)(N_c^2-6)}{3 N_c^2}\,\bm{T}_1\cdot\bm{T}_2 \,, \\
   \bm{S}_4 &= \frac{N_c^2-1}{N_c^2}\,\bm{1} + \frac{N_c^2-4}{N_c}\,\sigma_1\spac\sigma_2\,\bm{T}_1\cdot\bm{T}_2 \,.
\end{aligned}
\end{equation}
In deriving these results we have used the trace relation~\eqref{eq:traces2} as well as the identities  
\begin{equation}
   d_{aab} = 0 \,, \qquad d_{abc} d_{abd} = \frac{N_c^2-4}{N_c}\,\delta_{cd} \,.
\end{equation}
It follows that we encounter only the following linear combinations of coefficients,
\begin{equation}
\begin{aligned}
   c_1^{(r)} - \frac{1}{N_c}\,c_3^{(r)} - \frac{N_c^2-1}{N_c}\,c_4^{(r)} 
   &= \left( 2 N_c \right)^r ,\\
   - \frac{N_c}{2}\,c_1^{(r)} + \frac{N_c^2-4}{N_c}\,c_2^{(r)} + c_3^{(r)}
   &= - 2^{r-1} N_c \left( 2 N_c \right)^r , \\
   - \frac{N_c^2-1}{N_c}\,d_1^{(r)} + \frac{N_c^2-1}{N_c^2}\,d_4^{(r)} + d_6^{(r)}
   &= 2\spac C_F \left( 1 - \delta_{r0} \right) \left( 2 N_c \right)^r , \\
   \frac{(N_c^2-4)(N_c^2-6)}{3 N_c^2}\,d_3^{(r)} + d_5^{(r)}
   &= 2 \left( 2^r - 1 \right) \left( 2 N_c \right)^r , \\
   - \frac{N_c^2-4}{2}\,d_1^{(r)} + \left( \frac{N_c^2-4}{N_c} \right)^2 d_2^{(r)} + \frac{N_c^2-4}{N_c}\,d_4^{(r)}
   &= 0 \,,
\end{aligned}
\end{equation}
where we have used that $C_1=C_2=C_F=(N_c^2-1)/(2N_c)$. The master formula then takes the very simple form shown in~\eqref{eq:oursimpleresult} and was first derived in~\cite{Becher:2021zkk}.

In the diagonal basis, only the first three operators (associated with the lowest three eigenvalues $v_0=0$, $v_1=\frac12$ and $v_2=1$) are non-zero, and we find 
\begin{equation}\label{eq:Qifundamental}
\begin{aligned}
   \bm{Q}_0 &= 2 J_{12}\, C_F\spac\bm{1} \,, \\
   \bm{Q}_1 &= \sum_{j=3}^{M+2}\spac J_j\,(\bm{T}_1-\bm{T}_2)\cdot\bm{T}_j - 2 J_{12}\,\big( C_F\spac\bm{1} - \bm{T}_1\cdot\bm{T}_2 \big) \,, \\
   \bm{Q}_2 &= \sum_{j=3}^{M+2}\spac J_j \left( - \frac{N_c}{2} \right) \left( \sigma_1 - \sigma_2 \right) 
    d_{abc}\,\bm{T}_1^a\spac\bm{T}_2^b\spac\bm{T}_j^c - 2 J_{12}\spac\bm{T}_1\cdot\bm{T}_2 \,, \\
   \bm{Q}_i &= 0 \,; \quad i=3,4,5,6 \,.
\end{aligned}
\end{equation}

\section{Numerical results for simple partonic scattering processes}
\label{sec:simpleprocesses}

Using color conservation, the master formulas~\eqref{eq:oursimpleresult},~\eqref{eq:MasterFormula} and~\eqref{eq:finalmaster} can be simplified for scattering processes involving at most two hard final-state partons at Born level ($M\le 2$). Since these processes are of great phenomenological importance, it is worthwhile to study these simplifications in some detail and analyze the size of the SLLs for such processes. To this end, we provide compact expressions for coefficients $C_{rn}$ for different partonic subprocesses and then plot resummed and order-by-order results for the partonic cross sections.

\subsection{Evaluation of the angular integrals}

To get numerical results, we need to evaluate the matrix elements of the basic color structures and the angular integrals which were defined in~\eqref{eq:Jints} as
\begin{align}
   J_j &= \int\frac{d\Omega(n_k)}{4\pi} \left( W_{1j}^k - W_{2j}^k \right) \Theta_{\rm veto}(n_k) \,.
\end{align}
For concreteness, we will consider in the following a veto region $y_a < y_k < y_b$, where the rapidity $y$ is defined with respect to the beam directions and particle 1 has rapidity $y_1 = +\infty$. With this definition of the veto region, the angular integral takes the form
\begin{align}
   J_j &= \int_{y_a}^{y_b}\!dy_k \int_0^{2\pi} \!\frac{d\phi_k}{2\pi}\, \frac{\sinh(y_k-y_j)}{\cosh(y_k-y_j)-\cos(\phi_k-\phi_j)}\,,
\end{align}
where $y_k$ and $\phi_k$ are the rapidity and azimuthal angle of the emission and $y_j$ and $\phi_j$ the ones of the hard parton along $n_j$. Carrying out the integrations leads to the result
\begin{equation}
   J_j = - \left(y_b - y_a\right)  {\rm sign}(y_j - y_b) \,,
\end{equation}
since the jets cannot be inside the veto region, i.e. $y_j\notin(y_a,y_b)$. This makes it clear that the integrals $J_j$ are invariant under boosts along the beam direction and only depend on the rapidity difference $\Delta Y = y_b - y_a$, so that they are the same in the laboratory frame and the partonic center-of-mass frame.

Below we will consider $2\to M$ scattering processes with $M=0,1,2$ color-charged partons in the final state. For the $2\to 0$ case, for which all final-state particles are color-neutral, only the integral
\begin{equation}
   J_{12} = J_2 = \Delta Y
\end{equation}
is relevant. For forward scattering in a $2 \to 2$ process, the hard final-state particles have $y_3>y_b$ and $y_4<y_a$, which yields
\begin{equation}
   J_3 = - \Delta Y \,, \qquad J_4 = + \Delta Y \,.
\end{equation}
For backward scattering, these signs are opposite. Symmetric $2\to 1$ scattering channels such as $gg\to g$ and $q\bar{q}\to g$ only involve the integral $J_{12}$, but for the $qg\to q$ channel also the integral $J_3$ arises, with $ J_3=-\Delta Y$ for forward scattering and the opposite sign for the backward case.

\subsection[\texorpdfstring{${2\to 0}$}{2->0} hard-scattering processes]{$\bm{2\to 0}$ hard-scattering processes}

The case where the final state does not contain any color-charged hard partons is particularly interesting since it applies to  phenomenologically relevant processes such as the production of one or several electroweak bosons ($H$, $\gamma$, $W^\pm$, $Z^0$) via $q\bar q$ scattering or gluon-gluon fusion. Especially for Higgs and diboson production it is experimentally often necessary to impose a jet veto to suppress background. Such jet-veto cross sections have been studied extensively, but so far no analysis has addressed the effects of the SLLs. Without a hard particle in the final state, the sum over $j$ in~\eqref{eq:MasterFormula} is absent, and color conservation implies that $\bm{T}_1+\bm{T}_2=0$. 

For the case of $q\bar q\to 0$ scattering, we then immediately obtain from~\eqref{eq:oursimpleresult} the simple expression 
\begin{equation}
  C_{rn} = - 512\pi^2\,C_F \left( 4 N_c \right)^n 
    \left( 1 - 2^{1-r} \right) \left( 1 - \delta_{r0} \right) J_{12}\, 
    \big\langle \bm{\mathcal{H}}_{q\bar q\to 0} \big\rangle \,,
\end{equation}
which vanishes for $r=0,1$. It follows that the super-leading terms start at five-loop order in this case ($n\ge 2$), in accordance with the argument given near the end of Section~\ref{subsec:difficult}. It is instructive to also consider the all-order result for the SLL contribution to the partonic cross section in the strict double-logarithmic approximation, as given in~\eqref{eq:resummedsigma}. We obtain 
\begin{equation}\label{eq:SLLqqbarSigma}
   \hat{\sigma}_{q\bar q\to 0}^{\rm SLL} = - \hat\sigma_{q\bar q\to 0}\spac\frac{4C_F}{3} \left( \frac{\alpha_s}{\pi} \right)^3 
    \pi^2 L^3\spac \Delta Y \left[ \Sigma(0,w) - 2\Sigma(\mbox{$\frac12$},w) + \Sigma(1,w) \right] ,
\end{equation}
where $\alpha_s\equiv\alpha_s(\bar\mu)$, $L=\ln(\mu_h/\mu_s)$, $w=\frac{N_c\spac\alpha_s(\bar\mu)}{\pi}\spac L^2$ from~\eqref{eq:wdef}, and we have used that according to~\eqref{eq:trace1} the color trace of the hard function $\bm{\mathcal{H}}_{q\bar q\to 0}$ is equal to the Born-level cross section $\hat\sigma_{q\bar q\to 0}$. The condition that the first two terms in the Taylor expansion of the expression in rectangular brackets must vanish can be formulated as (cf.~\eqref{eq:Sigmaseries})
\begin{equation}\label{eq:sumrules}
   \sum_{i=0}^6\spac k_i = 0 \,, \qquad
   \sum_{i=0}^6\spac k_i\spac v_i = 0 \,, 
\end{equation}
which is obviously satisfied in the present case.

The analysis is significantly more involved if the initial-state particles do not transform in the (anti-)fundamental representation of $SU(N_c)$. In this case, one needs to simplify the color structures $\bm{S}_i$ in~\eqref{eq:Sibasis} after replacing $\bm{T}_2\to-\bm{T}_1$. After a lengthy calculation, using the relations derived in Section~\ref{subsec:QCD}, we find
\begin{equation}
\begin{aligned}
   C_{rn} &= 512\pi^2\,N_c^2 \left( 4 N_c \right)^{n-r}\spac 2^r \left( 1 - \delta_{r0} \right) J_{12}\, 
    \big\langle \bm{\mathcal{H}}_{gg\to 0} \big\rangle \\[1mm]
   &\quad\times \bigg\{ 
    \frac{N_c+3}{N_c+1}\,\Big[ 2^{2r-1} \left( N_c+1 \right)^r - \left( 3N_c+2 \right)^r \Big] \\
   &\hspace{1.1cm} - \frac{N_c-3}{N_c-1}\,\Big[ 2^{2r-1} \left( N_c-1 \right)^r - \left( 3N_c-2 \right)^r \Big]
    + \frac{4 N_c^{r+1}}{N_c^2-1} \bigg\} \,,
\end{aligned}
\end{equation}
which also vanishes for $r=0,1$. In this case, the all-order result for the SLL contribution to the partonic cross section in the strict double-logarithmic approximation is given by
\begin{equation}\label{eq:SLLggSigma}
\begin{aligned}
   \hat{\sigma}_{gg\to 0}^{\rm SLL} 
   &= - \hat\sigma_{gg\to 0}\spac\frac{2 N_c}{3} \left( \frac{\alpha_s}{\pi} \right)^3 
    \pi^2 L^3\spac \Delta Y\,\bigg\{ \frac{8 N_c^2}{N_c^2-1} \left[ \Sigma(0,w) - \Sigma(\mbox{$\frac12$},w) \right] \\
   &\qquad - \frac{N_c(N_c+3)}{N_c+1} \left[ \Sigma(0,w) - 2\Sigma(\mbox{$\frac{3N_c+2}{2N_c}$},w) 
    + \Sigma(\mbox{$\frac{2(N_c+1)}{N_c}$},w) \right] \\
   &\qquad + \frac{N_c(N_c-3)}{N_c-1} \left[ \Sigma(0,w) - 2\Sigma(\mbox{$\frac{3N_c-2}{2N_c}$},w) 
    + \Sigma(\mbox{$\frac{2(N_c-1)}{N_c}$},w) \right] \bigg\} \\
   &= - \hat\sigma_{gg\to 0} \left( \frac{\alpha_s}{\pi} \right)^3 9\pi^2 L^3\spac \Delta Y 
    \left[ \Sigma(0,w) - 2\Sigma(\mbox{$\frac12$},w) + 2\Sigma(\mbox{$\frac{11}{6}$},w) - \Sigma(\mbox{$\frac83$},w) \right] ,
\end{aligned}
\end{equation}
where the last relation holds for $N_c=3$. It can easily be checked that the sum rules~\eqref{eq:sumrules} are again satisfied. 

As discussed in the introduction, due to the appearance of the Glauber phases the SLL contribution are of subleading order in large-$N_c$ counting, which is indicated by the prefactors $C_F=(N_c^2-1)/(2N_c)$ and $N_c$ (instead of $N_c^3$) in~\eqref{eq:SLLqqbarSigma} and~\eqref{eq:SLLggSigma}. The terms shown in the last two lines of~\eqref{eq:SLLggSigma} are multiplied by coefficients which appear to upset this counting. However, as they should, the leading terms cancel out in the relevant combination of $\Sigma(v,w)$ functions.

\begin{figure}
    \centering
    \includegraphics[scale=1]{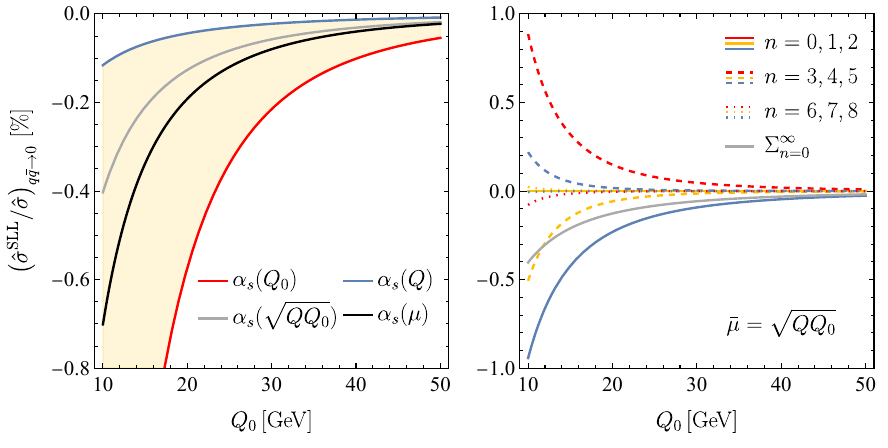}
    \includegraphics[scale=1]{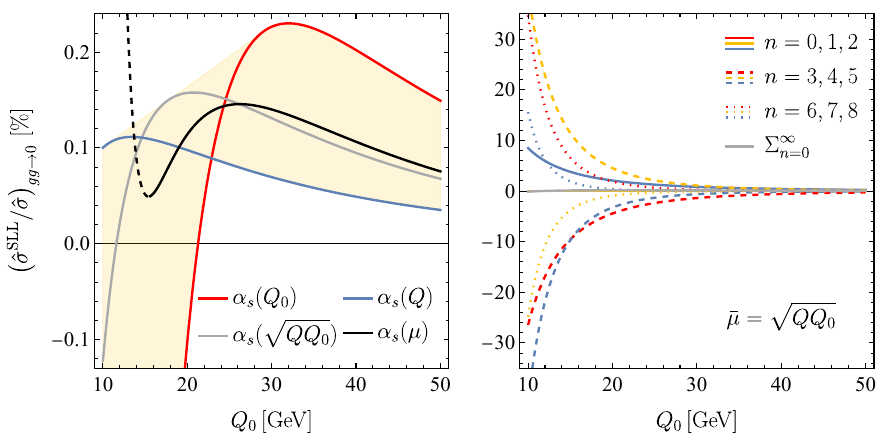}
    \caption{Numerical results for super-leading contributions to partonic $q\bar q\to 0$ scattering (top row) and $gg\to 0$ scattering (bottom row) as a function of the jet-veto scale $Q_0$, at fixed partonic center-of-mass energy $\sqrt{\hat s}=Q=1$\,TeV and for a central rapidity gap with $\Delta Y=2$. The left plots show the resummed all-order contribution of the SLLs for three different scale choices in  $\alpha_s(\bar\mu)$ (red, blue and gray lines) and with a running coupling $\alpha_s(\mu)$ used inside the scale integrals (black line), see~\eqref{eq:Intsresult}. The right plots show the individual contributions at $(3+n)$-th order in perturbation theory (always summed for $r\leq n$) obtained with the intermediate scale choice $\bar\mu=\sqrt{Q\spac Q_0}$. The terms with $n=0,1$ vanish for $2\to 0$ processes. The gray line depicts the infinite sum over all contributions and is the same in the left and right panels.}
    \label{fig:2_to_0}
\end{figure}

In Figure~\ref{fig:2_to_0} we study the SLL contributions to the total cross sections for $q\bar q\to 0$ (top row) and $gg\to 0$ (bottom row) as a function of the jet-veto scale $Q_0$ for fixed partonic center-of-mass energy $Q=1$\,TeV and a gap region in rapidity with $\Delta Y=2$. In both cases, the right panel shows the individual SLL contributions arising at $(3+n)$-th order in perturbation theory, with the strong coupling $\alpha_s(\bar\mu)$ evaluated at the intermediate scale \mbox{$\bar\mu=\sqrt{Q\spac Q_0}$}. We use the two-loop running coupling with the normalization \mbox{$\alpha_s(m_Z)=0.118$} and include the two-loop cusp anomalous dimension according to relation~\eqref{eq:inclusion_2-loop_cusp}. Recall that the first contributions to $2\to 0$ processes arise at five-loop order, corresponding to $n=2$. The gray line shows the all-order sum of the SLL terms. Notice the alternating-sign behavior of the perturbative series, which is a general feature of the SLLs. For $q\bar q\to 0$ scattering, the individual contributions are rather small, reaching at most 1\% at $Q_0=10$\,GeV. The five- and six-loop terms ($n=2,3$) give the biggest contributions but cancel each other to a large extent. For $gg\to 0$ scattering, on the other hand, the individual contributions are larger by more than an order of magnitude, with the biggest terms arising at six-, seven- and eight-loop order ($n=3,4,5$). This reflects the well-known fact that the color factors in gluon-initiated scattering processes are typically much larger than those in quark-initiated processes. 

The left panels in the figure show the infinite sum of the SLL terms for three different choices of the renormalization scale in the running coupling $\alpha_s(\bar\mu)$: the high scale $\bar\mu=Q$ (blue line), the low scale $\bar\mu=Q_0$ (red line), and the intermediate scale $\bar\mu=\sqrt{Q\spac Q_0}$ (gray line). The shaded band between these lines serves as an estimator of the residual scale ambiguity. The running of the coupling is a single-logarithmic effect, which is beyond the accuracy of our calculation. Any choice for $\bar\mu$ between $Q$ and $Q_0$, the two physical scales of the cross section, should be considered a valid choice. Yet, it makes sense to obtain a ``best guess'' for the appropriate scale by performing the scale integrals using the running coupling $\alpha_s(\mu)$ in the integrands, as shown in~\eqref{eq:Intsresult}. In this case we sum up the first 13 terms in the series ($0\le n\le 12$), corresponding to 15-loop order. The corresponding result is shown by the black line, which we consider this to be our best estimate of the total contribution of the SLLs.

In the case of $q\bar q\to 0$ scattering one finds significant cancellations when summing up the individual terms obtained at different orders, so that the resummed result is smaller by roughly a factor of 2 compared with the lowest-order (five-loop) contribution. For $gg\to 0$ scattering, on the other hand, one finds extreme cancellations, and the resummed result is almost two orders of magnitude smaller than the dominant contributions arising at six- to eight-loop order. In fact, the cancellations are so strong that the curves in the left-hand plot belonging to different scale choices cross each other at scales below $Q_0\approx 25$\,GeV. The black line in the left plot, which is obtained by summing up the terms up to $n=12$, stops being accurate below $Q_0\approx 16$\,GeV (as marked by the dashing), indicating that higher-order contributions are still important for such low scale choice.

Our results suggest that SLLs play a subdominant role in processes where electroweak bosons are produced without additional jets.
We stress, however, that a careful study of subleading logarithmic effects will be necessary to corroborate this conclusion. This is left for future work. We also emphasize that for the gluon case one would be led to a rather different conclusion if one estimated the SLL contributions using the five-loop result, in which the super-leading terms first arise. For $Q_0\approx 15$\,GeV, for example, one would then find a correction to the Born-level cross section of approximately $-10\%$, which is a huge five-loop effect.

In our analysis of other partonic channels below we observe a similar pattern. The individual contributions at fixed $n$, but also the cancellations among them, will always be larger in gluon-initiated processes compared with quark-initiated ones. Due to the sum rules in~\eqref{eq:sumrules}, the cancellations are strongest for the $2\to 0$ processes studied above. 

\subsection[\texorpdfstring{${2\to 1}$}{2->1} hard-scattering processes]{$\bm{2\to 1}$ hard-scattering processes}

These processes are also of great phenomenological importance since they include some benchmark Standard Model reactions such as $pp\to V+\text{jet}$ or $pp\to H+\text{jet}$. In this case, the sum over $j$ in~\eqref{eq:MasterFormula} includes only a single term, and color conservation implies that $\bm{T}_3\to-\bm{T}_1-\bm{T}_2$. The relevant partonic scattering reactions are $q\bar q\to g$, $gg\to g$, $qg\to q$, and $\bar qg\to\bar q$. Only for the first of these both initial-state particles transform in the (anti-)fundamental representation of $SU(N_c)$. In this case, it is not difficult to show from~\eqref{eq:oursimpleresult} that the term with $j=3$ gives no contribution, and we obtain 
\begin{equation}\label{eq:qqbarg}
  C_{rn} = 256\pi^2 \left( 4 N_c \right)^n \left( 1 - \delta_{r0} \right) 
   \left( 2^{-r} N_c + \frac{1-2^{1-r}}{N_c} \right)
   J_{12}\,\big\langle \bm{\mathcal{H}}_{q\bar q\to g} \big\rangle \,.
\end{equation}
Note that this result vanishes for $r=0$, and hence the super-leading terms for $q\bar q\to g$ scattering start at four-loop order ($n\ge 1$). For the all-order SLL contribution to the partonic cross section in the double-logarithmic approximation we obtain 
\begin{equation}\label{eq:SLLqqbargSigma}
\begin{aligned}
   \hat{\sigma}_{q\bar q\to g}^{\rm SLL} 
   &= - \hat\sigma_{q\bar q\to g}\spac\frac{2 N_c}{3} \left( \frac{\alpha_s}{\pi} \right)^3 \pi^2 L^3\spac \Delta Y \\
   &\quad\times \left\{ \Sigma(0,w) - \Sigma(\mbox{$\frac12$},w) 
    - \frac{1}{N_c^2} \left[ \Sigma(0,w) - 2\Sigma(\mbox{$\frac12$},w) + \Sigma(1,w) \right] \right\} .
\end{aligned}
\end{equation}
In this case only the first sum rule in~\eqref{eq:sumrules} is satisfied, which ensures that the first contribution to the cross section arises at four-loop order.

The analysis of the remaining partonic channels is again more involved. We obtain
\begin{equation}\label{eq:Crn_2_to_1}
\begin{aligned}
    C_{rn} 
    &= 256\pi^2 \left( 4 N_c \right)^{n-r} \Bigg\{
     J_{12}\,\bigg[ d_1^{(r)}\spac\big\langle \bm{\mathcal{H}}_{2\to 1}\spac\bm{S}_1 \big\rangle 
     + d_2^{(r)}\spac\big\langle \bm{\mathcal{H}}_{2\to 1}\spac\bm{S}_2 \big\rangle 
     + d_4^{(r)}\spac\big\langle \bm{\mathcal{H}}_{2\to 1}\spac\bm{S}_4 \big\rangle \\
    &\hspace{4.16cm} + \left(\bar{d}_5^{(r)} \,\frac{C_3-C_1-C_2}{2} + d_6^{(r)}\right) \big\langle \bm{\mathcal{H}}_{2\to 1} \big\rangle \bigg] \\
    &\quad + J_3 \left[ c_2^{(r)}\spac (R_1-R_2) + 2 \left( c_3^{(r)} + c_4^{(r)} \right) (C_1-C_2) \right]\,
    \frac{C_3-C_1-C_2}{2}\,\big\langle \bm{\mathcal{H}}_{2\to 1} \big\rangle \Bigg\} \,,
\end{aligned}
\end{equation}
where the coefficients $R_i$ have been defined in~\eqref{eq:Rivalues}, and the coefficient $\bar d_5^{(r)}$ has been given in~\eqref{eq:bard5}. The terms shown in the last line of~\eqref{eq:Crn_2_to_1} are present only for the mixed channels $qg\to q$ and $\bar qg\to\bar q$. All coefficients involved vanish for $r=0$, so that the super-leading terms start at four-loop order ($n\ge 1$).

\begin{figure}
    \centering
    \includegraphics[scale=1]{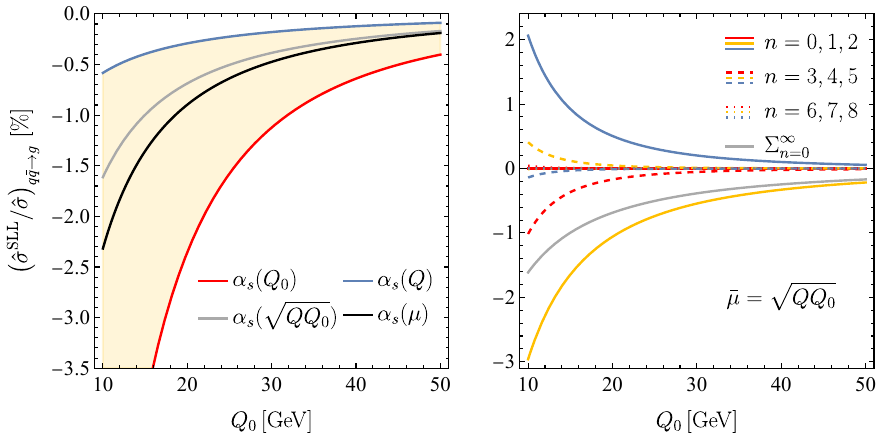}
    \includegraphics[scale=1]{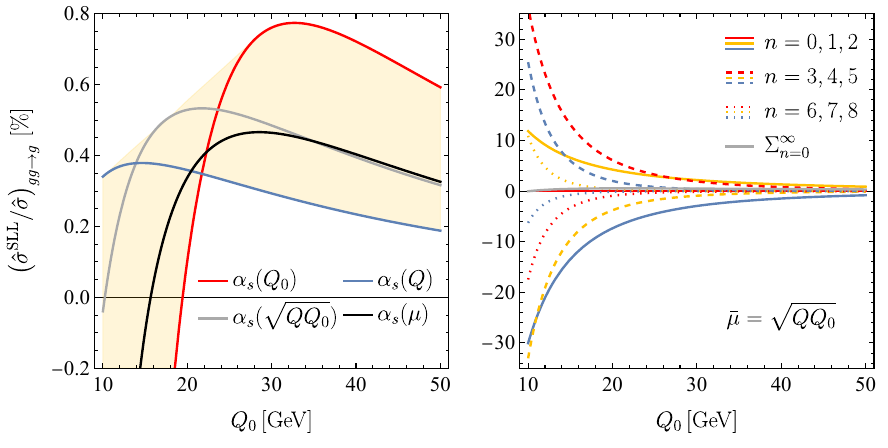}
    \caption{Numerical results for super-leading contributions to partonic $q\bar q\to g$ scattering (top row) and $gg\to g$ scattering (bottom row) as a function of the jet-veto scale $Q_0$, at fixed partonic center-of-mass energy $Q=1$\,TeV and for a central rapidity gap with $\Delta Y=2$. The meaning of the curves is the same as in Figure~\ref{fig:2_to_0}. The term with $n=0$ vanishes for $2\to 1$ processes.}
    \label{fig:2_to_1_diagonal}
\end{figure}

It is straightforward to evaluate the relevant color traces for the different partonic channels. For the case of $q\bar q\to g$ scattering the above expression reduces to the simple result shown in~\eqref{eq:qqbarg}. For $gg\to g$ scattering, we obtain
\begin{align}
  C_{rn} 
  &= - 256\pi^2 \left( 4 N_c \right)^{n-r} \left( 1 - \delta_{r0} \right) 2^r\spac N_c \nonumber\\*
  &\quad\times \bigg[
   \frac{(N_c-2)(N_c+3)}{2(N_c+1)} \left( 3 N_c+2 \right)^r 
   - \frac{(N_c+2)(N_c-3)}{2(N_c-1)} \left( 3 N_c-2 \right)^r \\*\nonumber
  &\hspace{1.0cm} + 4^r\spac(N_c+3) \left( N_c+1 \right)^{r-1} 
   + 4^r\spac(N_c-3) \left( N_c-1 \right)^{r-1} - \frac{6\spac N_c^{2+r}}{N_c^2-1} \bigg]\,
   J_{12}\,\big\langle \bm{\mathcal{H}}_{gg\to g} \big\rangle \,,
\end{align}
as well as
\begin{align}\label{eq:SLLgggSigma}
   \hat{\sigma}_{gg\to g}^{\rm SLL} 
   &= - \hat\sigma_{gg\to g}\spac\frac{2 N_c}{3} \left( \frac{\alpha_s}{\pi} \right)^3 
    \pi^2 L^3\spac \Delta Y\,\bigg\{ \frac{6 N_c^2}{N_c^2-1} \left[ \Sigma(0,w) - \Sigma(\mbox{$\frac12$},w) \right] \nonumber\\*
   &\qquad - \frac{(N_c-2)(N_c+3)}{2(N_c+1)} \left[ \Sigma(0,w) - \Sigma(\mbox{$\frac{3N_c+2}{2N_c}$},w) \right] \nonumber\\*
   &\qquad + \frac{(N_c+2)(N_c-3)}{2(N_c-1)} \left[ \Sigma(0,w) - \Sigma(\mbox{$\frac{3N_c-2}{2N_c}$},w) \right] \\*\nonumber
   &\qquad - \frac{N_c+3}{N_c+1} \left[ \Sigma(0,w) - \Sigma(\mbox{$\frac{2(N_c+1)}{N_c}$},w) \right] 
    - \frac{N_c-3}{N_c-1} \left[ \Sigma(0,w) - \Sigma(\mbox{$\frac{2(N_c-1)}{N_c}$},w) \right] \bigg\} \\*\nonumber
   &= - \hat\sigma_{gg\to g} \left( \frac{\alpha_s}{\pi} \right)^3 
    9\pi^2 L^3\spac \Delta Y \left[ \Sigma(0,w) - \frac32\spac\Sigma(\mbox{$\frac12$},w) 
    + \frac16\spac\Sigma(\mbox{$\frac{11}{6}$},w) + \frac13\spac\Sigma(\mbox{$\frac83$},w) \right] .
\end{align}
The SLL contribution to the total cross section for $q\bar q\to g$ (top row) and $gg\to g$ (bottom row) are shown in Figure~\ref{fig:2_to_1_diagonal}, where the meaning of the curves is the same as before. We observe a similar pattern as in the case of $2\to 0$ scattering, see Figure~\ref{fig:2_to_0}, where now the dominant contributions in perturbation theory arise at four- and five-loop order ($n=1,2$) for $q\bar{q}\to g$ and at five-, six-, seven- and eight-loop order ($n=2,3,4,5$) for $gg\to g$. The presence of initial-state gluons leads to larger color factors for $gg\to g$ and, therefore, the individual contributions arising at fixed order in perturbation theory are quite large. The level of cancellations seen after resumming the infinite series of the SLLs is milder compared to the $2\to0$ case. As a result, the resummed contribution for $q\bar{q}\to g$ amounts to a negative correction to the Born-level cross section, which can reach a few percent for low values of $Q_0$, about a factor 3 larger than in the case of $q\bar q\to 0$. For $gg\to g$ the correction are also about a factor 3 larger compared to $gg\to0$ but still remain at the level of a few permille.

\begin{figure}
    \centering
    \includegraphics[scale=1]{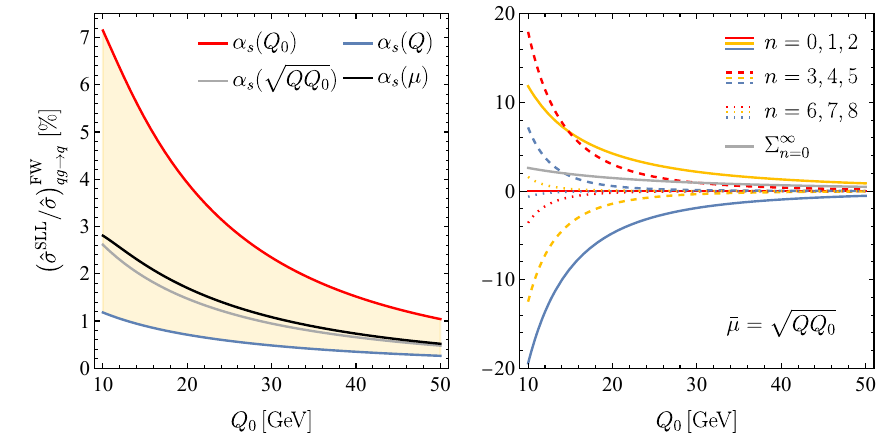}
    \includegraphics[scale=1]{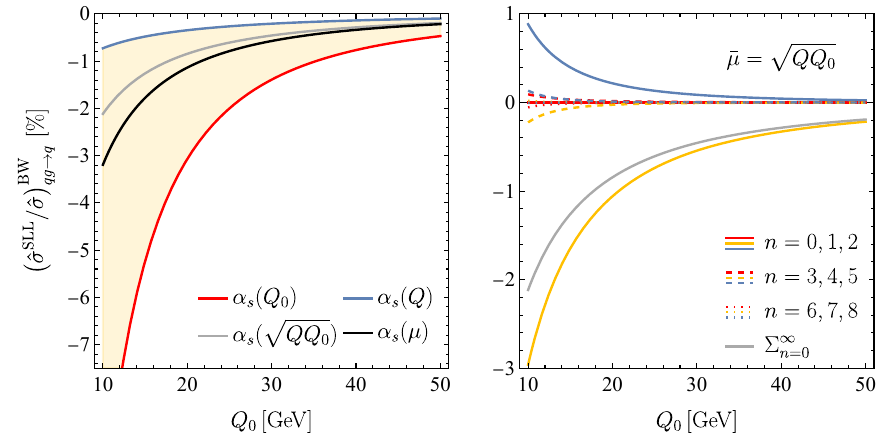}
    \caption{Numerical results for super-leading contributions to partonic $qg\to q$ scattering in forward (top row) and backward scattering (bottom row) as a function of the jet-veto scale $Q_0$, at fixed partonic center-of-mass energy $Q=1$\,TeV and for a central rapidity gap with $\Delta Y=2$. The meaning of the curves is the same as in Figure~\ref{fig:2_to_0}. The term with $n=0$ vanishes for $2\to 1$ processes.}
    \label{fig:qg_q}
\end{figure}

Finally, for the case of $qg\to q$ (and $\bar q g\to\bar q$) scattering, we find 
\begin{align}
  C_{rn} 
  &= - 128\pi^2 \left( 4 N_c \right)^{n-r} \left( 1 - \delta_{r0} \right) 2^r\spac N_c \nonumber\\*
  &\quad\times \Bigg\{ \left( J_{12} - J_3 \right)
   \bigg[ \frac{N_c\spac(N_c+3)}{2(N_c+1)} \left( 3 N_c+2 \right)^r 
   - \frac{N_c\spac(N_c-3)}{2(N_c-1)} \left( 3 N_c-2 \right)^r - \frac{N_c^2+1}{N_c^2-1}\,N_c^r \bigg] \nonumber\\*
  &\hspace{1.1cm} + \left( J_{12} + J_3 \right) \left( 2 N_c \right)^r
   - 6 J_{12}\spac N_c^r \Bigg\}\,
   \big\langle \bm{\mathcal{H}}_{qg\to q} \big\rangle \,.
\end{align}
Here the angular integral $J_3=\mp\Delta Y$ contributes, where the upper (lower) sign refers to forward (backward) scattering. For forward scattering we obtain 
\begin{align}\label{eq:SLLqgqSigma}
   \hat{\sigma}_{qg\to q}^{\rm SLL} 
   &= - \hat\sigma_{qg\to q}\spac\frac{2 N_c}{3} \left( \frac{\alpha_s}{\pi} \right)^3 
    \pi^2 L^3\spac \Delta Y\,\bigg\{ \frac{4 N_c^2-2}{N_c^2-1} \left[ \Sigma(0,w) - \Sigma(\mbox{$\frac12$},w) \right] \nonumber\\*
   &\qquad - \frac{N_c(N_c+3)}{2(N_c+1)}\!\left[ \Sigma(0,w) - \Sigma(\mbox{$\frac{3N_c+2}{2N_c}$},w) \right] 
    + \frac{N_c(N_c-3)}{2(N_c-1)}\!\left[ \Sigma(0,w) - \Sigma(\mbox{$\frac{3N_c-2}{2N_c}$},w) \right] \!\bigg\} \nonumber\\*
   &= - \hat\sigma_{qg\to q} \left( \frac{\alpha_s}{\pi} \right)^3 
    4\pi^2 L^3\spac \Delta Y \left[ \Sigma(0,w) - \frac{17}{8}\spac\Sigma(\mbox{$\frac12$},w) 
    + \frac98\spac\Sigma(\mbox{$\frac{11}{6}$},w) \right] \bigg\} \,,
\end{align}
while for backward scattering 
\begin{equation}
   \hat{\sigma}_{qg\to q}^{\rm SLL} 
   = - \hat\sigma_{qg\to q}\spac\frac{2 N_c}{3} \left( \frac{\alpha_s}{\pi} \right)^3 
    \pi^2 L^3\spac \Delta Y
    \left[ 2\Sigma(0,w) - 3\Sigma(\mbox{$\frac12$},w) + \Sigma(1,w) \right] .
\end{equation}
Figure~\ref{fig:qg_q} shows the SLL contributions to the total cross sections for the $qg\to q$ process in the case of forward scattering (top row) and backward scattering (bottom row). Even though the resummed contributions are similar in both cases and reach a few percent for small values of $Q_0$, we observe two crucial differences. First, for forward scattering the cross section is enhanced by the SLLs, whereas for backward scattering it is reduced. Secondly, the individual contributions arising at fixed order in perturbation theory show a very different pattern in the two cases. For forward scattering, they are comparable to the corrections to $gg\to g$ scattering (see the lower panels in Figure~\ref{fig:2_to_1_diagonal}) and quite large, but in the process of resummation strong cancellations take place. For backward scattering, on the other hand, the individual contributions are much smaller and the resummed result is determined to a large extend by the first non-vanishing term ($n=1$) in perturbation theory. This is the first example showing that in general SLLs can affect the differential cross section for a process in a non-trivial way and do not simply result in an overall $K$-factor.

\subsection[\texorpdfstring{${2\to 2}$}{2->2} hard-scattering processes]{$\bm{2\to 2}$ hard-scattering processes}

\begin{figure}
    \centering
    \begin{tabular}{ccccc}
        \includegraphics[scale=1]{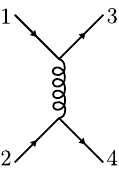} && \includegraphics[scale=1]{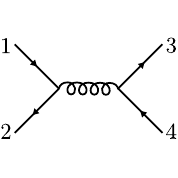} && \includegraphics[scale=1]{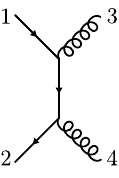}
        \\[0.5cm]
        \includegraphics[scale=1]{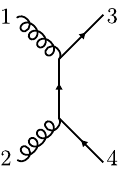} && \includegraphics[scale=1]{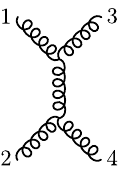} && \includegraphics[scale=1]{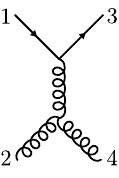}
    \end{tabular}
    \caption{Tree-level Feynman diagrams dominating in small-angle scattering for the $2\to2$ partonic scattering processes studied in this paper. The second diagram is only relevant for $q\bar{q}\to q'\bar{q}'$, if the initial- and final-state quarks have different flavor. In all other cases the $t$-channel diagrams dominate.}
    \label{fig:Feynman_diagrams}
\end{figure}

We now proceed to study $2\to 2$ partonic scattering processes, for which the color structure of the SLLs and their dependence on the scattering kinematics is far more complicated. As we will show, while strong cancellations between the SLLs arising at different orders in perturbation theory still persist, the effects obtained after resummation turn out to be significantly larger in these cases, such that they should be included in future precision calculations of multi-jet LHC cross sections.

We begin with $2\to 2$ small-angle scattering, in which case the Born-level scattering amplitude is dominated by a single Feynman diagram, as shown in Figure~\ref{fig:Feynman_diagrams}. As a result, the SLLs give a correction to the Born-level cross section that takes the form of a constant $K$-factor, as in the cases of $2\to 0$ and $2\to 1$ scattering. Using the relation $\sum_{i=1}^4\spac\bm{T}_i=0$ implied by color conservation in the master formulas~\eqref{eq:oursimpleresult} and~\eqref{eq:MasterFormula}, or in the equivalent formula~\eqref{eq:finalmaster}, we find that the results for the partonic channels $qg\to qg$ and $\bar q g\to\bar q g$ depend on the three angular integrals $J_{12}$, $J_3$ and $J_4$, whereas for all other $2\to 2$ channels the last two integrals only enter in the combination $(J_3-J_4)$.

\begin{figure}
    \centering
    \includegraphics[scale=1]{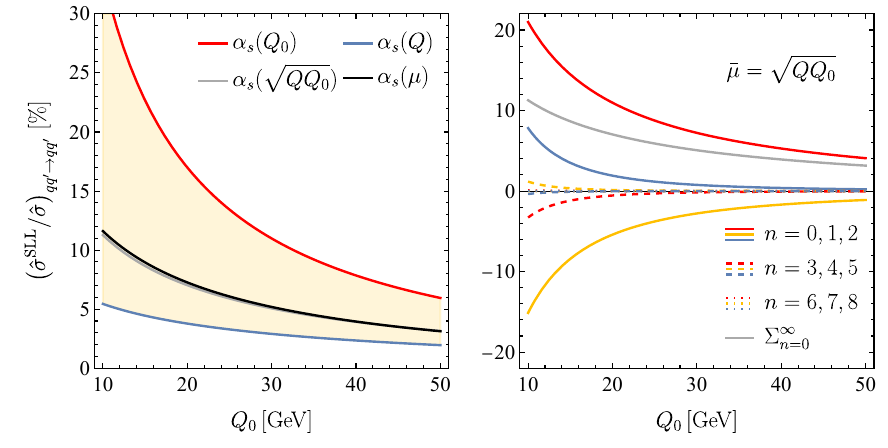}
    \includegraphics[scale=1]{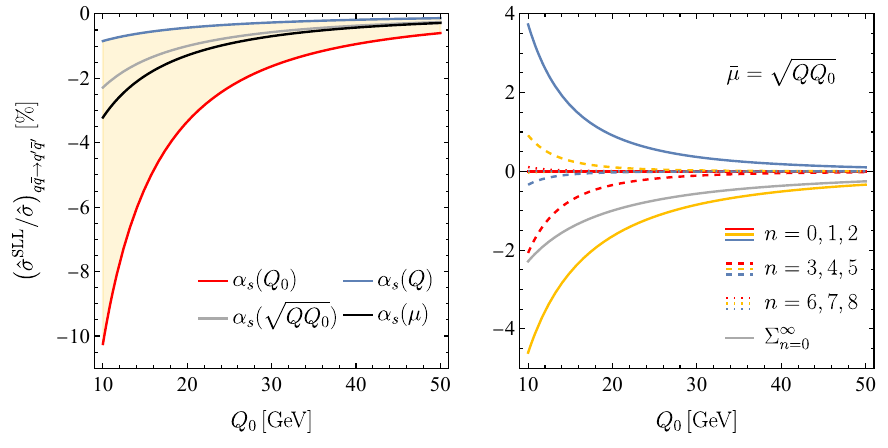}
    \caption{Numerical results for super-leading contributions to partonic $qq^{\prime}\to qq^{\prime}$ (top row) and $q\bar{q}\to q^{\prime}\bar{q}^{\prime}$ (bottom row) small-angle scattering as a function of the jet-veto scale $Q_0$, at fixed partonic center-of-mass energy $Q=1$\,TeV and for a central rapidity gap with $\Delta Y=2$. The meaning of the curves is the same as in Figure~\ref{fig:2_to_0}.}
    \label{fig:2_to_2_quark_quark}
\end{figure}

\begin{figure}
    \centering
    \includegraphics[scale=1]{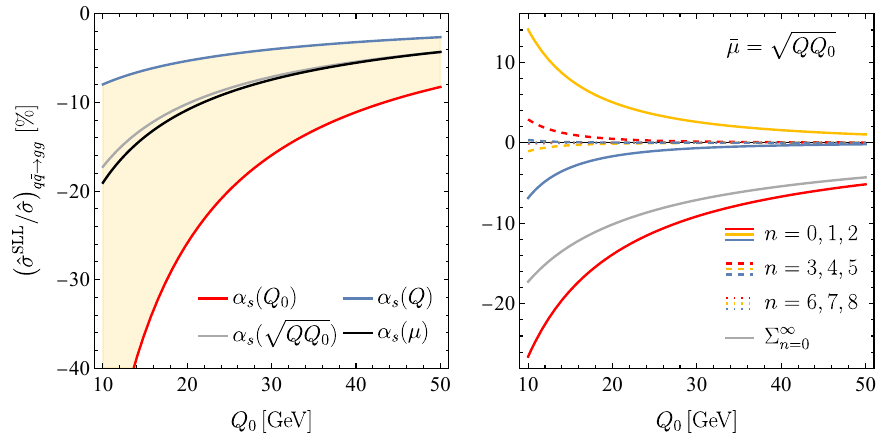}
    \includegraphics[scale=1]{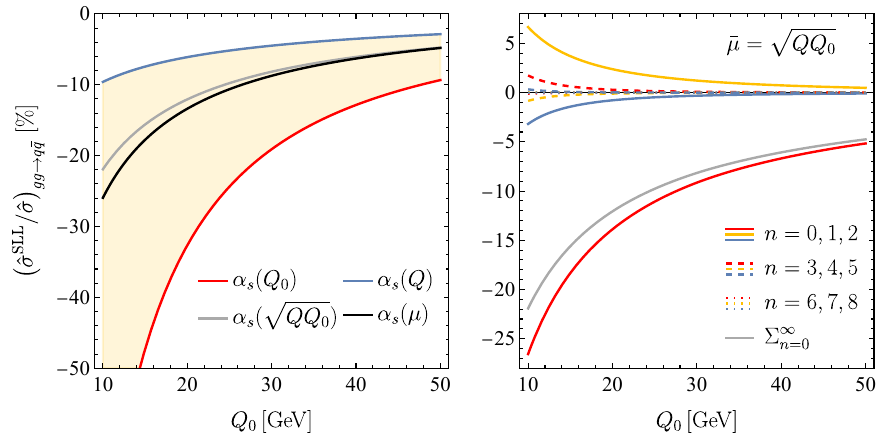}
    \caption{Numerical results for super-leading contributions to partonic $q\bar{q}\to gg$ (top row) and $gg\to q\bar{q}$ (bottom row) small-angle scattering as a function of the jet-veto scale $Q_0$, at fixed partonic center-of-mass energy $Q=1$\,TeV and for a central rapidity gap with $\Delta Y=2$. The meaning of the curves is the same as in Figure~\ref{fig:2_to_0}.}
    \label{fig:2_to_2_quark_gluon}
\end{figure}

\begin{figure}
    \centering
    \includegraphics[scale=1]{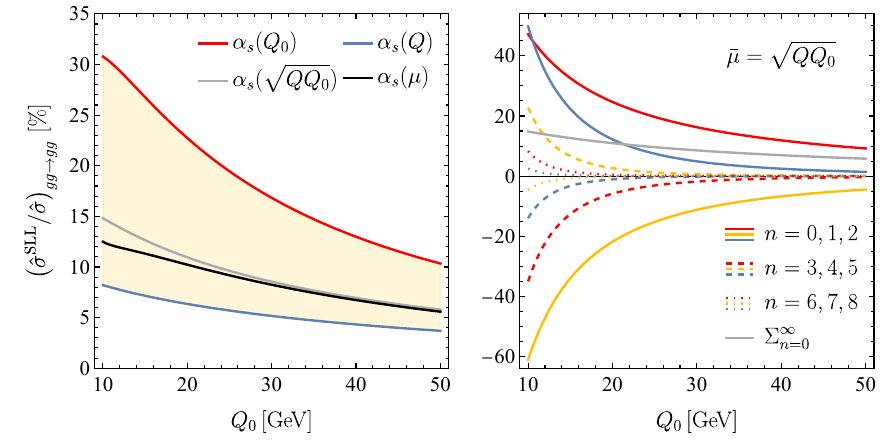}
    \includegraphics[scale=1]{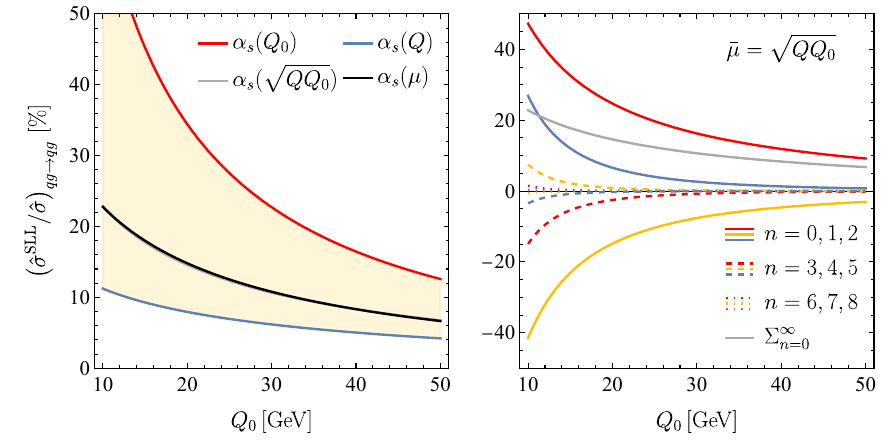}
    \caption{Numerical results for super-leading contributions to partonic $gg\to gg$ (top row) and $qg\to qg$ (bottom row) small-angle scattering  as a function of the jet-veto scale $Q_0$, at fixed partonic center-of-mass energy $Q=1$\,TeV and for a central rapidity gap with $\Delta Y=2$. The meaning of the curves is the same as in Figure~\ref{fig:2_to_0}.}
    \label{fig:2_to_2_gluon_gluon_mixed}
\end{figure}

In Figures~\ref{fig:2_to_2_quark_quark},~\ref{fig:2_to_2_quark_gluon}, and~\ref{fig:2_to_2_gluon_gluon_mixed} we show the SLL contributions to the (anti-)quark-initiated, gluon-initiated, and quark-gluon-initiated $2\to 2$ scattering cross sections, respectively, for the case of forward scattering. In all cases, the corrections to the Born-level cross section found after resummation are significant and exceed the 10\% level for small values of $Q_0$, even if the coupling $\alpha_s(\bar\mu)$ is evaluated at the intermediate scale $\bar\mu=\sqrt{Q\spac Q_0}$ or if a running coupling is used. (For the choice $\bar\mu=Q_0$, the effects are still much larger.) The only exception is $q\bar{q}\to q^{\prime}\bar{q}^{\prime}$ scattering, shown in the lower panels in Figure~\ref{fig:2_to_2_quark_quark}, for which the SLL contribution amounts to just a few percent. This can be understood by looking at the result
\begin{equation}
\begin{aligned}
    \hat{\sigma}_{q\bar{q}\to q^{\prime}\bar{q}^{\prime}}^{\rm SLL} 
    &= - \hat{\sigma}_{q\bar{q}\to q^{\prime}\bar{q}^{\prime}}\,\frac{2 N_c}{3} \left( \frac{\alpha_s}{\pi} \right)^3 \pi^2 L^3 \spac \Delta Y
    \\
    &\quad\times \left\{ \Sigma(0,w) - \Sigma(1,w) - \frac{1}{N_c^2} \left[ \Sigma(0,w) + 2\Sigma(\mbox{$\frac12$},w) -3 \Sigma(1,w) \right] \right\} .
\end{aligned}
\end{equation}
By coincidence, this particular $2\to 2$ process fulfills the first sum rule in~\eqref{eq:sumrules} and, therefore, starts at four-loop order, which makes it comparable to the $2\to1$ scattering processes, see e.g.\ \eqref{eq:SLLqqbargSigma}. The pattern of the SLL contributions in individual orders of perturbation theory can differ quite substantially between different channels. For example, in both $q\bar{q}\to gg$ and $gg\to q\bar{q}$ scattering (Figure~\ref{fig:2_to_2_quark_gluon}), the three-loop contribution ($n=0$) yields the dominant correction to the cross sections.\footnote{In the strict sense of the word, these $n=0$ terms are not a ``super-leading'' effect, even though they result from two Glauber exchanges.} In other cases, such as $gg\to gg$ and, to a lesser extent, $qg\to qg$ scattering (Figure~\ref{fig:2_to_2_gluon_gluon_mixed}), also higher-loop contributions can be very large, and significant cancellations among them take place, so our resummation formalism is crucial to obtain reliable results.

Once we leave the kinematic region of small-angle scattering, the calculation of the SLL terms becomes more complicated. An interesting new feature of $2\to 2$ hard-scattering processes is that there are in general several different color configurations which contribute to a given process. Choosing an orthonormal basis $\{|\mathcal{B}_I\rangle\}$ of color configurations, the amplitudes in a given channel can be decomposed as
\begin{equation} \label{eq:M_color_decomposition}
    |\mathcal{M}_4 \rangle = \sum_I \mathcal{M}_4^{(I)} \, | \mathcal{B}_I \rangle \, , 
\end{equation}
where the coefficients $\mathcal{M}_4^{(I)}$ are functions of the kinematic invariants. The ``unintegrated'' hard function~\eqref{eq:hard_function_definitionHat} can then be represented by a matrix
\begin{align}
    (\widetilde{\bm{\mathcal{H}}}_4)_{IJ} = \mathcal{M}_4^{(I)} \mathcal{M}_4^{(J)\ast} \,.
\end{align}
The one-loop hard functions for all $2\to 2$ parton processes have been compiled in~\cite{Kelley:2010fn}. The authors of~\cite{Broggio:2014hoa} have extended these results to two-loop order and also provided a {\sc Mathematica} notebook to access the results. Due to the simple kinematics for $2\to2$ scattering in the partonic center-of-mass system, we find from~\eqref{eq:factorization_formula} and~\eqref{eq:Wlead} that in the leading-logarithmic approximation the differential cross section is given by%for the partonic cross section
\begin{align} \label{eq:partonic_cross_section}
   \left( \frac{d\hat\sigma}{dr} \right)_{\!2\to2} 
   = \frac{1}{16\pi\hat s}\,\big\langle\widetilde{\bm{\mathcal{H}}}_4\spac\bm{1}\big\rangle \,,
\end{align}
where $r=-\hat{t}/\hat{s}=\sin^2(\theta/2)$. 

In the literature so far, SLLs were analyzed only for processes such as $qq'\to qq'$, which at lowest order only involve a single color structure, shown in the first diagram in Figure~\ref{fig:Feynman_diagrams}. To illustrate the interference effects arising in the presence of multiple color structures we study the process $qq\to qq$, for which the amplitude receives contributions from two color structures. They can be can chosen as
\begin{equation} \label{eq:ONB_color_structures}
    \mathcal{B}_1 \equiv  \langle \{\underline{\alpha}\}| \mathcal{B}_1 \rangle  = \frac{1}{N_c} \, \delta_{\alpha_3 \alpha_2}\, \delta_{\alpha_4 \alpha_1}  \, , \qquad
    \mathcal{B}_2 \equiv  \langle \{\underline{\alpha}\}| \mathcal{B}_2 \rangle  = \frac{2}{\sqrt{N_c^2-1}} \, t^c_{\alpha_3 \alpha_2}\, t^c_{\alpha_4 \alpha_1} \, .
\end{equation}
These color structures are normalized such that $\langle\mathcal{B}_I|\mathcal{B}_J\rangle=\delta_{IJ}$. In this basis, the tree-level, spin-averaged hard function for the $qq\to qq$ process is given by~\cite{Broggio:2014hoa}\footnote{These authors did not normalize their color structures, and hence they found slightly different entries in the matrix, see Appendix~\ref{app:quark_scattering}.}
\begin{align} \label{eq:ONB_hard_function}
\begin{aligned}
    &\frac{1}{4} \sum_{\text{spins}} \widetilde{\bm{\mathcal{H}}}_{qq\to qq} \\
    ={} & (4\pi\alpha_s)^2\, \frac{2C_F}{N_c \spac r^2}
    \begin{pmatrix}
        N_c\, C_F (r^2-2r+2)  &  \frac{\sqrt{N_c^2-1}}{2} \left(\frac{r^3-3r^2+(N_c+4)r-2}{1-r}\right) \\
        \frac{\sqrt{N_c^2-1}}{2} \left(\frac{r^3-3r^2+(N_c+4)r-2}{1-r}\right) & \frac{(N_c^2+1)r^4-4r^3+(N_c^2+2N_c+7)r^2-2(N_c+3)r+2}{2(1-r)^2}
    \end{pmatrix} .
\end{aligned}
\end{align}
(Let us note parenthetically that the color space for $qg\to qg$ is three-dimensional, while that for $gg\to gg$ is nine-dimensional. The color bases and hard functions for all $2\to 2$ processes can be found in~\cite{Broggio:2014hoa}.)
Not only the hard functions, but also the operators $\bm{O}_i^{(j)}$ and $\bm{S}_i$ are represented by matrices in this basis. The full set of matrices is listed in Appendix~\ref{app:quark_scattering}, including a discussion on how to work with color structures that are not normalized. The off-diagonal elements of these matrices correspond to interference effects, i.e.\ contributions in which the color structures in the amplitude and the conjugate amplitude differ. To obtain the coefficients $C_{rn}$ in~\eqref{eq:MasterFormula}, one must evaluate the traces of the products of the hard-function matrix in~\eqref{eq:ONB_hard_function} with the operator matrices given in Appendix~\ref{app:quark_scattering}.

\begin{figure}
    \centering
    \includegraphics[scale=1]{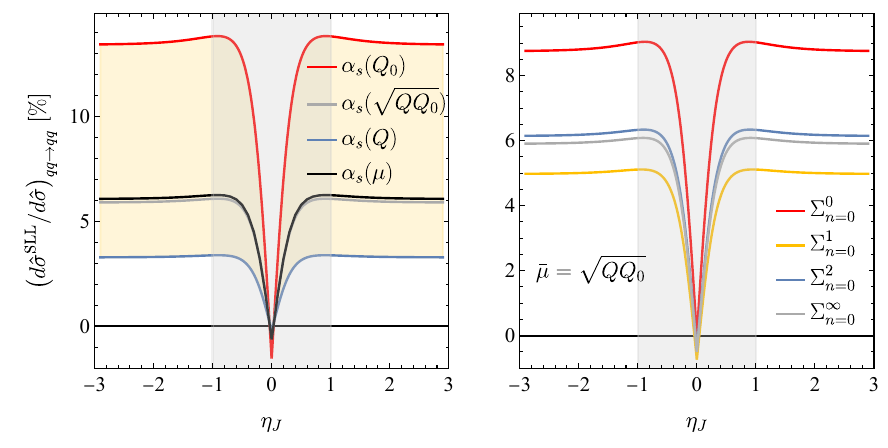}
    \caption{Numerical results for super-leading contributions to partonic $qq\to qq$ scattering as a function of the jet rapidity $\eta_J$, at fixed partonic center-of-mass energy $Q=1\,{\rm TeV}$ and jet-veto scale $Q_0=25\,{\rm GeV}$, and for a central rapidity gap with $\Delta Y=2$ (gray area). The left plot shows the resummed all-order contribution of the SLLs for three different scale choices in $\alpha_s(\bar\mu)$ (red, blue and gray lines) and with a running coupling $\alpha_s(\mu)$ used inside the scale integrals (black line). The right plot shows the SLL contributions summed up to $(3+n)$-th order in perturbation theory for different values of $n$, obtained with the intermediate scale choice $\bar\mu=\sqrt{Q\spac Q_0}$.}
    \label{fig:qqScattering}
\end{figure}

Due to the matrix structure of the operators, the kinematic dependence of the SLL contributions differs from that of the Born-level cross section. In Figure~\ref{fig:qqScattering}, we show the SLL corrections to the differential cross section for $qq \to qq$ scattering, normalized to the Born-level cross section, for a partonic center-of-mass energy of $Q=1\,{\rm TeV}$, a jet-veto scale $Q_0=25\,{\rm GeV}$, and a central rapidity gap with $\Delta Y =2$. Instead of the scattering angle $\theta$, we use the rapidity of the jet, $\eta_J=\ln\cot(\theta/2)$, as the kinematic variable. Because of the central gap (gray area in the plot), the hard jet must be restricted to $|\eta_J|>1 $. For illustrative purposes, however, we also plot our results inside the gap region, even though the associated integrals $J_j$ are only meaningful if the jet is outside this region. The right plot shows the partial sum of the SLL terms up to order $\alpha_s^{3+n}$ for different values of $n$, obtained with the intermediate scale choice $\bar\mu=\sqrt{Q\spac Q_0}$. The left plot shows the effect of changing the scale in $\alpha_s(\bar\mu)$. Even though it is a small effect, the kinematic dependence on the SLL contributions on the jet rapidity is clearly visible in the two plots.

\section{Discussion}\label{sec:discussion}

In this work, we have developed a full description of the leading double-logarithmic corrections to arbitrary non-global observables at hadron colliders. While the existence of SLLs had been known since the seminal 2006 paper~\cite{Forshaw:2006fk}, a systematic all-order understanding of the subtle quantum effects responsible for the breakdown of color coherence in non-global observables had been lacking. The insights we have obtained change this situation in both a qualitative and a quantitative way and raise the hope that it will be possible to develop a complete theoretical description of such effects. 

Yet, much remains to be accomplished before our findings can be turned into accurate predictions for physical cross sections for a large class of relevant collider processes. The most pressing challenge is, perhaps, the extension of our results beyond the leading double-logarithmic approximation. This will be necessary in order to remove the significant $\mathcal{O}(1)$ scale uncertainties seen in our numerical estimates for different partonic scattering processes. At the technical level, reaching single-logarithmic accuracy will require accounting for arbitrarily many insertions of all terms in the one-loop anomalous dimension in~\eqref{eq:gammaH}, not just those enhanced by the cusp logarithm in~\eqref{eq:cuspterms}. These include, in particular, the collinear parts in~\eqref{eq:collineargammas}, which introduce new and non-trivial color structures.\footnote{In addition, one needs to account for the two-loop contributions to the cusp anomalous dimension and the $\beta$-function. This is not difficult to implement and has been sketched in~\eqref{eq:inclusion_2-loop_cusp_run} and~\eqref{eq:inclusion_2-loop_cusp_run_Glauber}.} 
Another important term is the Glauber-phase operator $\bm{V}^G$, which gives rise to $\pi^2$-enhanced contributions to the cross sections, which for realistic choices of parameters are numerically not much smaller than the double-logarithmic corrections. The structure of these terms will be discussed briefly in Section~\ref{subsec:Glauber}. 

Perhaps the most challenging task will be to account for multiple wide-angle soft emissions, which build up the NGLs in the cross section. In the large-$N_c$ limit, higher-order insertions of the corresponding operator $\overline{\bm{\Gamma}}$ can be computed using Monte Carlo methods~\cite{Balsiger:2018ezi}, a technique pioneered in~\cite{Dasgupta:2001sh}. Recently even the subleading NGLs generated by two-loop contributions to the soft anomalous dimension~\cite{Becher:2021urs} have been implemented into a parton shower~\cite{Becher:2023vrh}, along with the one-loop corrections to the hard functions and low-energy matrix elements~\cite{Balsiger:2019tne}. Accounting for SLLs in a Monte Carlo approach will require methods to compute these terms for $N_c=3$. As discussed in the introduction, such methods are currently being developed, and first numerical results are available since a few years~\cite{Hatta:2013iba,Hagiwara:2015bia,Hatta:2020wre,Platzer:2013fha,AngelesMartinez:2018cfz,Forshaw:2019ver,DeAngelis:2020rvq}. However, the Glauber phases have so far only been included in a framework with an approximate treatment of color~\cite{Nagy:2019bsj,Nagy:2019rwb}.

On a more fundamental level, it will be important to understand the asymptotic behavior of non-global cross sections in a detailed way, in order to include subleading logarithmic corrections in a systematic way. While the systematics of RG-improved perturbation theory for standard Sudakov problems is well understood, this is not the case for double-logarithmic non-global observables, where the resummation of large logarithmic corrections cannot be simply organized as an expansion in the exponent. This crucial complication will be discussed in Section~\ref{subsec:systematics}.

In order to obtain results relevant for phenomenology it will be necessary to go from the partonic cross sections studied in this paper to hadronic cross sections, which include a sum over partonic channels, convolutions with the relevant parton distribution functions, and realistic cuts. Finally, it will be crucial to understand 
if the loss of color coherence, which is responsible for the existence of double-logarithmic corrections to non-global observables, also leads to a violation of conventional collinear PDF factorization, which is the foundation for all theoretical predictions of LHC cross sections. The authors of~\cite{Forshaw:2021fxs} have studied Glauber exchanges in partonic hard-scattering cross sections and stated that the loss of coherence extends also to many global observables. To answer the question whether PDF factorization holds, it is important to study Glauber exchanges involving also spectator partons. If these are non-vanishing, they amount to multi-parton interactions which violate factorization. At least for certain observables, such effects indeed seem to be present~\cite{Bauer:2010cc,Gaunt:2014ska,Zeng:2015iba}. A phenomenological study of factorization-violating effects due to multi-parton interactions was performed in the recent paper~\cite{Bijl:2023dux}. A detailed study of these important open issues is left for future work.

\subsection{Glauber series}
\label{subsec:Glauber}

We have seen that the leading-logarithmic corrections to the cross sections for non-global observables at hadron colliders form a series of the form
\begin{equation}
   \sigma \sim \sum_{n=0}^\infty \left[ c_{0,n} \left( \frac{\alpha_s}{\pi}\,L \right)^n   
    + c_{1,n} \left( \frac{\alpha_s}{\pi}\,L \right) \left( \frac{\alpha_s}{\pi}\,i\pi\spac L \right)^2 
    \left( \frac{\alpha_s}{\pi}\,L^2 \right)^n + \dots \right] ,
\end{equation}
with $L\sim\ln(Q/{Q_0})\gg 1$. It requires two insertions of Glauber phases to obtain a double-logarithmic series, which therefore starts at three-loop order in perturbation theory. It is possible to generalize the color traces in~\eqref{eq:colortraces} by including a higher (even) number of insertions of the Glauber operator $\bm{V}^G$. This generates additional double-logarithmic contributions at higher orders. Defining the variables
\begin{equation}
   w = \frac{N_c\spac\alpha_s(\bar\mu)}{\pi}\,L^2 \,, \qquad
   w_\pi = \frac{N_c\spac\alpha_s(\bar\mu)}{\pi}\,\pi^2 \,,
\end{equation}
we can write the leading logarithms in the presence of an arbitrary number of Glauber insertions in the form
\begin{equation}
\begin{aligned}
   \sigma 
   &\sim \sum_{n=0}^\infty \left[ c_{0,n} \left( \frac{\alpha_s}{\pi}\,L \right)^n  
    + c_{1,n} \left( \frac{\alpha_s}{\pi}\,L \right) w^n \left( w_\pi\spac w \right)
    + c_{2,n} \left( \frac{\alpha_s}{\pi}\,L \right) w^n \left( w_\pi\spac w \right)^2 + \dots \right] \\
   &= \sum_{n=0}^\infty \left[ c_{0,n} \left( \frac{\alpha_s}{\pi}\,L \right)^n  
    + \left( \frac{\alpha_s}{\pi}\,L \right) \sum_{\ell=1}^\infty\,c_{\ell,n}\,w^{n+\ell}\,w_\pi^{\ell} \right] .
\end{aligned}
\end{equation}
For realistic values of parameters, such as $Q=1$\,TeV and $Q_0=25$\,GeV, the quantities $w\approx 1.4$ and $w_\pi\approx 1.0$ both take values of $\mathcal{O}(1)$ for our default choice $\bar\mu=\sqrt{Q\spac Q_0}$, and hence the infinite set of terms involving these parameters is expected to give effects which can be as large as the one-loop prefactor. In other words, the imaginary part of the cusp logarithm can be numerically of similar order as the real part. Referring to both the real and imaginary parts as logarithmic terms, we see that the double-logarithmic behavior of the sum of SLLs starts at three-loop order ($n=0$, $\ell=1$). We refer to the expression shown in the second line of the above result as the ``Glauber series''. We leave a detailed analysis of the higher-order terms in $w_\pi$ for later work.

\subsection{Systematics of the double-logarithmic approximation}
\label{subsec:systematics}

It is a difficult open problem to understand the systematics of the resummation of large double-logarithmic terms for non-global hadron-collider observables. The RG evolution equation~\eqref{eq:hardRG} is of the Sudakov type, meaning that the anomalous dimension $\bm{\Gamma}^H$ contains ``cusp terms'', accompanied by one power of the logarithm $\ln(\mu^2/\mu_h^2)$ and proportional to the cusp anomalous dimension $\gamma_{\rm cusp}(\alpha_s)$, as well as ``ordinary terms'' without such a logarithmic enhancement. The complication is, of course, that in our case each term in the anomalous dimension is an operator in color space as well as in the infinite space of parton multiplicities. Consequently, the path-ordered exponential~\eqref{eq:Uexp} does not give rise to an exponential in the usual sense. 

\subsubsection*{\boldmath Conventional counting scheme with $\alpha_s\spac L=\mathcal{O}(1)$}

If one adopts the conventional counting scheme, where $\alpha_s\spac L=\mathcal{O}(1)$, then the Sudakov double logarithms $\alpha_s\spac L^2\sim 1/\alpha_s$ are formally larger than $\mathcal{O}(1)$. In simpler applications of SCET, the resummation of Sudakov logarithms produces a perturbative series in the exponent, such that after RG resummation the expression for an observable takes the form
\begin{equation}
\begin{aligned}
   \sigma &\sim \sigma_0\,\exp\left[ - \frac{1}{\alpha_s(\mu_h)}\,g_0(x_s) + g_1(x_s) + \alpha_s(\mu_h)\,g_2(x_s) + \dots \right] \\
   &= \sigma_0\,\exp\left[ - \frac{1}{\alpha_s(\mu_h)}\,g_0(x_s) + g_1(x_s) \right] 
    \Big[ 1 + \alpha_s(\mu_h)\,g_2(x_s) + \dots \Big] \,,
\end{aligned}
\end{equation}
with $x_s=\alpha_s(\mu_h)/\alpha_s(\mu_s)$ as in~\eqref{eq:Intsresult}. The function $g_0$ involves the one-loop cusp anomalous dimension and $\beta$-function, $g_1$ involves the two-loop cusp anomalous dimension and $\beta$-function and the one-loop coefficients of the remaining terms in the anomalous dimension, and so on. Keeping only $g_0$ in the exponent correctly reproduces the leading double-logarithmic terms, which are formally larger than $\mathcal{O}(1)$, but the term proportional to $g_1$ still gives rise to a modification of the cross section by an $\mathcal{O}(1)$ amount. The next correction term, proportional to $g_2$, can however be treated as a parametrically small correction to the cross section. For this conclusion, the fact that the exact solution of the RG equation yields a perturbative expansion in the exponent is crucial. A series expansion of the first term in the exponential generates arbitrarily many inverse powers of the strong coupling, and in this expanded form 
\begin{equation}
   \sigma \sim \sigma_0\,\sum_{n_1=0}^\infty \sum_{n_2=0}^\infty \sum_{n_3=0}^\infty\spac \frac{1}{n_1!\spac n_2!\spac n_3!} 
    \left[ - \frac{g_0(x_s)}{\alpha_s(\mu_h)} \right]^{n_1} \left[ g_1(x_s) \right]^{n_2}
    \left[ \alpha_s(\mu_h)\,g_2(x_s) \right]^{n_3}, 
\end{equation}
it would not seem that the terms involving $g_2$ give a small correction to the cross section, since for $n_3<n_1$ these contributions are enhanced by inverse powers of $\alpha_s(\mu_h)$.

Our problem resembles this latter situation. The integrals $I_{rn}$ in~\eqref{eq:Intsresult} involve more and more inverse powers of $\alpha_s$ in higher order, but the resummation of these terms does not produce a simple exponential. In the strict double-logarithmic limit, where one neglects the running of the coupling, the resummation yields the functions $\Sigma_0(w)$ and $\Sigma(v_i,w)$ introduced in~\eqref{eq:sum0} and~\eqref{eq:Feriet}. These correspond to the exponential of the $g_0$ term above. Taking into account that $w\sim 1/\alpha_s$ in this counting scheme, it follows from~\eqref{eq:resummedsigma} that the SLL contribution to a cross section scales like $\alpha_s\ln\alpha_s$, where the logarithmic enhancement results from the presence of the function $\Sigma_0(w)$. The problem is that we see no reason to expect that higher-order terms (for instance, those involving the two-loop cusp anomalous dimension and $\beta$-function, the effects of the running of the coupling, or multiple insertions of $\overline{\bm{\Gamma}}$) give a multiplicative correction to the leading result. Rather, compared with~\eqref{eq:resummedsigma} we expect a result of the form
\begin{equation}\label{eq:subldg}
   \left( \frac{\alpha_s(\bar\mu)}{4\pi} \right)^4 \ln^4\!\bigg(\frac{\mu_h}{\mu_s}\bigg)\,G(w) 
   \sim G_\infty(w) \,,
\end{equation}
with some unknown function $G(w)$, whose asymptotic form for $w\to \infty$ we denote by $G_\infty(w)$. At any fixed order in perturbation theory, this expression features one large logarithm less than the corresponding expression~\eqref{eq:resummedsigma} for the SLLs. However, how large this contribution is after resummation, relative to the sum of the SLLs, depends on the properties of this function $G_\infty(w)$. The best one can do in this case is to adopt the same approximation scheme for the anomalous dimension $\bm{\Gamma}^H$ as in the standard Sudakov case and hope for the best (i.e., assume that the asymptotic properties of the resummed series of the leading SLLs are similar to those of the resummed series of subleading SLLs). The series of the leading double logarithms is then expected to receive corrections of $\mathcal{O}(1)$ from subleading logarithmic effects, like in the Sudakov case. To calculate these effects one would need to use the two-loop expressions for the cusp anomalous dimension and the $\beta$-function, allow for an arbitrary number of insertions of $\overline{\bm{\Gamma}}$ as well as $\bm{V}^G$, and also include the collinear anomalous dimensions in~\eqref{eq:collinear_part_anomalous_dimension}. In addition, one would have to resum the rapidity logarithms in the low-energy matrix elements $\bm{\mathcal{W}}_m$. Similar to the case of transverse-momentum resummation, such logarithms must be present for consistency of the RG equations~\cite{Becher:2010tm}, but their structure has not been explored so far.

\subsubsection*{\boldmath Counting scheme with $\alpha_s\spac L^2=\mathcal{O}(1)$}

The analysis changes if one adopts a counting scheme where $\alpha_s\spac L^2=\mathcal{O}(1)$. Then it is consistent to count $\alpha_s\spac L\sim\sqrt{\alpha_s}$, and hence subleading logarithmic contributions such as that in~\eqref{eq:subldg} are parametrically suppressed. Note that, since $w=\mathcal{O}(1)$ in this case, the function $G(w)$ also counts as $\mathcal{O}(1)$. In other words, one is not in an asymptotic regime, where this function could develop a non-trivial behavior. In this counting scheme, the sum of the leading double-logarithmic corrections presents the parametrically leading contribution to the cross sections, and subleading logarithmic effects are parametrically suppressed by at least a factor of $\sqrt{\alpha_s}$. It follows from~\eqref{eq:resummedsigma} that the SLL contribution to a cross section scales like $\alpha_s^{3/2}$, whereas the subleading terms in~\eqref{eq:subldg} scale like $\alpha_s^2$.

Given that the relevant parameter $w=\frac{N_c\spac\alpha_s(\bar\mu)}{\pi}\,L^2\approx 1.4$ for representative values $\mu_h=1$\,TeV and $\mu_s=25$\,GeV, it is not clear which counting scheme is more appropriate. On the one hand, counting $w=\mathcal{O}(1)$ seems not unreasonable. On the other hand, we have found that the scale ambiguity from the choice of $\bar\mu$ turns out to be an $\mathcal{O}(1)$ effect in our case and does not appear to be suppressed with a factor of $\sqrt{\alpha_s(\bar\mu)}$.

\section{Conclusions}

In this paper, we have presented a factorization formula for non-global hadron-collider observables. It separates cross sections into hard functions and soft-collinear matrix elements. We have provided a detailed derivation of the one-loop anomalous dimension that governs the scale evolution of the hard functions, which forms the basis for the resummation of large logarithms in these cross sections. In contrast to the case of non-global observables at lepton colliders, the anomalous dimension contains non-trivial complex phases, which spoil the cancellation of soft\,+\,collinear contributions between the real and virtual parts. As a consequence, the renormalization-group evolution produces double-logarithmic terms, and even the purely collinear part of the evolution kernel has a non-trivial color structure, in contrast to the standard DGLAP evolution of parton distribution functions.

The double-logarithmic terms in the cross section are called super-leading logarithms (SLLs). We have presented general results for the SLL terms arising in arbitrary $2\to M$ scattering processes, extending our earlier results for quark-initiated processes reported in~\cite{Becher:2021zkk}. Our results have been obtained by solving the relevant evolution equations order by order, after identifying the subset of color traces giving rise to the leading double-logarithmic effects. We find that the structure of the resulting terms is simple enough that they can be resummed to all orders and expressed in closed form in terms of Kamp\'e de F\'eriet functions. Our results open the door for a full phenomenological analysis of the contributions of SLLs to some benchmark LHC cross sections. Studying the numerical size of the effects for a number of partonic channels, we have observed several interesting features. The SLL terms at different orders in $\alpha_s$ alternate in sign, and the individual terms can give rise to contributions that are much larger than the all-order sum. The same is true for the pattern of double logarithms arising in the standard Sudakov form factor. What is quite different in our case is the asymptotic behavior: in the Sudakov regime the cross section gets an exponential suppression in the double-logarithmic variable $w\sim \alpha_s L^2$, while the SLLs are only suppressed by $\ln w/w$. To find the asymptotic behavior beyond the double-logarithmic approximation is an interesting open problem. In standard Sudakov problems, the double-logarithmic terms give rise to a global prefactor in the cross section, but given the complicated structure of the SLLs this will not be the case here. 

For a generic $2\to M$ partonic process, the Glauber-phase terms first contribute at third order in perturbation theory, since one needs two insertions of the phase terms and a soft emission into the veto region to obtain a non-zero result. Once the phases are present, the first SLL arises at four-loop order. For $2\to 0$ and $2\to 1$ processes, such as Higgs-boson production in association with $M\leq 1$ jets, the pattern is different. For $2\to 0$ parton scattering, the third- and fourth-order terms vanish and the SLLs start at the level of five loops. For $2\to 1$ processes, the third-order term vanishes. In our all-order result, this involves a cancellation among terms with similar all-order behavior, and hence we find that the SLLs are numerically suppressed also in higher orders. No such cancellations are present if the Born-level hard process involves two or more partons in the final state, which then generically leads to much larger effects. Another interesting feature that was not discussed in the literature so far is quantum interference. In general, the SLL effects mix different color structures in the amplitude, and as a consequence the kinematic dependence of the effect can be different from the Born-level process, as we have illustrated using $qq\to qq$ scattering as an example.

For $2\to 2$ partonic scattering processes, the resummed contribution of the SLLs can amount to a correction at the level of a few to several tens of percent on the cross sections, depending on the choice of the variables $Q$ and $Q_0$ as well as the scale $\bar\mu$ used in the running coupling. The next step should be a full analysis of the dijet gap-between-jets cross section, including all partonic channels and the interference between different color configurations. The basic ingredients for this analysis are the $2\to 2$ hard functions, given in terms of color matrices in a basis of the color space of the particular channel. The color matrices for the $2\to 2$ hard functions can be found in the literature, and after evaluating the basic color structures in the basis of a given partonic channel, the result for the cross section can be assembled. Such an analysis is interesting and will allow us to realistically assess the importance of the leading SLLs for the first time, albeit with large perturbative uncertainties. 

While the systematics of the ``exponentiation'' of SLLs is still an open issue, we expect that (as in standard Sudakov problems) theoretical predictions obtained in the double-logarithmic approximation suffer from an $\mathcal{O}(1)$ uncertainty. Indeed, already the uncertainty from choosing the scale in the coupling -- which is a single-logarithmic effect -- is of this size. We have included the running of the coupling  to 15-th order in the perturbative expansion of the cross section and find that the effect is similar to evaluating the coupling at the intermediate scale $\bar\mu=\sqrt{Q\spac Q_0}$. However, to achieve single-logarithmic accuracy all pieces of the one-loop anomalous dimension in the path-ordered exponential~\eqref{eq:U} should be iterated to all orders. This includes higher powers of the Glauber terms, additional emissions into the gap, and iterating the purely collinear part of the anomalous dimension. The additional emissions into the gap produce a complicated pattern of single-logarithmic non-global logarithms (NGLs), which interleaves with the double-logarithmic effects of the SLLs. It is well known that the numerical impact of the global and non-global single logarithms is sizeable~\cite{Banfi:2010pa,DuranDelgado:2011tp,Hatta:2013qj,Balsiger:2018ezi}. However, it is an interesting open question how the two effects combine. Since the NGLs involve successively more complicated angular integrals, it will probably not be possible to analyze them analytically, but as a first step one could analyze the two-emission case. To achieve single-logarithmic accuracy, also a better understanding of the low-energy theory becomes mandatory. The consistency of the RG evolution equations implies that the low-energy theory must involve large rapidity logarithms due to a collinear anomaly, which should be resummed as well. 

There are many applications and extensions of the effective-theory formalism developed here. An interesting phenomenological application would be a study of SLLs for Higgs-boson production in vector-boson fusion, where a central jet veto is used to suppress the QCD background. Another extension of the formalism be a study of small-radius jets, for which the non-global structure associated with the jets factors off~\cite{Becher:2015hka,Becher:2016mmh}. However, the SLLs are associated with radiation along the beams and would still be present. A third interesting extension are jets at high rapidity, for which the resummation of logarithms associated with forward scattering would become relevant. The resummation of these logarithms is obtained from BFKL evolution~\cite{Fadin:1975cb,Kuraev:1976ge,Kuraev:1977fs,Balitsky:1978ic} and has been implemented into a dedicated numerical code in~\cite{Andersen:2009nu,Andersen:2009he,Andersen:2011hs,Andersen:2019yzo,Andersen:2023kuj}. It would clearly be very useful to combine BFKL, NGL, and SLL resummations in one coherent framework. We look forward to addressing these interesting open questions in the future and to further develop the theory and phenomenology of non-global observables at hadron colliders.

\pdfbookmark[1]{Acknowledgements}{Acknowledgements}
\begin{acknowledgments}
The authors thank Dominik Schwienbacher for discussions and comments on the manu\-script. 
The research of TB was supported by the Swiss National Science Foundation (SNF) under grant 200020\_182038. The work of MN and MS was supported by the Cluster of Excellence PRISMA$^+$ (Precision Physics, Fundamental Interactions, and Structure of Matter, EXC~2118/1) funded by the German Research Foundation (DFG) under Germany's Excellence Strategy (Project ID 390831469). DYS is supported by the National Science Foundations of China under Grant No.~12275052 and No.~12147101 and the Shanghai Natural Science Foundation under Grant No.~21ZR1406100. This research also received funding from the European Research Council (ERC) under the European Union's Horizon 2022 Research and Innovation Programme (ERC Advanced Grant, Agreement No.~101097780, EFT4jets). TB would like to thank the Pauli Center at ETHZ, the CERN Theory Department and the Munich Institute for Astro-, Particle and BioPhysics (MIAPbP) for hospitality and support. MIAPbP is funded by the German Research Foundation under Germany's Excellence Strategy (EXC~2094, Project ID 390783311). 
\end{acknowledgments}

\begin{appendix}

\section{Operator matrix elements for \texorpdfstring{\bm{$qq\to qq$}}{qq->qq}} \label{app:quark_scattering}

Even though it seems natural to choose an orthonormal basis of color structures for the decomposition of the amplitude in~\eqref{eq:M_color_decomposition}, as for example in~\eqref{eq:ONB_color_structures}, it is sometimes more convenient to work with just an orthogonal basis $\{|\mathcal{C}_I\rangle\}$ so as to avoid square roots of normalization factors. One then needs to carefully keep track of those non-trivial normalization factors, which we write out explicitly in the following. In such a basis, the ``unintegrated'' hard function for generic $2\to2$ scattering is given by the matrix
\begin{align} \label{eq:OB_hard_function}
    \langle\mathcal{C}_I|\widetilde{\bm{\mathcal{H}}}_4|\mathcal{C}_J\rangle = \mathcal{M}_4^{(I)} \mathcal{M}_4^{(J)\ast} \, \langle\mathcal{C}_I|\mathcal{C}_I\rangle \langle\mathcal{C}_J|\mathcal{C}_J\rangle \,,
\end{align}
and the partonic cross section in~\eqref{eq:partonic_cross_section} evaluates to
\begin{align}
    \left(\frac{d\hat{\sigma}}{dr}\right)_{\!2\to2} &= \frac{1}{16\pi \hat{s}} \, \frac{1}{\mathcal{N}_1\mathcal{N}_2} \,\sum_{\text{spins}}\sum_{I,J} \frac{\langle\mathcal{C}_I|\widetilde{\bm{\mathcal{H}}}_4|\mathcal{C}_J\rangle}{\langle\mathcal{C}_I|\mathcal{C}_I\rangle \langle\mathcal{C}_J|\mathcal{C}_J\rangle} \,  \langle\mathcal{C}_J|\bm{1}|\mathcal{C}_I\rangle \,,
\end{align}
with multiplicity factors $\mathcal{N}_i$ given in~\eqref{eq:spin_color_average_factors}. The normalization factors in the denominator cancel the ones on the right-hand side of~\eqref{eq:OB_hard_function}.

For $qq\to qq$ scattering an orthogonal basis could be~\cite{Broggio:2014hoa}
\begin{equation} \label{eq:OB_color_structures}
    \mathcal{C}_1 \equiv  \langle \{\underline{\alpha}\}| \mathcal{C}_1 \rangle  = \delta_{\alpha_3 \alpha_2}\,\delta_{\alpha_4 \alpha_1}  \, , \qquad
    \mathcal{C}_2 \equiv  \langle \{\underline{\alpha}\}| \mathcal{C}_2 \rangle  = t^c_{\alpha_3 \alpha_2}\,t^c_{\alpha_4 \alpha_1} \,,
\end{equation}
where $\langle\mathcal{C}_1|\mathcal{C}_1\rangle=N_c^2$ and $\langle\mathcal{C}_2|\mathcal{C}_2\rangle=C_FN_c/2$. In this basis the low-energy matrix element takes the form
\begin{align}
    \langle\mathcal{C}_J|\bm{1}|\mathcal{C}_I\rangle =
    \begin{pmatrix}
        N_c^2 & 0 \\
        0 & \frac{C_F N_c}{2}
    \end{pmatrix}_{\!\!JI} ,
\end{align}
and the tree-level, spin-averaged hard function is given by~\cite{Broggio:2014hoa}
\begin{align} \label{eq:OB_hard_function_quark_scattering}
\begin{aligned}
    &\frac{1}{4} \sum_{\text{spins}} \frac{\langle\mathcal{C}_I|\widetilde{\bm{\mathcal{H}}}_{qq\to qq}|\mathcal{C}_J\rangle}{\langle\mathcal{C}_I|\mathcal{C}_I\rangle \langle\mathcal{C}_J|\mathcal{C}_J\rangle} \\
    ={} & (4\pi\alpha_s)^2\, \frac{2C_F}{N_c^2 \spac r^2}
    \begin{pmatrix}
        C_F (r^2-2r+2)  &  \frac{r^3-3r^2+(N_c+4)r-2}{1-r} \\
        \frac{r^3-3r^2+(N_c+4)r-2}{1-r} & \frac{(N_c^2+1)r^4-4r^3+(N_c^2+2N_c+7)r^2-2(N_c+3)r+2}{C_F \,(1-r)^2}
    \end{pmatrix}_{\!\!IJ} \,.
\end{aligned}
\end{align}
To evaluate the coefficients $C_{rn}$ in~\eqref{eq:MasterFormula}, we also need the matrix representations
\begin{align}
    \big(\bm{O}_i^{(j)}\big)_{IJ} \equiv \langle\mathcal{C}_I|\bm{O}_i^{(j)}|\mathcal{C}_J\rangle \,, \qquad
    \big(\bm{S}_i\big)_{IJ} \equiv \langle\mathcal{C}_I|\bm{S}_i|\mathcal{C}_J\rangle
\end{align}
in the color basis~\eqref{eq:OB_color_structures}. For the operators $\bm{O}_i^{(j)}$ defined in~\eqref{eq:Oibasis} we find
\begin{align}\label{eq:O_mat}
\begin{aligned}
    \sum_{j=3}^4 \bm{O}_1^{(j)} J_j &=\frac{C_F N_c J_{43}}{2} \left(
\begin{array}{cc}
 -2 N_c & 1 \\
 1 & C_F \\
\end{array}
\right) , \quad &
\sum_{j=3}^4 \bm{O}_2^{(j)} J_j &=
\left(
\begin{array}{ccc}
 0 && 0 \\
 0 && 0 \\
\end{array}
\right), \\
\sum_{j=3}^4 \bm{O}_3^{(j)} J_j &=\frac{C_F J_{43}}{2} \left(
\begin{array}{cc}
 2 N_c & -1 \\
 -1 & -C_F \\
\end{array}
\right),&
\sum_{j=3}^4 \bm{O}_4^{(j)} J_j &=C_F^2 N_c J_{43} \left(
\begin{array}{cc}
 2 N_c & -1 \\
 -1 & -C_F \\
\end{array}
\right) ,
\end{aligned}
\end{align}
where $J_{43} = J_4 - J_3$. The operators~\eqref{eq:Sibasis} multiplying the integral $J_{12}$ are given by
\begin{align}\label{eq:S_mat}
\bm{S}_1&=C_F \left(
\begin{array}{ccc}
 -2 N_c^2 && N_c-\frac{N_c^3}{4} \\
 N_c-\frac{N_c^3}{4} && -\frac{1}{4}\!\left(N_c^2+2\right) \\
\end{array}
\right) , &
\bm{S}_2&=C_F\left(
\begin{array}{ccc}
 0 && \frac{\left(N_c^2-4\right)^2}{2 N_c} \\
 \frac{\left(N_c^2-4\right)^2}{2 N_c} &&
   -\frac{\left(N_c^2-4\right)^2}{2 N_c^2} \\
\end{array}
\right),\nonumber \\
\bm{S}_3&=C_F \left(
\begin{array}{ccc}
 0 && \frac{N_c^3}{6}-\frac{5 N_c}{3}+\frac{4}{N_c} \\
 \frac{N_c^3}{6}-\frac{5 N_c}{3}+\frac{4}{N_c} &&
   -\frac{N_c^2}{6}-\frac{4}{N_c^2}+\frac{5}{3} \\
\end{array}
\right) , \quad &
\bm{S}_4&=C_F \left(
\begin{array}{ccc}
 2 N_c && \frac{1}{2}\!\left(N_c^2-4\right) \\
 \frac{1}{2}\!\left(N_c^2-4\right) && \frac{3}{2 N_c} \\
\end{array}
\right),\nonumber \\
\bm{S}_5&=C_F \left(
\begin{array}{ccc}
 0 && \frac{N_c}{2} \\
 \frac{N_c}{2} && -\frac{1}{2} \\
\end{array}
\right), &
\bm{S}_6&=C_F \left(
\begin{array}{ccc}
 \frac{N_c^2}{C_F} && 0 \\
 0 && \frac{N_c}{2} \\
\end{array}
\right).
\end{align}
We note that the combinations of matrix elements in~\eqref{eq:O_mat} and~\eqref{eq:S_mat} are invariant under the crossing $3 \leftrightarrow 4$ of the final-state legs. In~\eqref{eq:O_mat} both the 
integral $J_{43} \to J_{34}=-J_{43}$ and the color matrix multiplying it pick up sign factors, which compensate in the product.

\end{appendix}

\newpage

\pdfbookmark[1]{References}{Refs}
\bibliography{refs.bib}

\providecommand{\href}[2]{#2}\begingroup\raggedright\begin{thebibliography}{10}

\bibitem{Dasgupta:2001sh}
M.~Dasgupta and G.P.~Salam, \emph{{Resummation of nonglobal QCD observables}},
  \href{https://doi.org/10.1016/S0370-2693(01)00725-0}{\emph{Phys. Lett. B}
  {\bfseries 512} (2001) 323}
  [\href{https://arxiv.org/abs/hep-ph/0104277}{{\ttfamily hep-ph/0104277}}].

\bibitem{Banfi:2002hw}
A.~Banfi, G.~Marchesini and G.~Smye, \emph{{Away from jet energy flow}},
  \href{https://doi.org/10.1088/1126-6708/2002/08/006}{\emph{JHEP} {\bfseries
  08} (2002) 006} [\href{https://arxiv.org/abs/hep-ph/0206076}{{\ttfamily
  hep-ph/0206076}}].

\bibitem{Weigert:2003mm}
H.~Weigert, \emph{{Nonglobal jet evolution at finite $N_c$}},
  \href{https://doi.org/10.1016/j.nuclphysb.2004.03.002}{\emph{Nucl. Phys. B}
  {\bfseries 685} (2004) 321}
  [\href{https://arxiv.org/abs/hep-ph/0312050}{{\ttfamily hep-ph/0312050}}].

\bibitem{Jalilian-Marian:1997qno}
J.~Jalilian-Marian, A.~Kovner, A.~Leonidov and H.~Weigert, \emph{{The BFKL
  equation from the Wilson renormalization group}},
  \href{https://doi.org/10.1016/S0550-3213(97)00440-9}{\emph{Nucl. Phys. B}
  {\bfseries 504} (1997) 415}
  [\href{https://arxiv.org/abs/hep-ph/9701284}{{\ttfamily hep-ph/9701284}}].

\bibitem{Weigert:2000gi}
H.~Weigert, \emph{{Unitarity at small Bjorken x}},
  \href{https://doi.org/10.1016/S0375-9474(01)01668-2}{\emph{Nucl. Phys. A}
  {\bfseries 703} (2002) 823}
  [\href{https://arxiv.org/abs/hep-ph/0004044}{{\ttfamily hep-ph/0004044}}].

\bibitem{Ferreiro:2001qy}
E.~Ferreiro, E.~Iancu, A.~Leonidov and L.~McLerran, \emph{{Nonlinear gluon
  evolution in the color glass condensate: II}},
  \href{https://doi.org/10.1016/S0375-9474(01)01329-X}{\emph{Nucl. Phys. A}
  {\bfseries 703} (2002) 489}
  [\href{https://arxiv.org/abs/hep-ph/0109115}{{\ttfamily hep-ph/0109115}}].

\bibitem{Balitsky:1995ub}
I.~Balitsky, \emph{{Operator expansion for high-energy scattering}},
  \href{https://doi.org/10.1016/0550-3213(95)00638-9}{\emph{Nucl. Phys. B}
  {\bfseries 463} (1996) 99}
  [\href{https://arxiv.org/abs/hep-ph/9509348}{{\ttfamily hep-ph/9509348}}].

\bibitem{Kovchegov:1999ua}
Y.V.~Kovchegov, \emph{{Unitarization of the BFKL pomeron on a nucleus}},
  \href{https://doi.org/10.1103/PhysRevD.61.074018}{\emph{Phys. Rev. D}
  {\bfseries 61} (2000) 074018}
  [\href{https://arxiv.org/abs/hep-ph/9905214}{{\ttfamily hep-ph/9905214}}].

\bibitem{Hatta:2013iba}
Y.~Hatta and T.~Ueda, \emph{{Resummation of non-global logarithms at finite
  $N_c$}}, \href{https://doi.org/10.1016/j.nuclphysb.2013.06.021}{\emph{Nucl.
  Phys. B} {\bfseries 874} (2013) 808}
  [\href{https://arxiv.org/abs/1304.6930}{{\ttfamily 1304.6930}}].

\bibitem{Hagiwara:2015bia}
Y.~Hagiwara, Y.~Hatta and T.~Ueda, \emph{{Hemisphere jet mass distribution at
  finite $N_c$}},
  \href{https://doi.org/10.1016/j.physletb.2016.03.028}{\emph{Phys. Lett. B}
  {\bfseries 756} (2016) 254}
  [\href{https://arxiv.org/abs/1507.07641}{{\ttfamily 1507.07641}}].

\bibitem{Hatta:2020wre}
Y.~Hatta and T.~Ueda, \emph{{Non-global logarithms in hadron collisions at $N_c
  = 3$}}, \href{https://doi.org/10.1016/j.nuclphysb.2020.115273}{\emph{Nucl.
  Phys. B} {\bfseries 962} (2021) 115273}
  [\href{https://arxiv.org/abs/2011.04154}{{\ttfamily 2011.04154}}].

\bibitem{Platzer:2013fha}
S.~Pl\"atzer, \emph{{Summing Large-$N$ Towers in Colour Flow Evolution}},
  \href{https://doi.org/10.1140/epjc/s10052-014-2907-2}{\emph{Eur. Phys. J. C}
  {\bfseries 74} (2014) 2907}
  [\href{https://arxiv.org/abs/1312.2448}{{\ttfamily 1312.2448}}].

\bibitem{AngelesMartinez:2018cfz}
R.~\'Angeles~Mart\'\i{}nez, M.~De~Angelis, J.R.~Forshaw, S.~Pl\"atzer and
  M.H.~Seymour, \emph{{Soft gluon evolution and non-global logarithms}},
  \href{https://doi.org/10.1007/JHEP05(2018)044}{\emph{JHEP} {\bfseries 05}
  (2018) 044} [\href{https://arxiv.org/abs/1802.08531}{{\ttfamily
  1802.08531}}].

\bibitem{Forshaw:2019ver}
J.R.~Forshaw, J.~Holguin and S.~Pl\"atzer, \emph{{Parton branching at amplitude
  level}}, \href{https://doi.org/10.1007/JHEP08(2019)145}{\emph{JHEP}
  {\bfseries 08} (2019) 145}
  [\href{https://arxiv.org/abs/1905.08686}{{\ttfamily 1905.08686}}].

\bibitem{DeAngelis:2020rvq}
M.~De~Angelis, J.R.~Forshaw and S.~Pl\"atzer, \emph{{Resummation and Simulation
  of Soft Gluon Effects beyond Leading Color}},
  \href{https://doi.org/10.1103/PhysRevLett.126.112001}{\emph{Phys. Rev. Lett.}
  {\bfseries 126} (2021) 112001}
  [\href{https://arxiv.org/abs/2007.09648}{{\ttfamily 2007.09648}}].

\bibitem{Nagy:2015hwa}
Z.~Nagy and D.E.~Soper, \emph{{Effects of subleading color in a parton
  shower}}, \href{https://doi.org/10.1007/JHEP07(2015)119}{\emph{JHEP}
  {\bfseries 07} (2015) 119}
  [\href{https://arxiv.org/abs/1501.00778}{{\ttfamily 1501.00778}}].

\bibitem{Nagy:2019bsj}
Z.~Nagy and D.E.~Soper, \emph{{Effect of color on rapidity gap survival}},
  \href{https://doi.org/10.1103/PhysRevD.100.074012}{\emph{Phys. Rev. D}
  {\bfseries 100} (2019) 074012}
  [\href{https://arxiv.org/abs/1905.07176}{{\ttfamily 1905.07176}}].

\bibitem{Nagy:2019rwb}
Z.~Nagy and D.E.~Soper, \emph{{Exponentiating virtual imaginary contributions
  in a parton shower}},
  \href{https://doi.org/10.1103/PhysRevD.100.074005}{\emph{Phys. Rev. D}
  {\bfseries 100} (2019) 074005}
  [\href{https://arxiv.org/abs/1908.11420}{{\ttfamily 1908.11420}}].

\bibitem{Banfi:2021xzn}
A.~Banfi, F.A.~Dreyer and P.F.~Monni, \emph{{Higher-order non-global logarithms
  from jet calculus}},
  \href{https://doi.org/10.1007/JHEP03(2022)135}{\emph{JHEP} {\bfseries 03}
  (2022) 135} [\href{https://arxiv.org/abs/2111.02413}{{\ttfamily
  2111.02413}}].

\bibitem{Banfi:2021owj}
A.~Banfi, F.A.~Dreyer and P.F.~Monni, \emph{{Next-to-leading non-global
  logarithms in QCD}},
  \href{https://doi.org/10.1007/JHEP10(2021)006}{\emph{JHEP} {\bfseries 10}
  (2021) 006} [\href{https://arxiv.org/abs/2104.06416}{{\ttfamily
  2104.06416}}].

\bibitem{Becher:2021urs}
T.~Becher, T.~Rauh and X.~Xu, \emph{{Two-loop anomalous dimension for the
  resummation of non-global observables}},
  \href{https://doi.org/10.1007/JHEP08(2022)134}{\emph{JHEP} {\bfseries 08}
  (2022) 134} [\href{https://arxiv.org/abs/2112.02108}{{\ttfamily
  2112.02108}}].

\bibitem{Becher:2023vrh}
T.~Becher, N.~Schalch and X.~Xu, \emph{{Resummation of Next-to-Leading
  Nonglobal Logarithms at the LHC}},
  \href{https://doi.org/10.1103/PhysRevLett.132.081602}{\emph{Phys. Rev. Lett.}
  {\bfseries 132} (2024) 081602}
  [\href{https://arxiv.org/abs/2307.02283}{{\ttfamily 2307.02283}}].

\bibitem{Forshaw:2006fk}
J.R.~Forshaw, A.~Kyrieleis and M.H.~Seymour, \emph{{Super-leading logarithms in
  non-global observables in QCD}},
  \href{https://doi.org/10.1088/1126-6708/2006/08/059}{\emph{JHEP} {\bfseries
  08} (2006) 059} [\href{https://arxiv.org/abs/hep-ph/0604094}{{\ttfamily
  hep-ph/0604094}}].

\bibitem{Catani:2011st}
S.~Catani, D.~de~Florian and G.~Rodrigo, \emph{{Space-like (versus time-like)
  collinear limits in QCD: Is factorization violated?}},
  \href{https://doi.org/10.1007/JHEP07(2012)026}{\emph{JHEP} {\bfseries 07}
  (2012) 026} [\href{https://arxiv.org/abs/1112.4405}{{\ttfamily 1112.4405}}].

\bibitem{Forshaw:2012bi}
J.R.~Forshaw, M.H.~Seymour and A.~Siodmok, \emph{{On the Breaking of Collinear
  Factorization in QCD}},
  \href{https://doi.org/10.1007/JHEP11(2012)066}{\emph{JHEP} {\bfseries 11}
  (2012) 066} [\href{https://arxiv.org/abs/1206.6363}{{\ttfamily 1206.6363}}].

\bibitem{Schwartz:2017nmr}
M.D.~Schwartz, K.~Yan and H.X.~Zhu, \emph{{Collinear factorization violation
  and effective field theory}},
  \href{https://doi.org/10.1103/PhysRevD.96.056005}{\emph{Phys. Rev. D}
  {\bfseries 96} (2017) 056005}
  [\href{https://arxiv.org/abs/1703.08572}{{\ttfamily 1703.08572}}].

\bibitem{Forshaw:2008cq}
J.R.~Forshaw, A.~Kyrieleis and M.H.~Seymour, \emph{{Super-leading logarithms in
  non-global observables in QCD: Colour basis independent calculation}},
  \href{https://doi.org/10.1088/1126-6708/2008/09/128}{\emph{JHEP} {\bfseries
  09} (2008) 128} [\href{https://arxiv.org/abs/0808.1269}{{\ttfamily
  0808.1269}}].

\bibitem{Keates:2009dn}
J.~Keates and M.H.~Seymour, \emph{{Super-leading logarithms in non-global
  observables in QCD: Fixed order calculation}},
  \href{https://doi.org/10.1088/1126-6708/2009/04/040}{\emph{JHEP} {\bfseries
  04} (2009) 040} [\href{https://arxiv.org/abs/0902.0477}{{\ttfamily
  0902.0477}}].

\bibitem{Forshaw:2009fz}
J.~Forshaw, J.~Keates and S.~Marzani, \emph{{Jet vetoing at the LHC}},
  \href{https://doi.org/10.1088/1126-6708/2009/07/023}{\emph{JHEP} {\bfseries
  07} (2009) 023} [\href{https://arxiv.org/abs/0905.1350}{{\ttfamily
  0905.1350}}].

\bibitem{Becher:2015hka}
T.~Becher, M.~Neubert, L.~Rothen and D.Y.~Shao, \emph{{Effective Field Theory
  for Jet Processes}},
  \href{https://doi.org/10.1103/PhysRevLett.116.192001}{\emph{Phys. Rev. Lett.}
  {\bfseries 116} (2016) 192001}
  [\href{https://arxiv.org/abs/1508.06645}{{\ttfamily 1508.06645}}].

\bibitem{Becher:2016mmh}
T.~Becher, M.~Neubert, L.~Rothen and D.Y.~Shao, \emph{{Factorization and
  Resummation for Jet Processes}},
  \href{https://doi.org/10.1007/JHEP11(2016)019}{\emph{JHEP} {\bfseries 11}
  (2016) 019} [\href{https://arxiv.org/abs/1605.02737}{{\ttfamily
  1605.02737}}].

\bibitem{Bauer:2001yt}
C.W.~Bauer, D.~Pirjol and I.W.~Stewart, \emph{{Soft collinear factorization in
  effective field theory}},
  \href{https://doi.org/10.1103/PhysRevD.65.054022}{\emph{Phys. Rev. D}
  {\bfseries 65} (2002) 054022}
  [\href{https://arxiv.org/abs/hep-ph/0109045}{{\ttfamily hep-ph/0109045}}].

\bibitem{Bauer:2002nz}
C.W.~Bauer, S.~Fleming, D.~Pirjol, I.Z.~Rothstein and I.W.~Stewart, \emph{{Hard
  scattering factorization from effective field theory}},
  \href{https://doi.org/10.1103/PhysRevD.66.014017}{\emph{Phys. Rev. D}
  {\bfseries 66} (2002) 014017}
  [\href{https://arxiv.org/abs/hep-ph/0202088}{{\ttfamily hep-ph/0202088}}].

\bibitem{Beneke:2002ph}
M.~Beneke, A.P.~Chapovsky, M.~Diehl and T.~Feldmann, \emph{{Soft collinear
  effective theory and heavy to light currents beyond leading power}},
  \href{https://doi.org/10.1016/S0550-3213(02)00687-9}{\emph{Nucl. Phys. B}
  {\bfseries 643} (2002) 431}
  [\href{https://arxiv.org/abs/hep-ph/0206152}{{\ttfamily hep-ph/0206152}}].

\bibitem{Balsiger:2018ezi}
M.~Balsiger, T.~Becher and D.Y.~Shao, \emph{{Non-global logarithms in jet and
  isolation cone cross sections}},
  \href{https://doi.org/10.1007/JHEP08(2018)104}{\emph{JHEP} {\bfseries 08}
  (2018) 104} [\href{https://arxiv.org/abs/1803.07045}{{\ttfamily
  1803.07045}}].

\bibitem{Becher:2021zkk}
T.~Becher, M.~Neubert and D.Y.~Shao, \emph{{Resummation of Super-Leading
  Logarithms}},
  \href{https://doi.org/10.1103/PhysRevLett.127.212002}{\emph{Phys. Rev. Lett.}
  {\bfseries 127} (2021) 212002}
  [\href{https://arxiv.org/abs/2107.01212}{{\ttfamily 2107.01212}}].

\bibitem{Magnea:1990zb}
L.~Magnea and G.F.~Sterman, \emph{{Analytic continuation of the Sudakov
  form-factor in QCD}},
  \href{https://doi.org/10.1103/PhysRevD.42.4222}{\emph{Phys. Rev. D}
  {\bfseries 42} (1990) 4222}.

\bibitem{Ahrens:2008qu}
V.~Ahrens, T.~Becher, M.~Neubert and L.L.~Yang, \emph{{Origin of the Large
  Perturbative Corrections to Higgs Production at Hadron Colliders}},
  \href{https://doi.org/10.1103/PhysRevD.79.033013}{\emph{Phys. Rev. D}
  {\bfseries 79} (2009) 033013}
  [\href{https://arxiv.org/abs/0808.3008}{{\ttfamily 0808.3008}}].

\bibitem{Ahrens:2009cxz}
V.~Ahrens, T.~Becher, M.~Neubert and L.L.~Yang, \emph{{Renormalization-Group
  Improved Prediction for Higgs Production at Hadron Colliders}},
  \href{https://doi.org/10.1140/epjc/s10052-009-1030-2}{\emph{Eur. Phys. J. C}
  {\bfseries 62} (2009) 333} [\href{https://arxiv.org/abs/0809.4283}{{\ttfamily
  0809.4283}}].

\bibitem{Catani:1996vz}
S.~Catani and M.H.~Seymour, \emph{{A General algorithm for calculating jet
  cross-sections in NLO QCD}},
  \href{https://doi.org/10.1016/S0550-3213(96)00589-5}{\emph{Nucl. Phys. B}
  {\bfseries 485} (1997) 291}
  [\href{https://arxiv.org/abs/hep-ph/9605323}{{\ttfamily hep-ph/9605323}}].

\bibitem{Hill:2002vw}
R.J.~Hill and M.~Neubert, \emph{{Spectator interactions in soft collinear
  effective theory}},
  \href{https://doi.org/10.1016/S0550-3213(03)00116-0}{\emph{Nucl. Phys. B}
  {\bfseries 657} (2003) 229}
  [\href{https://arxiv.org/abs/hep-ph/0211018}{{\ttfamily hep-ph/0211018}}].

\bibitem{ATLAS:2011yyh}
{\scshape ATLAS} collaboration, \emph{{Measurement of dijet production with a
  veto on additional central jet activity in $pp$ collisions at $\sqrt{s}=7$
  TeV using the ATLAS detector}},
  \href{https://doi.org/10.1007/JHEP09(2011)053}{\emph{JHEP} {\bfseries 09}
  (2011) 053} [\href{https://arxiv.org/abs/1107.1641}{{\ttfamily 1107.1641}}].

\bibitem{ATLAS:2014lzu}
{\scshape ATLAS} collaboration, \emph{{Measurements of jet vetoes and azimuthal
  decorrelations in dijet events produced in $pp$ collisions at
  $\sqrt{s}=7\,\mathrm{TeV}$ using the ATLAS detector}},
  \href{https://doi.org/10.1140/epjc/s10052-014-3117-7}{\emph{Eur. Phys. J. C}
  {\bfseries 74} (2014) 3117}
  [\href{https://arxiv.org/abs/1407.5756}{{\ttfamily 1407.5756}}].

\bibitem{Rothstein:2016bsq}
I.Z.~Rothstein and I.W.~Stewart, \emph{{An Effective Field Theory for Forward
  Scattering and Factorization Violation}},
  \href{https://doi.org/10.1007/JHEP08(2016)025}{\emph{JHEP} {\bfseries 08}
  (2016) 025} [\href{https://arxiv.org/abs/1601.04695}{{\ttfamily
  1601.04695}}].

\bibitem{Collins:1983ju}
J.C.~Collins, D.E.~Soper and G.F.~Sterman, \emph{{All Order Factorization for
  {Drell-Yan} Cross-sections}},
  \href{https://doi.org/10.1016/0370-2693(84)90684-1}{\emph{Phys. Lett. B}
  {\bfseries 134} (1984) 263}.

\bibitem{Collins:1988ig}
J.C.~Collins, D.E.~Soper and G.F.~Sterman, \emph{{Soft Gluons and
  Factorization}},
  \href{https://doi.org/10.1016/0550-3213(88)90130-7}{\emph{Nucl. Phys. B}
  {\bfseries 308} (1988) 833}.

\bibitem{Collins:1989gx}
J.C.~Collins, D.E.~Soper and G.F.~Sterman, \emph{{Factorization of Hard
  Processes in QCD}},
  \href{https://doi.org/10.1142/9789814503266_0001}{\emph{Adv. Ser. Direct.
  High Energy Phys.} {\bfseries 5} (1989) 1}
  [\href{https://arxiv.org/abs/hep-ph/0409313}{{\ttfamily hep-ph/0409313}}].

\bibitem{Becher:2010tm}
T.~Becher and M.~Neubert, \emph{{Drell-Yan Production at Small $q_T$,
  Transverse Parton Distributions and the Collinear Anomaly}},
  \href{https://doi.org/10.1140/epjc/s10052-011-1665-7}{\emph{Eur. Phys. J. C}
  {\bfseries 71} (2011) 1665}
  [\href{https://arxiv.org/abs/1007.4005}{{\ttfamily 1007.4005}}].

\bibitem{Chiu:2012ir}
J.-Y.~Chiu, A.~Jain, D.~Neill and I.Z.~Rothstein, \emph{{A Formalism for the
  Systematic Treatment of Rapidity Logarithms in Quantum Field Theory}},
  \href{https://doi.org/10.1007/JHEP05(2012)084}{\emph{JHEP} {\bfseries 05}
  (2012) 084} [\href{https://arxiv.org/abs/1202.0814}{{\ttfamily 1202.0814}}].

\bibitem{Becher:2009cu}
T.~Becher and M.~Neubert, \emph{{Infrared singularities of scattering
  amplitudes in perturbative QCD}},
  \href{https://doi.org/10.1103/PhysRevLett.102.162001}{\emph{Phys. Rev. Lett.}
  {\bfseries 102} (2009) 162001}
  [\href{https://arxiv.org/abs/0901.0722}{{\ttfamily 0901.0722}}].

\bibitem{Gardi:2009qi}
E.~Gardi and L.~Magnea, \emph{{Factorization constraints for soft anomalous
  dimensions in QCD scattering amplitudes}},
  \href{https://doi.org/10.1088/1126-6708/2009/03/079}{\emph{JHEP} {\bfseries
  03} (2009) 079} [\href{https://arxiv.org/abs/0901.1091}{{\ttfamily
  0901.1091}}].

\bibitem{Becher:2009qa}
T.~Becher and M.~Neubert, \emph{{On the Structure of Infrared Singularities of
  Gauge-Theory Amplitudes}},
  \href{https://doi.org/10.1088/1126-6708/2009/06/081}{\emph{JHEP} {\bfseries
  06} (2009) 081} [\href{https://arxiv.org/abs/0903.1126}{{\ttfamily
  0903.1126}}].

\bibitem{Dixon:2009ur}
L.J.~Dixon, E.~Gardi and L.~Magnea, \emph{{On soft singularities at three loops
  and beyond}}, \href{https://doi.org/10.1007/JHEP02(2010)081}{\emph{JHEP}
  {\bfseries 02} (2010) 081} [\href{https://arxiv.org/abs/0910.3653}{{\ttfamily
  0910.3653}}].

\bibitem{Ahrens:2012qz}
V.~Ahrens, M.~Neubert and L.~Vernazza, \emph{{Structure of Infrared
  Singularities of Gauge-Theory Amplitudes at Three and Four Loops}},
  \href{https://doi.org/10.1007/JHEP09(2012)138}{\emph{JHEP} {\bfseries 09}
  (2012) 138} [\href{https://arxiv.org/abs/1208.4847}{{\ttfamily 1208.4847}}].

\bibitem{Becher:2019avh}
T.~Becher and M.~Neubert, \emph{{Infrared singularities of scattering
  amplitudes and N$^{3}$LL resummation for $n$-jet processes}},
  \href{https://doi.org/10.1007/JHEP01(2020)025}{\emph{JHEP} {\bfseries 01}
  (2020) 025} [\href{https://arxiv.org/abs/1908.11379}{{\ttfamily
  1908.11379}}].

\bibitem{Catani:2003vu}
S.~Catani, D.~de~Florian and G.~Rodrigo, \emph{{The Triple collinear limit of
  one loop QCD amplitudes}},
  \href{https://doi.org/10.1016/j.physletb.2004.02.039}{\emph{Phys. Lett. B}
  {\bfseries 586} (2004) 323}
  [\href{https://arxiv.org/abs/hep-ph/0312067}{{\ttfamily hep-ph/0312067}}].

\bibitem{Catani:1999ss}
S.~Catani and M.~Grazzini, \emph{{Infrared factorization of tree level QCD
  amplitudes at the next-to-next-to-leading order and beyond}},
  \href{https://doi.org/10.1016/S0550-3213(99)00778-6}{\emph{Nucl. Phys. B}
  {\bfseries 570} (2000) 287}
  [\href{https://arxiv.org/abs/hep-ph/9908523}{{\ttfamily hep-ph/9908523}}].

\bibitem{Haber:2019sgz}
H.E.~Haber, \emph{{Useful relations among the generators in the defining and
  adjoint representations of SU(N)}},
  \href{https://doi.org/10.21468/SciPostPhysLectNotes.21}{\emph{SciPost Phys.
  Lect. Notes} {\bfseries 21} (2021) 1}
  [\href{https://arxiv.org/abs/1912.13302}{{\ttfamily 1912.13302}}].

\bibitem{Kelley:2010fn}
R.~Kelley and M.D.~Schwartz, \emph{{1-loop matching and NNLL resummation for
  all partonic 2 to 2 processes in QCD}},
  \href{https://doi.org/10.1103/PhysRevD.83.045022}{\emph{Phys. Rev. D}
  {\bfseries 83} (2011) 045022}
  [\href{https://arxiv.org/abs/1008.2759}{{\ttfamily 1008.2759}}].

\bibitem{Broggio:2014hoa}
A.~Broggio, A.~Ferroglia, B.D.~Pecjak and Z.~Zhang, \emph{{NNLO hard functions
  in massless QCD}}, \href{https://doi.org/10.1007/JHEP12(2014)005}{\emph{JHEP}
  {\bfseries 12} (2014) 005} [\href{https://arxiv.org/abs/1409.5294}{{\ttfamily
  1409.5294}}].

\bibitem{Balsiger:2019tne}
M.~Balsiger, T.~Becher and D.Y.~Shao, \emph{{NLL${'}$ resummation of jet
  mass}}, \href{https://doi.org/10.1007/JHEP04(2019)020}{\emph{JHEP} {\bfseries
  04} (2019) 020} [\href{https://arxiv.org/abs/1901.09038}{{\ttfamily
  1901.09038}}].

\bibitem{Forshaw:2021fxs}
J.R.~Forshaw and J.~Holguin, \emph{{Coulomb gluons will generally destroy
  coherence}}, \href{https://doi.org/10.1007/JHEP12(2021)084}{\emph{JHEP}
  {\bfseries 12} (2021) 084}
  [\href{https://arxiv.org/abs/2109.03665}{{\ttfamily 2109.03665}}].

\bibitem{Bauer:2010cc}
C.W.~Bauer, B.O.~Lange and G.~Ovanesyan, \emph{{On Glauber modes in
  Soft-Collinear Effective Theory}},
  \href{https://doi.org/10.1007/JHEP07(2011)077}{\emph{JHEP} {\bfseries 07}
  (2011) 077} [\href{https://arxiv.org/abs/1010.1027}{{\ttfamily 1010.1027}}].

\bibitem{Gaunt:2014ska}
J.R.~Gaunt, \emph{{Glauber Gluons and Multiple Parton Interactions}},
  \href{https://doi.org/10.1007/JHEP07(2014)110}{\emph{JHEP} {\bfseries 07}
  (2014) 110} [\href{https://arxiv.org/abs/1405.2080}{{\ttfamily 1405.2080}}].

\bibitem{Zeng:2015iba}
M.~Zeng, \emph{{Drell-Yan process with jet vetoes: breaking of generalized
  factorization}}, \href{https://doi.org/10.1007/JHEP10(2015)189}{\emph{JHEP}
  {\bfseries 10} (2015) 189}
  [\href{https://arxiv.org/abs/1507.01652}{{\ttfamily 1507.01652}}].

\bibitem{Bijl:2023dux}
P.~Bijl, S.~Niedenzu and W.J.~Waalewijn, \emph{{Probing factorization violation
  with vector angularities}},
  \href{https://doi.org/10.1103/PhysRevD.109.014011}{\emph{Phys. Rev. D}
  {\bfseries 109} (2024) 014011}
  [\href{https://arxiv.org/abs/2307.02521}{{\ttfamily 2307.02521}}].

\bibitem{Banfi:2010pa}
A.~Banfi, M.~Dasgupta, K.~Khelifa-Kerfa and S.~Marzani, \emph{{Non-global
  logarithms and jet algorithms in high-pT jet shapes}},
  \href{https://doi.org/10.1007/JHEP08(2010)064}{\emph{JHEP} {\bfseries 08}
  (2010) 064} [\href{https://arxiv.org/abs/1004.3483}{{\ttfamily 1004.3483}}].

\bibitem{DuranDelgado:2011tp}
R.M.~Duran~Delgado, J.R.~Forshaw, S.~Marzani and M.H.~Seymour, \emph{{The dijet
  cross section with a jet veto}},
  \href{https://doi.org/10.1007/JHEP08(2011)157}{\emph{JHEP} {\bfseries 08}
  (2011) 157} [\href{https://arxiv.org/abs/1107.2084}{{\ttfamily 1107.2084}}].

\bibitem{Hatta:2013qj}
Y.~Hatta, C.~Marquet, C.~Royon, G.~Soyez, T.~Ueda and D.~Werder, \emph{{A QCD
  description of the ATLAS jet veto measurement}},
  \href{https://doi.org/10.1103/PhysRevD.87.054016}{\emph{Phys. Rev. D}
  {\bfseries 87} (2013) 054016}
  [\href{https://arxiv.org/abs/1301.1910}{{\ttfamily 1301.1910}}].

\bibitem{Fadin:1975cb}
V.S.~Fadin, E.A.~Kuraev and L.N.~Lipatov, \emph{{On the Pomeranchuk Singularity
  in Asymptotically Free Theories}},
  \href{https://doi.org/10.1016/0370-2693(75)90524-9}{\emph{Phys. Lett. B}
  {\bfseries 60} (1975) 50}.

\bibitem{Kuraev:1976ge}
E.A.~Kuraev, L.N.~Lipatov and V.S.~Fadin, \emph{{Multi - Reggeon Processes in
  the Yang-Mills Theory}}, {\emph{Sov. Phys. JETP} {\bfseries 44} (1976) 443}.

\bibitem{Kuraev:1977fs}
E.A.~Kuraev, L.N.~Lipatov and V.S.~Fadin, \emph{{The Pomeranchuk Singularity in
  Nonabelian Gauge Theories}}, {\emph{Sov. Phys. JETP} {\bfseries 45} (1977)
  199}.

\bibitem{Balitsky:1978ic}
I.I.~Balitsky and L.N.~Lipatov, \emph{{The Pomeranchuk Singularity in Quantum
  Chromodynamics}}, {\emph{Sov. J. Nucl. Phys.} {\bfseries 28} (1978) 822}.

\bibitem{Andersen:2009nu}
J.R.~Andersen and J.M.~Smillie, \emph{{Constructing All-Order Corrections to
  Multi-Jet Rates}}, \href{https://doi.org/10.1007/JHEP01(2010)039}{\emph{JHEP}
  {\bfseries 01} (2010) 039} [\href{https://arxiv.org/abs/0908.2786}{{\ttfamily
  0908.2786}}].

\bibitem{Andersen:2009he}
J.R.~Andersen and J.M.~Smillie, \emph{{The Factorisation of the t-channel Pole
  in Quark-Gluon Scattering}},
  \href{https://doi.org/10.1103/PhysRevD.81.114021}{\emph{Phys. Rev. D}
  {\bfseries 81} (2010) 114021}
  [\href{https://arxiv.org/abs/0910.5113}{{\ttfamily 0910.5113}}].

\bibitem{Andersen:2011hs}
J.R.~Andersen and J.M.~Smillie, \emph{{Multiple Jets at the LHC with High
  Energy Jets}}, \href{https://doi.org/10.1007/JHEP06(2011)010}{\emph{JHEP}
  {\bfseries 06} (2011) 010} [\href{https://arxiv.org/abs/1101.5394}{{\ttfamily
  1101.5394}}].

\bibitem{Andersen:2019yzo}
J.R.~Andersen, T.~Hapola, M.~Heil, A.~Maier and J.~Smillie, \emph{{HEJ 2: High
  Energy Resummation for Hadron Colliders}},
  \href{https://doi.org/10.1016/j.cpc.2019.06.022}{\emph{Comput. Phys. Commun.}
  {\bfseries 245} (2019) } [\href{https://arxiv.org/abs/1902.08430}{{\ttfamily
  1902.08430}}].

\bibitem{Andersen:2023kuj}
J.R.~Andersen, B.~Duclou\'e, C.~Elrick, H.~Hassan, A.~Maier, G.~Nail et~al.,
  \emph{{HEJ 2.2: W boson pairs and Higgs boson plus jet production at high
  energies}},  \href{https://arxiv.org/abs/2303.15778}{{\ttfamily 2303.15778}}.

\end{thebibliography}\endgroup
   
\end{document}